\documentclass[fleqn,usenatbib]{mnras}

\usepackage{newtxtext,newtxmath}

\usepackage[T1]{fontenc}
\usepackage{ae,aecompl}
\usepackage{bm}
\usepackage{orcidlink}

\newcommand{\markus}[1]{{\color{black} #1}}
\newcommand{\mmref}[1]{{\color{black} #1}}
\newcommand{\mmreff}[1]{{\color{black} #1}}
\newcommand{\rev}[1]{{\color{black} #1}}
\usepackage[mathlines]{lineno} 
\let\oldequation\equation
\let\oldendequation\endequation

\renewenvironment{equation}
  {\linenomathNonumbers\oldequation}
  {\oldendequation\endlinenomath}

 \let\oldalign\align
\let\oldendalign\endalign

\renewenvironment{align}
  {\linenomathNonumbers\oldalign}
  {\oldendalign\endlinenomath}

\usepackage{graphicx}	
\usepackage{amsmath}	
\usepackage{enumitem}
\usepackage[ruled,vlined]{algorithm2e}
\usepackage{mathtools}
\usepackage{bbm}
\usepackage[ruled,vlined]{algorithm2e}

\usepackage{lipsum,eso-pic,xcolor}

\newcommand{\numberlist}[1][r]{%
  \begin{tabular}{#1}
    \strut
     1 \\  2 \\  3 \\  4 \\  5 \\  6 \\  7 \\  8 \\  9 \\ 10 \\
    11 \\ 12 \\ 13 \\ 14 \\ 15 \\ 16 \\ 17 \\ 18 \\ 19 \\ 20 \\
    21 \\ 22 \\ 23 \\ 24 \\ 25 \\ 26 \\ 27 \\ 28 \\ 29 \\ 20 \\
    31 \\ 32 \\ 33 \\ 34 \\ 35 \\ 36 \\ 37 \\ 38 \\ 39 \\ 30 \\
    41 \\ 42 \\ 43 \\ 44 \\ 45 \\ 46
  \end{tabular}%
}

\newcommand{\numberlistfont}{%
  \ttfamily\color{black!50}
}

\AddToShipoutPictureFG{%
  \AtTextUpperLeft{%
    \numberlistfont
    \makebox[0pt][r]{\raisebox{-\height}{%
      \numberlist[r]
    }%
    \hspace{50pt}
    }%
    \hspace*{\textwidth}
    \hspace{3em}
    \makebox[0pt][l]{\raisebox{-\height}{%
      \numberlist[l]
    }}%
  }%
}

\newcommand{\angstrom}{\text{\normalfont\AA}}
\DeclareMathOperator\erf{erf}

\title[Likelihood Inference under Photo-z error]{A Composite Likelihood Approach for Inference under Photometric Redshift Uncertainty}

\author[]{M.~M.~Rau$^{12}$\thanks{E-mail: markusmichael.rau@googlemail.com}\orcidlink{https://orcid.org/0000-0003-3709-1324}, C.~B.~Morrison$^{3}$, S.~J.~Schmidt$^{5}$\orcidlink{0000-0002-5091-0470}, S.~Wilson$^{4}$,  R.~Mandelbaum$^{1}$\orcidlink{0000-0003-2271-1527}, \newauthor Y.-Y.~Mao$^{6}$\orcidlink{0000-0002-1200-0820} for the LSST Dark Energy Science Collaboration
\\
$^{1}$McWilliams Center for Cosmology, Department of Physics, Carnegie Mellon University, Pittsburgh, PA 15213\\
$^{2}$High Energy Physics Division, Argonne National Laboratory, Lemont, IL 60439, USA\\
$^{3}$Department of Astronomy, University of Washington, Box 351580, Seattle, WA 98195, USA\\
$^{4}$School of Computer Science and Statistics,
Lloyd Institute,
Trinity College,
Dublin,
Ireland\\
$^{5}$Department of Physics, 
University of California, 
Davis, CA 95616, 
USA\\
$^{6}$Department of Physics and Astronomy, Rutgers, The State University of New Jersey, Piscataway, NJ 08854, USA \\
}

\date{Accepted XXX. Received YYY; in original form ZZZ}

\pubyear{2015}

\begin{document}
\label{firstpage}
\pagerange{\pageref{firstpage}--\pageref{lastpage}}
\maketitle

\begin{abstract}
Obtaining accurately calibrated redshift distributions of photometric samples is one of the great challenges in photometric surveys like LSST, Euclid, HSC, KiDS, and DES. We \rev{present an inference methodology that combines} the redshift information from the galaxy photometry with constraints from two-point functions, utilizing cross-correlations with spatially overlapping spectroscopic samples\rev{, and illustrate the approach on CosmoDC2 simulations}. Our likelihood framework is designed to integrate directly into a typical large-scale structure and weak lensing analysis based on two-point functions. We discuss efficient and accurate inference techniques that allow us to scale the method to the large samples of galaxies to be expected in LSST. We consider statistical challenges like the parametrization of redshift systematics, discuss and evaluate techniques to regularize the sample redshift distributions, and investigate techniques that can help to detect and calibrate sources of systematic error using posterior predictive checks. We evaluate and forecast photometric redshift performance using data from the CosmoDC2 simulations, within which we mimic a DESI-like spectroscopic calibration sample for cross-correlations. Using a combination of spatial cross-correlations and photometry, we show that we can provide calibration of the mean of the sample redshift distribution to an accuracy of at least $0.002(1+z)$, consistent with the LSST-Y1 science requirements for weak lensing and large-scale structure probes.   
\end{abstract}




\section{Introduction}
With ongoing and future large area photometric surveys like the Dark Energy Survey \citep[DES; e.g.,][]{2018ApJS..239...18A}, the Kilo-Degree Survey  \citep[KiDS; e.g.,][]{2017MNRAS.465.1454H}, the Hyper Suprime-Cam  \citep[HSC; e.g.,][]{2018PASJ...70S...4A}, the Rubin Observatory Legacy Survey of Space and Time  \citep[LSST; e.g.,][]{2019ApJ...873..111I}, the Roman Space Telescope \citep[e.g.][]{2015arXiv150303757S} and Euclid \citep[e.g.][]{2011arXiv1110.3193L} modern cosmology has entered the era of precision cosmology, where it becomes increasingly important to accurately account for sources of systematic bias and uncertainty \citep[e.g.][]{2018ARA&A..56..393M}. 
Large area photometric surveys constrain cosmological parameters and the growth of structure using two-point statistics of galaxy and shear fields \citep[see e.g.][]{2017MNRAS.465.1454H, UitertKids, 2018PhRvD..98d3526A, 2018MNRAS.474.4894J,  2019PASJ...71...43H, 2020arXiv200715632H}. Using only the broadband photometry of galaxies allows for a limited accuracy in the estimated redshifts. In photometric surveys, we therefore typically consider two-point statistics of density fields that have been projected along the line-of-sight, i.e., in the redshift direction. They are then subsequently compared with the corresponding weak lensing (WL) and large scale structure (LSS) theory predictions in a likelihood framework. These theory predictions have to account for the line-of-sight projection, and therefore depend on the redshift distribution of the galaxies in the sample that have to be accurately modelled and calibrated \citep[see e.g.][]{10.1111/j.1365-2966.2005.09782.x, Hoyle_2018, 2018PASJ...70S...9T, 2021A&A...647A.124H, 2020A&A...638L...1J}.  

A primary goal of large area photometric survey programs is to map the growth of structure and expansion history of the Universe, and thereby constrain the dark energy equation of state via the distance-redshift and growth-redshift relations \citep[see e.g.,][p. 31]{dark_energy_taskforce} which both enter the WL and LSS modelling. Note that these fundamental relationships within our cosmological model are redshift dependent, as are some key sources of theoretical uncertainty, such as the galaxy-dark matter bias model \citep[see e.g.][]{1997MNRAS.286..115M, 2015MNRAS.448.1389C, 2016MNRAS.459.3203C, 2018A&A...613A..15S, 2018MNRAS.473.1667P}. The inferred ensemble redshift distributions for samples of galaxies can therefore exhibit a degeneracy with cosmological or astrophysical parameters.   
Inaccurate distance, or redshift, measurements based on the photometry of the galaxies are therefore important modelling systematics in these surveys \citep[e.g.][]{2006ApJ...636...21M, 2010MNRAS.401.1399B}. 
We therefore must exploit all data sources that have the potential to break these degeneracies to perform efficient and accurate inference.

The two \mmref{methods} available to constrain the redshift of galaxies in the absence of accurate spectroscopic measurements are `template fitting' methods and empirical methods that `learn' the mapping between photometry and redshift \citep[for a recent review, see][]{2019NatAs...3..212S}. SED fitting methods fit the galaxy photometry with models of the galaxy spectral energy distribution \citep[SED; e.g.,][]{1999MNRAS.310..540A, 2000ApJ...536..571B,2006A&A...457..841I, 2006MNRAS.372..565F, 2015MNRAS.451.1848G, 2016MNRAS.460.4258L, 2020arXiv200712178M}. Machine Learning-based methods infer photometric redshifts by constructing a density estimate for the conditional distribution of the galaxy redshifts given their photometry  \citep{2003LNCS.2859..226T, 2004PASP..116..345C, 2010ApJ...715..823G, 2013MNRAS.432.1483C,  2015MNRAS.449.1043B, 2015MNRAS.452.3710R, 2016A&C....16...34H}. Combinations of both these techniques have also been investigated \citep{2015arXiv151008073S, 2015MNRAS.450..305H}. Unfortunately the accuracy of these techniques is limited since they suffer from different sources of systematic error. Template fitting approaches can be systematically biased, if fits are constructed using sets of spectral energy distributions that are not representative of all galaxies in the sample. In contrast, photometric redshift techniques that require a training set can produce systematically biased results due to incomplete spectroscopic training samples. It is particularly difficult to obtain representative spectroscopic data due to the long exposure times that are necessary to obtain accurate spectroscopic redshifts for faint sources \citep[see e.g.][]{10.1093/mnras/stu1424, 2015APh....63...81N}.

Instead of inferring photometric redshifts by fitting models for the spectral energy distribution, we can also infer photometric redshift infromation using spatial cross-correlations between photometric samples and spectroscopic samples \citep[e.g.][]{2008ApJ...684...88N, 2013arXiv1303.4722M, 2013MNRAS.433.2857M, 2016MNRAS.462.1683S, 10.1093/mnras/stx691, Morrison2016, 2017arXiv171002517D, 2018MNRAS.477.1664G, 2020A&A...642A.200V, 2021A&A...647A.124H}. Cross-correlation methods measure the spatial cross correlation between a reference sample with accurate redshift information, typically spectroscopic galaxy catalogs, and photometric samples that do not have accurate redshift information. Ignoring cosmic magnification effects \citep[see e.g.][]{2005ApJ...633..589S} the \mmref{expected} spatial cross correlation is only nonzero for samples at the same redshift. By cross correlating subsamples of spectroscopic samples that are selected in thin redshift slices with these photometric catalogs and comparing the resulting signals, we can reconstruct the redshift distribution of the unknown photometric sample. 

It is important to highlight the different sources of systematic uncertainty in these two approaches: the measurement of spatial cross correlations requires that the sample with unknown redshift information and the reference sample overlap spatially and cover the same redshift range. However, the spectroscopic calibration sample does not have to cover the same color/magnitude space as the unknown photometric sample. It is, however, important to accurately model the redshift-dependent galaxy-dark matter bias of the photometric sample and the spectroscopic calibration sample, since the redshift-dependent ratio between these two functions is completely degenerate with the photometric redshift distribution to be inferred. In contrast, template-based redshift inference requires a complete set of templates but no calibration sample. Checking a fitted model can also, in principle, use the color space alone, by comparing the photometry generated by the fitted templates with the measurements. In practice this approach has limitations. The generation of SED model templates is challenging and often requires spectroscopic reference data for some galaxies. Furthermore, degeneracies between galaxy type and galaxy redshift can make the aforementioned color-based approach ill-defined. Thus, while template fitting does not require spectroscopic data to infer redshifts of galaxies, in practise it is often necessary for building and evaluating models.  Finally, empirical techniques that construct photometric redshift estimates by `learning' from a spectroscopic calibration dataset require reference data that does not have to spatially overlap, but needs to be \mmref{representative} in color-redshift space.

\rev{Besides spatial correlations of galaxy clustering, we can also use other two-point statistics from e.g., weak gravitational lensing \citep[e.g.][]{2013MNRAS.431.1547B, 2020arXiv201207707S}, or geometrical probes like shear-ratios \citep[e.g.][]{2019MNRAS.487.1363P, 2021A&A...645A.105G, 2021arXiv210513542S} to extract redshift information. The modelling of any statistic extracted from a projected cosmological density field will depend on the sample redshift distribution and therefore can be used in a joint likelihood for `self-calibration' in the cosmological analysis.} There also exists a considerable literature in how photometric redshift uncertainty can be treated in the individual cosmological probes \citep{2017MNRAS.466.3558M, 2018arXiv180202581H} or how one can combine template fitting and cross correlation measurements \citep{2019arXiv191007127A, 2019MNRAS.483.2801S, 2019MNRAS.483.2487J, 2020MNRAS.491.4768R}. {\color{black} Shortly before this paper was submitted for publication, \citet{2020arXiv201208566M, 2020arXiv201208569G, 2020arXiv201212826C} presented the redshift inference scheme for the DES Y3 analyses, that combines a cross-correlation and shear ratio data vector with redshift information derived using an empirical mapping of broad band `Wide field' photometry to spatially smaller calibration fields with narrow-band photometric and spectroscopic redshift information. }

\rev{This paper presents a composite likelihood approach to enable consistent inference of photometric sample redshift distributions using both the available photometry and the clustering of galaxies. This provides opportunities to break degeneracies in the modelling of galaxy-dark matter bias, the modelling of Spectral Energy Distributions and cosmological parameters.} We focus on statistical challenges in this inference. In particular, the parts of the model that utilize the photometry of galaxies can pose computational challenges, since the likelihood depends on measurements of all galaxies in the sample. We therefore derive an efficient methodology that facilitates inference of redshift distributions within this computationally expensive part of the model. Redshift inference based on noisy photometry is an inverse problem and the inference scheme requires careful regularization to achieve good probability coverage. We therefore describe several regularization techniques and evaluate their respective merits in numerical experiments. Information from the spatial distribution of galaxies can then be incorporated within the composite likelihood framework by efficient MCMC sampling. We test our methodology \rev{by generating an LSST-like sample} using data from the CosmoDC2 \citep{cosmodc2:2019} simulated extragalactic catalog. While some of the inference techniques developed in this paper can also be used in the context of an empirical mapping to a small-area calibration field, our primary goal is to facilitate inference using physical SED modelling that utilizes a likelihood that jointly describes photometry and spatial information for all observed galaxies. Inference under spatial variations in photometry or redshift information will be addressed in the course of the paper and in \S~\ref{sec:future_work}. \rev{We finally note that our methodology can be applied in the context of both Machine Learning and Template Fitting methods. In the former case, the inference results will be conditional on a training set with exact redshifts and the parameters of the underlying conditional density estimate, in the latter case on the Spectral Energy Distribution models and redshift-galaxy type priors. }

The paper is structured as follows: \S~\ref{sec:DC2sim} describes the simulated galaxy samples used in this work, while \S~\ref{sec:towards_z_prob} gives a brief introduction into inverse problems and deconvolution \rev{in the context of photometric redshift inference.} The following sections describe our inference methodology in detail: \S~\ref{subsec:photometric_likelihood} starts with a description of the photometric likelihood, where we also discuss several regularization schemes, and \S~\ref{sec:clustering_likelihood} formulates the cross-correlation likelihood. Both of these parts are then combined in a composite likelihood framework in \S~\ref{subsec:composite_like}. \S~\ref{sec:model_eval_param_sys} discusses aspects of model evaluation and parametrization of systematics. We then apply our methodology to the simulated data in \S~\ref{sec:forecast_ideal_data}. \S~\ref{sec:summary_and_conclus} summarizes our findings. \S~\ref{sec:future_work} closes the paper with a discussion of future work. 


\section{Simulated Galaxy Samples}
\label{sec:DC2sim}
We use data from the CosmoDC2 simulated extragalactic catalog \citep{cosmodc2:2019} in this work. CosmoDC2 is a mock extragalactic catalog based on a trillion particle N-body simulation with a box size of $4.225 \, {\rm Gpc}^3$, the `Outer Rim' run \citep{2019ApJS..245...16H}. \mmref{The simulated catalog \rev{contains $\sim 2.26$ billion galaxies,} covers $440 \, {\rm deg}^2$ of sky area and spans a redshift range $0<z\leq 3$ \citep{cosmodc2:2019}. Galaxies are assigned to the halo catalog and supplemented with additional galaxies based on the assumption of a power law extrapolation of a power law sub-halo mass function at lower masses. The resulting catalog exhibits a number count slope consistent with that of the Hyper SuprimeCam Deep survey \citep{2018PASJ...70S...4A} down to an r-band magnitude of r $\sim$ 28, well beyond the apparent magnitudes that will be utilized in this paper.} The galaxy catalog uses a combination of empirical and semi-analytic modelling, utilizing the Galacticus \citep{2012NewA...17..175B} and GalSampler codes \citep{2020MNRAS.495.5040H}. For more details on the catalog generation and properties we refer the reader to \citet{cosmodc2:2019}. 

\rev{In this work we consider two datasets: an LSST-like photometric sample and a DESI-like spectroscopic sample that is used to investigate how much the inclusion of spatial cross-correlations constrains the sample redshift distributions. We expect that the samples considered in this work are representative of what can be expected in the LSST analysis, albeit within the limits of the `idealized' setup we assume in this work, on which we will comment in the course of this work.   }

In \S~\ref{sec:pzsample} we will describe the particular selection of photometric data and the photometric redshift catalog used in this work. \S~\ref{para:spectroscophic_sample} describes the generation of the reference spectroscopic sample.

\begin{figure*}
   \centering  
   \includegraphics[scale=0.6]{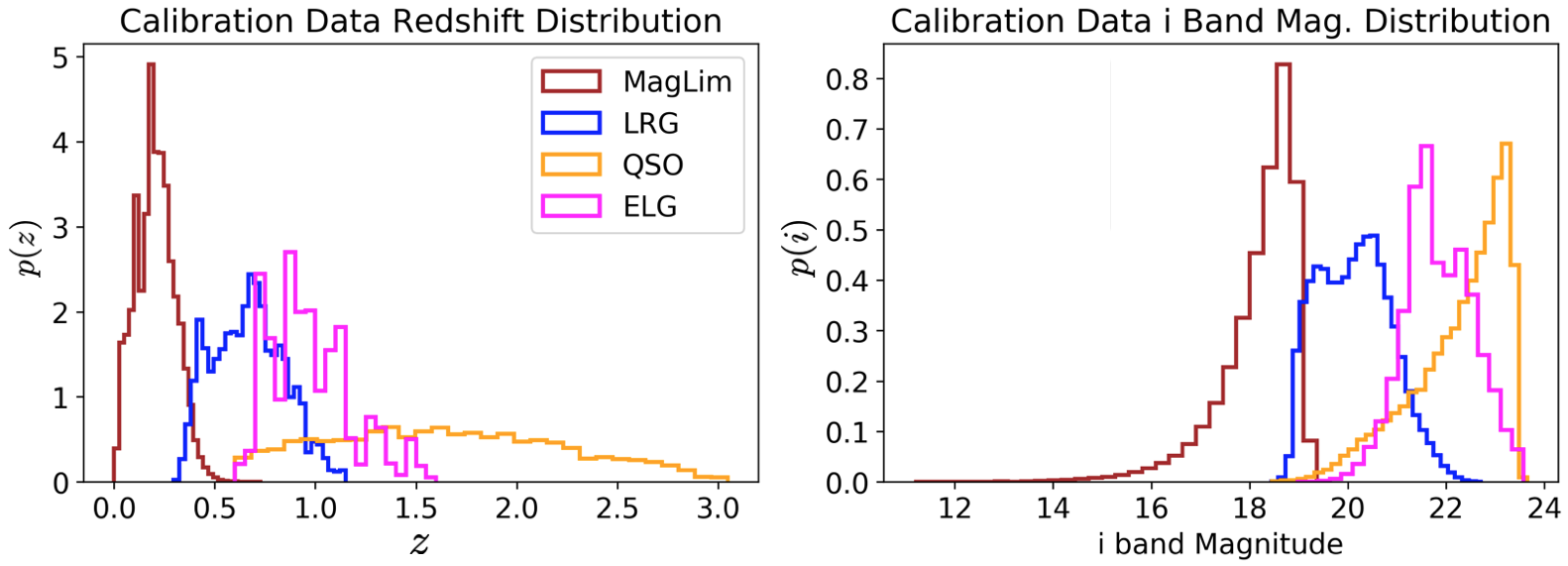}
  \caption{\label{fig:merged_calibration} \textit{Left:} \mmref{Redshift probability density functions} of the galaxy populations that constitute the DESI-like spectroscopic reference sample. \textit{Right:} Corresponding i-band magnitude distributions.}
\end{figure*}
\subsection{Photometric Sample and Photometric Redshift Catalog}
\label{sec:pzsample}


The photometric sample consists of mock galaxies from the LSST-DESC ``CosmoDC2" synthetic sky catalog \citep{cosmodc2:2019}. The catalogs do not contain stars or AGN, so star-galaxy separation and non-thermal contamination are not an issue in this data set.  Observations consist of magnitudes in the six $ugrizy$ Rubin Observatory filters.  Simulated photometric errors were added to the six bands using a simple model designed to match the expected photometric S/N due to depth, seeing, airmass, and sky brightness at the completion of the full 10-year Wide Fast Deep survey \citep[]{2019ApJ...873..111I}.  All galaxies are assumed to be isolated, i.e.~blending effects are not modeled.
We restrict the sample to galaxies with an $i_{\rm LSST}$-band magnitude of $i_{\rm LSST}<25.0$ that corresponds to a \mmref{point source} $i_{\rm LSST}$-band signal-to-noise (S/N) of $\sim 20$.  We make this cut because redshift estimates for lower S/N objects degrade rapidly below this S/N level.  We reserve a small set of $\sim 100 000$ galaxies for training of the photo-z algorithms; this training set is a random subset of the $i_{\rm LSST}<25.0$ sample, and thus completely representative of the underlying galaxy distribution, so no modeling of spectroscopic incompleteness effects is necessary. \rev{We note that this sample size is optimistic in what can be expected for a representative spectroscopic sample for LSST photometric redshift calibration.}

\rev{As noted previously, our methodology can be applied to both Machine Learning and Template Fitting estimates. We therefore discuss both methodologies in this work. Discontinuities in the cosmoDC2 color-redshift mapping prevent us from producing realistic template fitting likelihoods for individual galaxies, so we use an analytical model to mimic template fitting estimates.}
\paragraph*{Template Fitting Redshifts}
We use the publicly available Bayesian photometric redshift code \textsc{BPZ}\footnote{available at: \url{
http://www.stsci.edu/~dcoe/BPZ/}} \citep{2000ApJ...536..571B} to compute redshift estimates for our simulated galaxies.  \textsc{BPZ} is a template-based redshift estimation code that estimates redshift by computing model fluxes from a set of template SEDs and evaluating the resulting $\chi^2$ when compared to observed fluxes.  \textsc{BPZ} includes the optional application of a bivariate Bayesian prior over the joint distribution of type/SED and apparent magnitude in the redshift estimation, though we do not employ the prior in this investigation.  

To construct a template set we begin with the empirical SED catalog of \citet{Brown:2014}.  We then use the \textsc{ESP} software package \citep{Kalmbach:2017}, which constructs a principal component basis set from the empirical SEDs and uses photometric training data to construct the final SED template via Gaussian Processes.  The final training set used by \textsc{BPZ} consists of the 129 empirical templates and 100 additional templates output from \textsc{ESP}.  These templates roughly, but not perfectly, span the observed range of colors for the sample.  
We compute the likelihoods for all SEDs by comparing the observed fluxes to model fluxes evaluated on a grid of redshift spanning $0<z<3$.
The 1-dimensional marginalized (over template type) posterior distributions for each galaxy comprise our final \mmref{template fitting} redshift estimate.

\paragraph*{Machine Learning-based Redshifts}
We use the python version of the publicly available \textsc{FlexCode}\footnote{available at \url{ https://github.com/tpospisi/flexcode}} \citep{Izbicki:17} combined with the \textsc{XGBoost} algorithm \citep{Chen:16} to compute photometric redshifts which we will refer to by the name \textsc{FlexZBoost}. \textsc{FlexZBoost} estimates the conditional density in redshift for each galaxy by fitting to an orthonormal set of basis functions (in this case cosines) via regression with \textsc{XGBoost}.  To further refine the estimates, 25 per cent of the training data is reserved as a validation set to determine optimal values for trimming extraneous low-level peaks in the likelihood, and a ``sharpening" parameter of the form $p(z) \propto p(z)^{\alpha}$ that adjusts the overall width of the density estimates to best match the data.  For this analysis we use 35 cosine basis functions, and a sharpening parameter, chosen via cross-validation, of $1.4$.  Given the representative training data used in this experiment, we expect very accurate redshift estimates from the \textsc{FlexZBoost} algorithm. 


\subsection{Spectroscopic Sample}
\label{para:spectroscophic_sample}

The simulated reference spectroscopic sample is selected to mimic, in broad strokes, the sample selections of the Dark Energy Spectrosopic Instrument (DESI, \citealt{2016arXiv161100036D}, \citealt{2020arXiv200106018Z}, \citealt{2020RNAAS...4..181Z}). 
This consists of a set of four samples with increasing mean redshift:  a magnitude-limited sample to $r_{\rm LSST}<19.5$; a Luminous Red Galaxy (LRG) sample; an Emission Line Galaxy (ELG) sample; and finally a high-redshift Quasar (QSO) sample. We show the redshift and $i_{\rm LSST}$-band magnitude distributions of these subsamples in Fig.~\ref{fig:merged_calibration}. The LRG, ELG, and QSO samples are selected such that their density per redshift matches that of the DESI samples (priv. comm. Rongpu Zhou and Jeffrey Newman). This sample is distinct from the redshift calibration data mentioned in the previous section.

\mmref{We construct a magnitude-limited sample, by imposing a magnitude cut of $r_{\rm LSST}<19.5$. To approximate the LRG, ELG and QSO galaxy samples, we use the values of the stellar mass, star formation rate, and black hole mass times Eddington ratio as proxies for objects that are LRG, ELG, and QSO-like respectively.  Our goal with these samples is to select galaxies that will have differing bias properties and mimic the complexities of the DESI sample in this regard, while matching the density and signal-to-noise we would expect with a DESI-like sample. We thus use these simple truth quantities from the simulation rather than recreate the full color selection of a true, simulated DESI sample.} 
The QSO, ELG, and LRG samples are all selected with $r_{\rm LSST}>19.5$, to be independent of the magnitude-limited sample. Additionally, QSOs and ELGs are selected to have $r_{\rm LSST}<23.4$ and LRGs have the cut \mmref{$z_{\rm LSST}<23.0$ }applied to them. QSOs are selected by ordering the candidate QSOs in a redshift bin by the product of their black hole mass and black hole Eddington ratio, cutting on the value when the density of QSOs matches the expected DESI density for a given redshift range. This process is repeated for ELGs using their star formation rate in the simulation as a proxy for ``ELG-ness''. We also impose the condition that the candidate ELGs have a black hole mass times Eddington ratio below what we cut on for the QSOs, to assure that the samples are independent. This process of rank-ordering and selecting the top galaxies until we achieve the expected DESI density is repeated again for the LRGs, this time with stellar mass as our proxy value. Both the ELG star formation and QSO selection are excluded from the LRG selection, ensuring that the samples are independent. We calculate values for these cuts on a $\sim$50 deg$^2$ test area in the CosmoDC2 simulations and apply them to the full 300 deg$^2$ area.

\mmref{\section{The photometric redshift problem: A complex inverse problem}
\label{sec:towards_z_prob}
\rev{The photometric redshift problem is an inverse problem: a true underlying redshift distribution of a sample of galaxies is inferred using noisy redshift information. Inverse problems are challenging because they can be ill-posed, i.e. small changes in the measured noisy redshift information can propagate into large differences in the inferred underlying sample redshift distribution. It is therefore typically necessary to restrict the function space of possible true sample redshift distributions based on assumptions about e.g. shape or smoothness of the true distribution. This process is referred to as `regularization' and will be discussed in the following. We refer the interested reader to Appendix~\ref{sec:methods} for an introduction to inverse problems.
}

} 

We structure the discussion of the likelihoods used in this work in practice based on the following roadmap: we first describe our likelihood framework for the photometry of galaxies given a set of templates in \S~\ref{subsec:photometric_likelihood} and the clustering likelihood in \S~\ref{sec:clustering_likelihood}. The considered regularization schemes are described in \S~\ref{subsubsec:regularization_ill_posed_inverse}. A particular challenge in the context of large area photometric surveys is the necessity to scale the inference to a large number of galaxies. In Appendices \S~\ref{par:ml_inference} and \S~\ref{par:derive_laplace} we derive an efficient inference framework based on the Laplace approximation that facilitates fast probabilistic deconvolution. We will use this deconvolution methodology in the following sections \S~\ref{subsec:photometric_likelihood} to \S~\ref{sec:forecast_ideal_data}. \rev{We list the  symbols and notation used in these sections in Tab.~\ref{tab:symbol_table}.}
\rev{
\begin{table}
    \centering
    \begin{tabular}{l|l}
    Symbol & Description \\\hline
       $\mathcal{F}(\lambda)$  & Transmission function of optical filter band \\
       ${\rm SED}_{\lambda}$  & Spectral Energy Distribution as a function of wavelength \\
       z & Redshift of galaxy \\
       $\alpha$ & Auxillary parameters in SED model \\
       $\hat{\mathbf{f}}_i$ & Flux of galaxy $i$ \\
       $\Sigma_i$ & Covariance of flux measurement of galaxy $i$ \\
       $\hat{\boldsymbol{F}}$ & Set of flux measurements of galaxies in the sample \\
       $N_{\rm gal}$ & Number of galaxies in the sample \\
       $\mathbf{n^{B/R}}$ & Set of histogram heights of the base sample z-distribution \\ 
       $\boldsymbol{\pi^{B/R}}$ & Normed histogram heights to sum to unity \\
       $[(z, \alpha) \in I_i]$ & unity if $(z, \alpha)$ fall into interval $I_i$, zero otherwise  \\
       $\Delta z$ & Histogram bin width \\ 
       $\mu_y$  & Mean of the logit-normal distribution \\
       $\Sigma_y$ & Covariance of logit-normal distribution \\
       $\mathbf{n_{-j}^{B}}$ & Vector of histogram heights, excluding bin j \\
       $\mathbf{H}$ & Hessian matrix \\
       $\Upsilon$ & Tikhonov matrix \\ 
       $\overline{w}_{\rm DM, i}$ & Annulus-integrated DM correlation function in z-bin $i$ \\
       $b_i^{R/B}$ & Reference/Base sample galaxy-DM bias in z-bin $i$ \\
       $\mathbf{\hat{w}^{RB/RR}}$ & Cross-/Self-correlations between ref.-base/ref.-ref. samples \\
       $\Gamma_i$ & Random variable of ratio $\mathbf{\hat{w}^{\rm RB}}$ and $\mathbf{\hat{w}^{\rm RR}}$ \\
       $C(z)$ & z-dependent ratio of galaxy-DM biases $\mathbf{b^{B}}/\mathbf{b^{R}}$ \\
       $\mathbf{\nu}$ & weighting factor in composite likelihood \\
       $\boldsymbol{\overline{n}^{\rm B, sys}}$ & Systematics corrected vector of histogram heights $\boldsymbol{\overline{n}^{B}}$ \\ 
       $\tau$ & Parameter vector of systematics kernel \\
       $\mathcal{D_{\rm rep}}$ & Sampled data from the posterior-predictive distribution \\

    \end{tabular}
    \caption{\rev{Symbols and notations used in this work. The first/second column list the symbol/description. The abbreviation $B/R$ denotes the quantity of interest of the base or reference sample.}}
    \label{tab:symbol_table}
\end{table}}

\section{Photometric Likelihood}
\label{subsec:photometric_likelihood}
The spectral energy distributions of distant galaxies are a complex superposition of spectral components from their stellar populations. 

The SED of the galaxy can be uniquely mapped to a given redshift $z$, which allows us to predict the galaxy flux as a function of redshift in a given optical filter band $\mathcal{F}(\lambda)$ by
\begin{align}
    f_i(z, \alpha) &= \int \mathcal{F}(\lambda) \, {\rm SED}_{\lambda}(\lambda, z, \alpha) \, \mathrm{d}\lambda  \, .
    \label{eq:map_z_color}
\end{align}
where ${\rm SED}_{\lambda}(\lambda, z, \alpha)$ is the Spectral Energy Distribution template in units of ${\rm erg} \big/ \mathrm{cm}^2 \big/ \mathrm{s} \big/ \angstrom$. The parameter $\alpha$ denotes additional free parameters in the SED template models, such as galaxy age, type, or red continuum slope. For a given set of photometric filters $\mathcal{F}(\lambda)$ we obtain a mapping between the redshift $z$ of the galaxy and a vector of fluxes $\mathbf{f}$. We will denote this mapping as $\mathcal{T}(z, \alpha)$.


Assuming that the measurements of photometry for different galaxies are independent\footnote{This assumption can be violated due to effects such as blending of nearby galaxy light profiles \mmref{on flux calibration errors.}.} we can make the ansatz for the joint likelihood of fluxes of a galaxy sample $\mathbf{\hat{F}}$
\begin{equation}
    p(\mathbf{\hat{F}} | \mathbf{z}, \boldsymbol{\alpha}) = \prod_{i = 1}^{N_{\rm gal}} \mathcal{N}(\mathbf{\hat{f}_i} | \mathcal{T}(z_i, \alpha_i), \boldsymbol{\Sigma}_i) \, , 
\label{eq:likelihood_photo}
\end{equation}
Here, $\boldsymbol{\Sigma_i}$ denotes the measurement covariance matrix of the flux measurements $\mathbf{\hat{f}_i}$, and $\mathbf{\hat{F}}$ denotes the set of all flux measurements of the galaxies. \mmref{We assume Gaussian uncertainties here, where $\mathcal{N}(x, \mu, \Sigma)$ denotes the Normal distribution.} The parameter $\alpha$ can either be a galaxy-specific index that selects a certain template from a pre-specified number of models, or a physical property of galaxies.

The prior on the parameters $z$ and $\alpha$ must account for their correlation. An example for a possible parametrization in the case of a galaxy-specific template index would be a two-dimensional histogram. However, other parametrizations are possible, especially if additional parameters that change the shape of the base templates are included in the template set. In this work, we will consider the simplest case, where we use a multidimensional histogram prior where each histogram cell denotes a combination of redshift bin and discretized $\alpha$ parameter value, that for example could indicate a template selection. The histogram index $i$ runs over all histogram bins $\{i : 0 < i \leq N_{\rm tot}\}$, where $N_{\rm tot} = N_{\rm bins} \times N_{\rm parameters}$. The \mmref{prior on the} corresponding histogram heights, denoted as \mmref{$n_i^{B}$} corresponding to the interval $I_i$ in the $z-\alpha$ parameter space, \mmref{reads}: 
\begin{equation}
    p(z, \alpha) = \sum_{i = 1}^{N_{\rm tot}}  n^{B}_{i} \, [ (z, \alpha) \in I_i]  \, .
    \label{eq:prior_def}
\end{equation}
\mmref{Here $[K]$ denotes the Iverson bracket, that is (0, 1) if the proposition $K$ is (false, true).}
\mmref{We note that $\mathbf{n^{B}}$ parametrizes the joint distribution of redshift histograms and $\boldsymbol{\alpha}$ parameter. For simplicity we will in the following omit the marginalization over $\boldsymbol{\alpha}$ and refer to $\mathbf{n^{B}}$ as the parameters of the sample redshift distribution. The reason is that in this paper we do not add additional parameters to parametrize the SEDs over which we need to marginalize.}
In applications like weak gravitational lensing and galaxy clustering we are mainly interested in estimating the redshift distribution of a sample of galaxies, here referred to as the base sample and parametrized by the vector $\mathbf{n}^{\rm B}$.   It is therefore useful to marginalize over the redshifts of individual galaxies. We note that if the posterior of individual galaxy redshifts is important, we can always post-sample using the final posterior on $\mathbf{n}^{\rm B}$, based on Eq.~\eqref{eq:composite_like}, that then also includes information from galaxy clustering.
\mmref{
The posterior distribution of the sample redshift distribution given $\mathbf{\hat{F}}$ is then:
\begin{equation}
    p(\mathbf{n^{\rm B}} | \mathbf{\hat{F}}) \propto p(\mathbf{n^{\rm B}}) \prod_{i = 1}^{N_{\rm gal}}  \int_{0}^{\infty} \mathrm{d}z_i \, p(z_i | \mathbf{n}^{\rm B}) \, \mathcal{N}(\mathbf{\hat{f}_i} | \mathcal{T}(z_i), \Sigma_i)  \, .
    \label{eq:def_post_nz_B}
\end{equation}

Discretizing the integral and using Eq.~\eqref{eq:prior_def} we obtain 
\begin{equation}
    p(\mathbf{\mathbf{n}^{\rm B}} | \mathbf{\hat{F}}) \propto p(\mathbf{n}^{\rm B}) \prod_{i = 1}^{N_{\rm gal}} \sum_{j = 1}^{N_{\rm tot}} n_{j}^{\rm B} \int_{z_{L}^j}^{z_{R}^j} \mathrm{d}z_i  \, p(\mathbf{\hat{f}_i} | \mathcal{T}(z_i), \Sigma_i) \, .
    \label{eq:post_eq_nb}
\end{equation}}
The histogram heights $n_{i}^{B} = \pi_{i}^{B} \big/ \Delta z$ can be expressed as the ratio between $\bm{\pi}$ and the histogram width $\Delta z$, assuming equal-sized redshift bins. The vector $\bm{\pi^{B}}$ has the properties $\sum_{i = 1}^{N_{\rm tot}} \pi_{i}^{B} = 1$ and $0 \leq \pi_i^{B} \leq 1$, and therefore lies on the simplex. Our first choice for a distribution on the simplex \mmref{for $p(\boldsymbol{\pi^{B}})$ (and therefore $p(\mathbf{n}^B)$) was the Dirichlet distribution\footnote{\mmref{The Dirichlet is the conjugate prior to the multinominal distribution, which can make sampling and inference easier. Concretely, if a Dirichlet prior is set on the probabilities of the multinomial likelihood (which are its parameters), the posterior over these probabilities is again a Dirichlet. However conjugacy does not imply that the prior is ideal in all circumstances.}}.} During the course of this project we have applied a mean field variational inference scheme that uses the Dirichlet as the variational distribution as well as a Gibbs sampling scheme based on the Dirichlet-Multinomial cojugacy for posterior inference. We found that the variational inference scheme yielded underestimated error bars, likely due to the restricted covariance structure of the Dirichlet \rev{that makes an independence assumption between neighboring bins}.  Moreover, the sampling approach did not scale well to the large galaxy samples expected for the first-year LSST observations. Specifically, the computational workload to update redshift variables for $10^{6}-10^{10}$ galaxies seems very large, and while subsampling techniques provide a possible mitigation, they can lead to biased inferences \citep{2018arXiv180708409Q}. \rev{Furthermore the application of sampling techniques requires a sufficiently large  trace to ensure convergence. For large sample sizes, this can be computationally very expensive, as each sampling step requires the evaluation of individual galaxy likelihoods for many galaxies. Also, in this regime, sources of systematics in the modelling of Spectral Energy Distributions will dominate the total error budget of the inference, so developing an efficient inference scheme for exploring a large number of models can be very useful. } In order to provide a more flexible distributional ansatz than the Dirichlet, while still maintaining the computational advantages of a mean field variational inference scheme, we decided to develop a scheme that is based on the logit-normal distribution \citep{10.1093/biomet/67.2.261}. \rev{ As we mentioned earlier, the Dirichlet is a very restrictive distribution on the simplex as it assumes that the marginal distribution of a set of datums equals the conditional distribution of that set given the complement. This independence assumption for a prior that models the sample redshift distribution, essentially a convolution of two smooth functions (the luminosity function and volume selection), seems questionable. As we will see in the following section, the logit-normal distribution has a much richer covariance structure and approximates the Dirichlet well \citep{10.1093/biomet/67.2.261}. Thus in circumstances where the Dirichlet is appropriate, we will likely still get reasonable inference assuming the logit-normal due to this closeness property. While the choice of a logit-normal is necessarily somewhat arbitrary, there are strong theoretical motivations and practical viability. 
} While these considerations motivate our choice of method, we note that this should not discredit alternative approaches based on sampling or variational inference in general. 
\subsection{Photometric Redshift inference}
\label{subsec:photoZestim}
The problem specified by Eq.~\eqref{eq:def_post_nz_B} is a deconvolution problem that extends the simple toy model considered in Appendix~\ref{sec:methods}. The `noise' PDF is now given by a joint likelihood that can depend on a complex set of parameters. Furthermore, while the discussion in Appendix~\ref{sec:methods} focused on deriving an estimator for the deconvolved density, the focus here is to infer posteriors using efficient inference techniques. We present the detailed description and derivation of the inference pipeline in Appendices~\ref{par:ml_inference} and~\ref{par:derive_laplace}. The final \mmref{form of} Eq.~\eqref{eq:def_post_nz_B} is then given in the form of a logit-normal posterior: 
\begin{equation}
\begin{split}
    &p(\mathbf{n^B} | \mathbf{\hat{F}}) \approx \frac{1}{\sqrt{|2 \pi \boldsymbol{\Sigma}_{\mathbf{y}}|}}  \frac{1}{ \Delta z^{N_{\rm bins}} \prod_{i = 1}^{N_{\rm bins}} n^B_i}\\ 
    &\exp{\left(-\frac{1}{2} \left(\log{\left(\frac{\boldsymbol{n^B}_{\rm -N_{\rm bins}}}{n^B_{N_{\rm bins}}}\right)} - \boldsymbol{\mu}_{\rm y, ML}\right) \boldsymbol{\Sigma}_{\mathbf{y}}^{-1} \left(\log{\left(\frac{\boldsymbol{n^B}_{\rm -N_{\rm bins}}}{n^B_{N_{\rm bins}}}\right)} - \boldsymbol{\mu}_{\rm y, ML}\right)\right)} \, ,
    \end{split}
    \label{eq:logit_normal_post}
\end{equation}
where $\Delta z$ denotes the histogram bin width. The estimation of the covariance $\boldsymbol{\Sigma}_{\mathbf{y}}$ and mean vector $\boldsymbol{\mu}_{\rm y, ML}$ are detailed in Appendices \ref{par:ml_inference} and \ref{par:derive_laplace}. However, we note that this formalism derives the hessian $\mathbf{H} = -\boldsymbol{\Sigma_y}^{-1}$ and obtaining the covariance matrix $\boldsymbol{\Sigma_y}$ requires matrix inversion. \mmref{The subscript `y' here refers to the variable transformation: 
\begin{equation}
    \mathbf{y}(\boldsymbol{\pi}) = \left[ \log{(\pi_1/\pi_{N_{\rm bins}})}, \dots, \log{(\pi_{N_{\rm bins} - 1}/\pi_{N_{\rm bins}})} \right] \, ,
\end{equation} 
that we discuss in more detail in Appendix \ref{par:derive_laplace}.} \mmref{The vector $\mathbf{n^{B}}_{-N_{\rm bins}}$ denotes here the vector of $\mathbf{n^{B}}$ excluding the last entry $n^{B}_{\rm N_{\rm bins}}$, where we assume equal sized redshift histogram bin width.}

Eq.~\eqref{eq:def_post_nz_B} describes an inverse problem. We can therefore expect, \rev{for non-negligible measurement error, that a reconstruction with high resolution in redshift can be ill-posed, i.e., small variations in color space or the SED modelling will produce comparatively large variations in estimates of the true underlying sample redshift distribution.}

Furthermore, as we have seen in Appendix~\ref{sec:methods}, the solutions of inverse problems do not have to be bounded\footnote{In our case we note that all parameter values $\boldsymbol{\pi} \in \Delta$ are bounded, because $\Delta$ (with a chosen metric) is a bounded metric space. However these solutions in logit space (see \S~\ref{par:derive_laplace}) do not have to be bounded.} (or even well defined), which implies that uncertainties can be arbitrarily large  \citep[see, e.g.,][]{Kuusela:220015}. 
Regularization, detailed in the following section, is therefore a key aspect in our inference pipeline. 

\subsection{Regularization}
\label{subsubsec:regularization_ill_posed_inverse}

In this section we describe techniques that we employ to regularize\footnote{\mmref{We will use the term `regularization' not only in the context of Bayesian statistics, where it's often implemented in the form of a prior, but in general to describe methods that restrict the complexity of parameters or functions.}} the deconvolution problem. As instabilities arise when the histogram width is of the same order as the uncertainty in the redshift of the individual galaxies, picking broader bins reduces these artifacts \citep[see e.g.][]{Kuusela:220015}. Considering our toy model in Eq.~\eqref{eq:kde_classical_convolution}, we see that if $K^{\rm ft}(t b)$ is narrower than $p_{\epsilon}^{\rm ft}(t)$, there can be values of $t$ where their ratio, and therefore the integrand in Eq.~\eqref{eq:kde_classical_convolution}, can become large or even unbounded. Subject to the aforementioned limitations, this behaviour generalizes to the deconvolution problem considered here.  In the following, we will denote this scheme as the \rev{`Oversmoothing'} method. As will be seen in the following section, this simple scheme can lead to posteriors \rev{that can be too narrow.} We therefore consider alternative approaches \rev{like the `Merging Bin Regularization', which involves combining initially narrow bins in parameter space to form larger bins with less noise.}
\begin{figure}
   \centering  
   \includegraphics[scale=0.5]{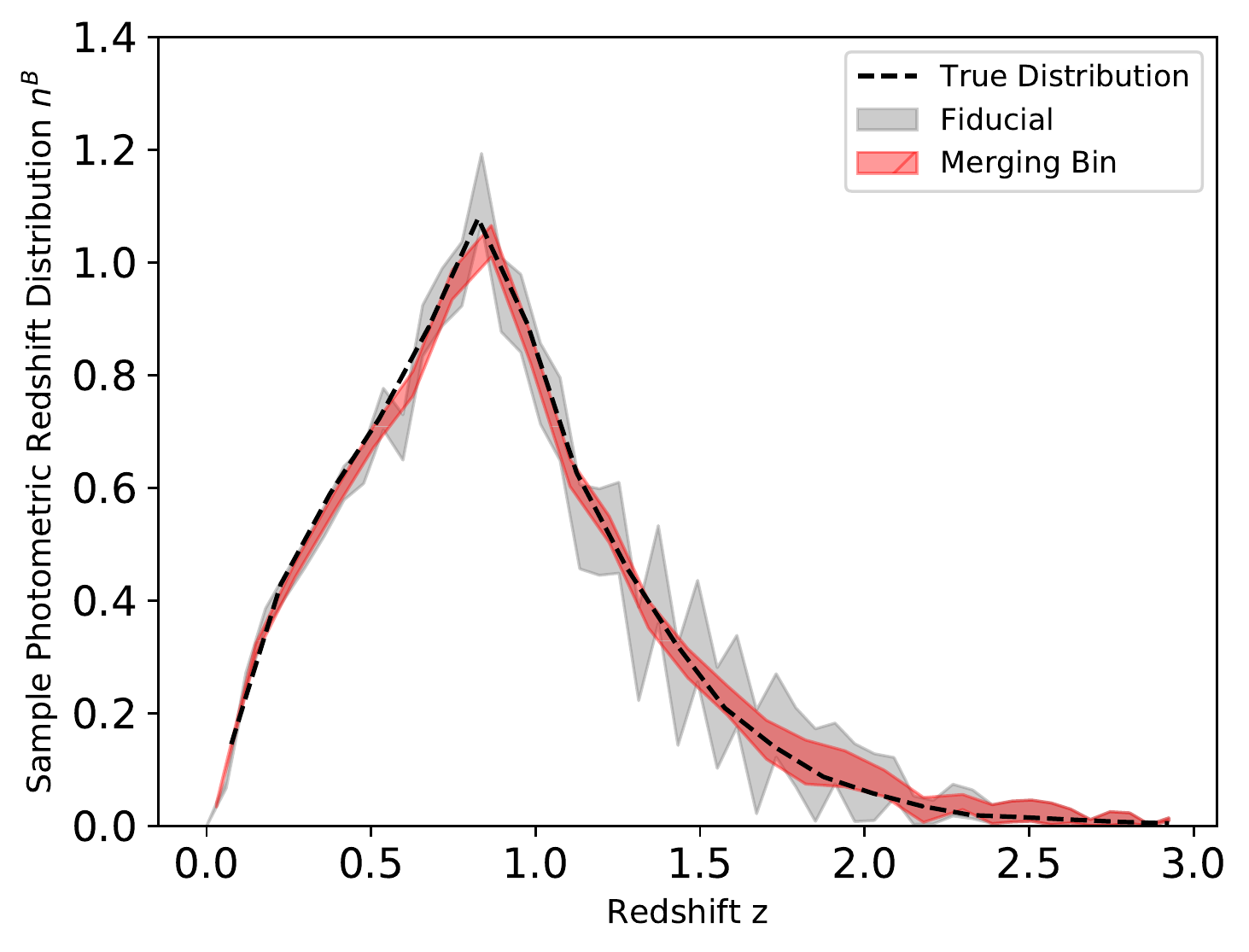}
  \caption{\label{fig:demo_regul} \mmref{Illustration of the impact of the `Merging Bin' regularization on the posterior of the sample photometric redshift distribution. We show $1 \sigma$ intervals. The x-axis shows the redshift value $z$, the y-axis the value of the $\mathbf{n^B}$ parameters. The errorbars are the $[16, 84]$ percentiles, which would correspond to $1 \sigma$ intervals for a normal distribution. The black dashed curve shows the spectroscopic redshift distribution. The red contour shows the result of the `Merging Bin' regularization with 30 bins applied to the `Fiducial' contours that are binned using 50 bins. We refer for a detailed explanation to \S~\ref{sec:forecast_ideal_data} and Fig.~\ref{fig:merged_graham_ideal}, which shows and discusses the `Fiducial' case as `Small Sample (50k)'. }    } 
\end{figure}
\subsubsection{Oversmoothing}
\label{subsubsec:oversmoothing}
\rev{`Oversmoothing' defines the likelihood on a coarser resolution in smeared space, i.e. the likelihood of individual galaxies is convolved with a coarser, i.e., lower resolution, sample redshift distribution. As a result, within-bin variations of the (unknown) true sample redshift distribution are not accounted for. As mentioned earlier, these variations can be large in the context of inverse problems, which in turn implies that credibility intervals constructed on the larger grid do not have to exhibit good coverage. }   
\subsubsection{Merging Bin Regularization}
\label{subsubsection:merging_bin}
\rev{In the previous section we discussed that `Oversmoothing' can lead to overly narrow credibility intervals for posterior sample redshift distributions. To remedy this issue, the `Merging Bin Regularization' \citep{Kuusela:220015} deconvolves the sample redshift distribution on a narrow, finely binned grid to reduce the aforementioned biasing effect and then recombine the deconvolved redshift distribution in parameter space. This avoids the aforementioned error, because the recombined, coarser binning will take into account within-bin variations in (deconvolved) parameter space. Deconvolving onto a thin initial redshift grid will result in the aforementioned typical instabilities of inverse problems. } However, we must ensure that the optimization of the maximum-likelihood solution converges to a global maximum. We therefore run multiple optimizations with different initial conditions and pick the best solution.  Furthermore, the hessian $\mathbf{H}$ can have a very high condition number. Even though it is possible to sample from the resulting posteriors without the matrix inverse using MCMC sampling (only the inverse covariance enters the $\chi^2$), sampling is more efficient using the standard Box-Mueller method \citep{box1958}, which requires an inverse. \rev{In the following paragraph we will describe Tikhonov Regularization as a method to obtain this inverse. However we note that Tikhonov Regularization can also be applied independently of the `Merging Bin' regularization scheme. In this case we would deconvolve onto a given grid resolution and apply the method without subsequently merging bins.}
\paragraph*{Tikhonov Regularization}
\label{subsub:tikhonov_regul}
We perform the matrix inversion of the hessian $\mathbf{H}$ using Tikhonov regularization \citep[see e.g.][pp. 86-90]{kress1998numerical}. Here, we treat the matrix inversion as a system of linear equations constructed from the hessian and the column-wise inverse hessian/unit matrix respectively. The \mmref{instability} of the problem can lead to very small entries in the hessian that imply large entries (in absolute values) in its inverse. As a regularization, one can add a penalty term and reformulate the problem as a minimization 
\begin{equation}
    {\rm min}\left\{ ||\mathbf{H} \mathbf{a}_i - \mathbf{1}_i ||_{2}^{2} + ||\mathbf{\Upsilon} \mathbf{a}_i ||_{2}^{2} \right\} \, ,
\end{equation}
where $i$ denotes column $i$ of the inverse hessian and the unit matrix respectively. The matrix $\mathbf{\Upsilon} = \alpha \mathbf{1}$ is the Tikhonov matrix, $\alpha$ the regularization parameter \mmref{and $\mathbf{1}$ the unit matrix}. The regularization term penalizes large values for $\mathbf{a_i}$, which regularizes the inversion and reduces the condition number. The analytic solution to this minimization problem is given as: 
\begin{equation}
    \mathbf{c}_i = \left(\mathbf{H}^{\rm T} \mathbf{H} + \mathbf{\Upsilon}^{\rm T} \mathbf{\Upsilon}\right)^{-1} \mathbf{H}^{\rm T} \mathbf{1_i} \, .
\end{equation}

\mmref{We note that we introduce the Tikhonov regularization here predominantly as a way to regularize the matrix inversion of the hessian. We recommend selecting the parameter $\alpha$ to be just as large as necessary to perform this inversion accurately.  Tikhonov regularization can be used as the main regularization in inverse problems; however, we find that the merging bin regularization scheme performs much better in terms of producing well-calibrated probability $\mathbf{n^{B}}$ posteriors.   }
We reiterate that the idea of the merging bin regularization scheme proposed by \citet{Kuusela:220015} is to deliberately start with histogram bins that are too small and lead to a noisy deconvolved density. We then exploit the characteristic noise pattern in the deconvolved distribution, where bins that overshoot, i.e. are larger than the true value, are immediately followed by those that undershoot. This results in an alternating or `zig-zag' pattern of the deconvolved density \rev{(see Fig.~\ref{fig:demo_regul})}. Merging these neighboring bins then helps to `stabilize' the deconvolved distribution. 
We therefore sample from a posterior obtained assuming a finely binned histogram and merge neighboring bins, which compensates the noise effect. We can then in principle directly use these samples in the cross-correlation likelihood. Nonetheless, it is computationally more efficient to remap these samples to a regular grid with the same or very similar resolution than the binning used for the cross-correlations, since treating the finely binned histogram heights as free parameters would not add more information due to the resolution loss. We found that the template fitting posterior after merging bin regularization can again be well-described by a logit normal distribution that we fit using Assumed Density filtering.

\paragraph*{Assumed Density Filtering}
To fit the logit normal distribution to the sampled and merged posterior samples, we work in logit space and make a gaussian ansatz
\begin{equation}
    q(\mathbf{y}) = \mathcal{N}\left(\mathbf{y} | \boldsymbol{\mu}, \mathbf{\Sigma}\right) \, .
\end{equation}
We can directly generate samples from the true distribution by sampling from the original, finely binned, logit-normal distribution with subsequent merging, i.e. averaging, of neighboring bins, and then  transforming back to logit space. We will denote this true distribution as $p_{\rm true}(\mathbf{y})$. Assumed density filtering then commences by minimizing the Kullback-Leibler divergence between our ansatz $q(\mathbf{y})$ and $p(\mathbf{y})$,
\begin{equation}
    {\rm KL}(p || q) = \int \mathrm{d}\mathbf{y} \,\left( p(\mathbf{y}) \log{p(\mathbf{y})} - p(\mathbf{y})  \log{q(\mathbf{y})}\right) \, .
\end{equation}
After optimizing $KL(p || q)$ for $\boldsymbol{\mu}$ and $\mathbf{\Sigma}$, we can show that the optimium is reached if
\begin{equation}
    \int \mathrm{d}\mathbf{y} \, p(\mathbf{y}) \,  \mathbf{y} = \boldsymbol{\mu} \, , 
\end{equation}
and 
\begin{equation}
    \mathbf{\Sigma} = \int \, \mathrm{d}\mathbf{y} \, p(\mathbf{y}) \, (\mathbf{y} - \boldsymbol{\mu}) (\mathbf{y} - \boldsymbol{\mu})^{\rm T}
\end{equation}
We see that assumed density filtering reduces to moment matching in logit space, when we apply the sample mean and sample covariance estimators to samples from $p_{\rm true}(\mathbf{y})$. We note that this is in general true for distributions of the exponential family \citep[see, e.g.,][]{Ranganathan04assumeddensity}.

To summarize, 
we perform the inference scheme described in the previous section using a fine histogram binning. Subsequently we sample from the posterior after regularizing the inverse hessian using Tikhonov regularization. We merge neighboring bins from each posterior draw until the noise is reduced and we obtain a smooth probability distribution. \mmref{We illustrate this process in Fig.~\ref{fig:demo_regul}, which illustrates the impact of the `Merging Bin' regularization scheme on the posterior of the sample photometric redshift distribution. Comparing the grey contours (`Fiducial') that uses 50 bins with the red contours (`Merging Bin') that merges neighboring bins to a binning of 30 bins, we see the much smoother shape and the elimination of the `zig-zag' pattern present in the grey contours. We defer a more thorough explanation of the methodology and sample to \S~\ref{sec:forecast_ideal_data} and Fig.~\ref{fig:merged_graham_ideal}, which discusses the result shown in the grey contours under the abbreviation `Small Sample (50k)'. }

Based on our experience we propose to initially merge neighboring bins until we obtain a bin size of the order of the average $\pm 2\sigma$ range of the individual galaxy redshift distributions. Subsequently we merge fewer bins until the aforementioned characteristic `zig-zag' noise pattern appears. This can be identified as the limiting resolution we can obtain. We note that it is important to distinguish patterns due to `real' line of sight structure and due to the aforementioned noise in the deconvolution. If the pattern appears gradually with increasing resolution (merging fewer bins), it is indicative of statistically significant line-of-sight structure. If the deconvolved density suddenly becomes unstable in a `zig-zag' pattern when fewer bins are merged, we have reached a resolution limit. Using assumed density filtering under the ansatz of a logit normal distribution, we finally reparametrize our model on the final redshift grid. 

The merging process described above is largely based on inspecting when the instabilities vanish. \rev{We refer to \S~\ref{sec:forecast_ideal_data} for a more quantitative description of the binning choices and the impact on the results. While the results are expected to be robust against slightly different initial binning choices, the number of neighboring bins can in practise be optimized either by Bayesian model comparison or by comparison with the clustering likelihood. In our experience, including a clustering likelihood into the selection of binning improves robustness of this inference, because the photometric likelihood is often subject to model misspecification and the clustering likelihood counteracts resulting biases.} Alternatively, in a classical deconvolution problem, like the one presented in Appendix~\ref{sec:methods}, one could use a bootstrap estimate of the bias and variance of the reconstruction with respect to the over-smoothed photometric redshift distribution. This is consistent with the approach taken by \citet{2005MNRAS.359..237P} based on the recommendation in \citet{1986ipag.book.....C}. In contrast to \citet{2005MNRAS.359..237P}, we use a joint likelihood between the photometry and spatial information to produce posteriors for the photometric sample redshift distribution and not a `point prediction'. Furthermore our `measured data' is the photometry and spatial information of galaxies. Accordingly our model selection, of which regularization is a part, must reproduce the measured photometry and spatial distribution, e.g., measured by the correlation functions of galaxies. In the Bayesian context, this would translate into the usage of posterior predictive checks (PPC) discussed in \S~\ref{subsubsec:post_pred_tests}. \rev{Irrespective of the chosen methodology, the necessity to regularize the photometric likelihood does not depend on the inference strategy, but on the conditioning of the problem. For example \citet{2005MNRAS.359..237P} uses a boostrap based loss function to balance the bias and variance in the reconstruction, while \citet{2019arXiv191007127A} use a prior constructed from a calibration sample. The inclusion of clustering information (as in \citealt{2019arXiv191007127A}, \citealt{2019MNRAS.483.2801S} or this work) helps in the regularization of the photometric likelihood.  } 
\section{Clustering Likelihood}
\label{sec:clustering_likelihood}


In order to include information \mmref{about} the spatial distribution of galaxies into the likelihood, we consider spatial cross-correlations between photometric and spectroscopic samples. Spatial correlations measure the excess probability over random to find two galaxies separated by a certain distance. 
This can be exploited to extract redshift information for galaxy samples \citep[see e.g.][]{2008ApJ...684...88N, 2013arXiv1303.4722M, 2013MNRAS.433.2857M, 2016MNRAS.462.1683S, 10.1093/mnras/stx691, Morrison2016, 2017arXiv171002517D, 2018MNRAS.477.1664G} for which we do not have accurate redshift information, i.e. photometric galaxy samples, using spatially overlapping spectroscopic catalogs. 

The idea is to select the spectroscopic samples in thin redshift slices and estimate the cross-correlation between these redshift-selected samples and the full photometric galaxy sample. As discussed in the following, the resulting signal will then be proportional to the photometric redshift distribution at that redshift.

\begin{figure}
   \centering  
   \includegraphics[scale=0.50]{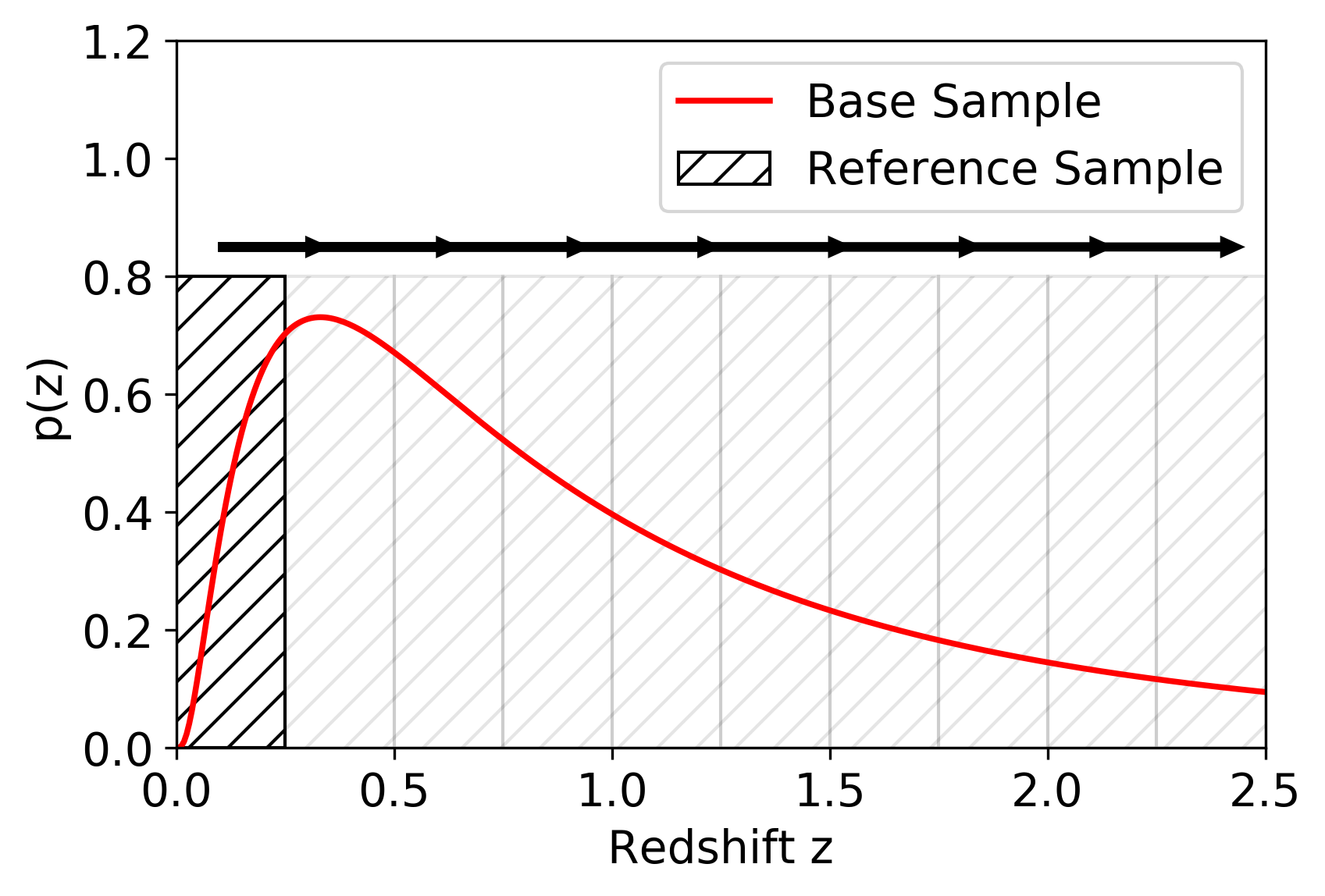}
  \caption{\label{fig:cross_corr} Illustration of the construction of the cross-correlation data vector. A `Reference Sample' that can be precisely selected in redshift is moved over a `Base Sample', which can either be a sample without redshift information (photometric sample) or the `Reference Sample' itself. In this illustration 10 cross-correlations would be estimated, constituting the cross-correlation data vector.  } 
\end{figure}

Fig.~\ref{fig:cross_corr} illustrates the basic idea of cross-correlation redshift inference.
We consider two galaxy samples: a reference sample `R' and a base galaxy sample `B'. The reference sample contains galaxies with accurate, often spectroscopic, redshift measurements; the base galaxy sample consists of galaxies observed in broad band photometric filters. As the base/reference samples are typically photometric/spectroscopic samples, we use these terms interchangably in text\footnote{\mmref{We note, however, that the reference sample does not have to be a spectroscopic dataset, as multi-band, narrow filter photometric observations \citep{ 2020arXiv200711132A}, or photometric redshifts of redMaGiC samples \citep[see e.g.][]{2018MNRAS.477.1664G}, also allow for reasonable redshift accuracy.}}. The redshift distribution of the base sample is illustrated by the red distribution, while the binned reference sample redshift distributions for simplicity are shown as tophat functions (unlike the simulated samples we use to test our methodology). A single cross-correlation is then obtained by cross-correlating a single tophat selection with the full base sample. Multiple measurements therefore `slice' through the redshift distribution of the base sample, illustrated here by the arrows and the grey \mmref{hatched} tophat slices.   


\mmref{As described in detail in \citet{2013MNRAS.431.3307S}, \citet{Morrison2016} and \citet{2013arXiv1303.4722M},} we measure the over-density, compared with a spatially random distribution of points, of photometric galaxies around each galaxy in the spectroscopic sample, within an annulus of physical scale $\Delta \chi = [\chi_{\rm min}, \chi_{\rm max}]$. The theoretical model for a cross-correlation function between the spectroscopic reference sample in tophat bin $i$ and the photometric base sample is given as:
\begin{equation}
    w_{i}^{\rm RB} \propto b_i^{R} b_i^{B} \left(\frac{\pi_i^{\rm B}}{z_H^{i} - z_L^{i}}\right) \, \overline{w}_{\rm DM, i} \, .
    \label{eq:angcorr_RB}
\end{equation}
Here, $b_i^{R}$ and $b_i^{B}$ denote the value of the redshift-dependent galaxy-dark matter bias of the reference (R) and base (B) samples. The normed histogram bin heights of the base, or photometric, sample redshift distribution are denoted as $\pi_i^{B}$, where $\sum_i \pi_i^{B} = 1$, and the size of a redshift bin is given as $z_H^{i} - z_L^{i}$. The term $\overline{w}_{\rm DM, i}$ denotes the contribution of dark-matter clustering to the cross-correlation signal, which depends on the cosmological model. 

We see that the modelling of the cross-correlation signal depends on the product of two redshift-dependent galaxy-dark matter bias functions that are completely degenerate with the set of parameters $\mathbf{\pi^{B}}$ that parametrizes the redshift distribution of the base sample. Furthermore, since $\overline{w}_{\rm DM, i}$ depends on the cosmological model, it will be computationally expensive to sample over these parameters. 

To reduce the impact of the cosmological model, we want to combine the cross-correlations $\mathbf{w^{\rm RB}}$ with the \mmref{correlations $\mathbf{w^{\rm RR}}$} of the spectroscopic sample. We therefore correlate the spectroscopic sample with itself in a manner analogous to what was just described, i.e., by correlating tophat selected spectroscopic samples with the full spectroscopic sample. 
The corresponding theory prediction then reads
\begin{equation}
        w_{i}^{\rm RR} \propto (b_i^{R})^2 \left(\frac{\pi_i^{R}}{z_H^{i} - z_L^{i}}\right) \overline{w}_{\rm DM, i} \, 
        \label{eq:angcorr_RR}
\end{equation}
where $\pi_i^{R}$ is the normalized histogram height of the spectroscopic (full) sample redshift distribution. Both \mmref{correlation function measurements} ($\hat{\mathbf{w}}^{\rm RB}$ and $\hat{\mathbf{w}}^{\rm RR}$) just described \mmref{are assumed to} individually follow a Gaussian likelihood\footnote{\mmref{In reality, we expect that the likelihood will deviate from the Gaussian assumption \citep[see e.g.][]{10.1093/mnras/stz558}}.} .

Based on these definitions and approximations and the considerations in the previous section, we construct a likelihood based on the ratio between $\mathbf{\hat{w}}^{\rm RB}$ and $\mathbf{\hat{w}}^{\rm RR}$. Under the assumption of a diagonal covariance matrix\footnote{\mmref{While the covariance matrix will be dominated by the diagonal, we can expect, that the cross-correlation measurements in different bins will be correlated. Thus the assumption of a diagonal covariance matrix is an approximation. }} for $\mathbf{\hat{w}}^{\rm RB}$ and $\mathbf{\hat{w}}^{\rm RR}$, we can construct the random variable $\boldsymbol{\Gamma}$ for bin $i$ with components
\begin{equation}
    \Gamma_i^{\rm meas} =  \left(\frac{\hat{w}_i^{RB}}{\hat{w}_i^{RR}}\right)  \, .
    \label{eq:angcorr_ratio}
\end{equation}
We reiterate that both $\hat{w}_i^{RB}$ and $\hat{w}_i^{RR}$ are described by a Gaussian Likelihood. Their respective means and standard deviations are given as $\mu_{\rm RB, i}$, $\mu_{\rm RR, i}$, $\sigma_{\rm RB, i}$, $\sigma_{\rm RR, i}$ respectively.
The theoretical prediction for the transformed random variable $\Gamma_i^{\rm meas}$ is then
\begin{equation}
    \Gamma_i^{\rm theo}(b_i^{B}, b_i^{R}, \pi_i^{B}, \pi_i^{R}) = \frac{b_i^{B}}{b_i^{R}} \frac{\pi_i^{B}}{\pi_i^{R}} \, ,
\end{equation}
and its likelihood:
\begin{equation}
\begin{split}
    & p(\Gamma_{i}^{\rm meas} | \Gamma_i^{\rm theo})  = \frac{b(\Gamma_{i}^{\rm theo}) d(\Gamma_{i}^{\rm theo})}{a^3(\Gamma_{i}^{\rm theo})} \frac{1}{\sqrt{2 \pi} \sigma_{\rm RB, i} \sigma_{\rm RR, i}} \times \\ &\left(\Phi\left(\frac{b(\Gamma_{i}^{\rm theo})}{a(\Gamma_{i}^{\rm theo})}\right) - \Phi\left(- \frac{b(\Gamma_{i}^{\rm theo})}{a(\Gamma_{i}^{\rm theo})}\right)\right) 
     \\ & +\frac{1}{a^2(\Gamma_{i}^{\rm theo}) \pi \sigma_{\rm RB, i} \sigma_{\rm RR, i}} \exp{\left(-\frac{c}{\Gamma_{i}^{\rm theo}}\right)} \, . 
    \label{eq:quasi_likelihood_ratio}
\end{split}
\end{equation}
Here $\Phi(z)$ denotes the cumulative distribution function of the zero mean unit variance normal distribution and
\begin{align}
    a(\Gamma_{i}^{\rm theo}) &= \sqrt{\frac{1}{\sigma_{\rm RB, i}^2} (\Gamma_{i}^{\rm theo})^2 + \frac{1}{\sigma_{\rm RR, i}^2}} \\
    b(\Gamma_{i}^{\rm theo}) &= \frac{\mu_{\rm RB, i}}{\sigma_{\rm RB, i}^2} \Gamma_{i} + \frac{\mu_{\rm RR, i}}{\sigma_{\rm RR, i}^2} \\
    d(\Gamma_{i}^{\rm theo}) &= \exp{\left(\frac{b(\Gamma_{i}^{\rm theo})^2 - c a(\Gamma_{i}^{\rm theo})^2}{2 a(\Gamma_{i}^{\rm theo})^2}\right)} \\
    c &= \frac{\mu_{\rm RB, i}^2}{\sigma_{\rm RB, i}^2} + \frac{\mu_{\rm RR, i}^2}{\sigma_{\rm RR, i}^2} \, .
\end{align}
For the following discussion we define $\mathbf{n} = \frac{\boldsymbol{\pi}}{z_H - z_L}$, i.e. the variables $\boldsymbol{\pi}$, $\mathbf{n}$ refer to histogram heights normalized to sum to unity and to unit area respectively. The variable $\mathbf{n}^{\rm B}$ and $\mathbf{n}^{\rm R}$ refer to the histogram heights of the base and reference samples. 
This likelihood assumes independence between neighboring redshift bins \rev{and linear bias}. We can, however, expect a degree of correlation especially for lower redshift bins due to magnification effects. \rev{Neglecting magnification biases is partially justified, since \citet{2020arXiv201208569G} recently found that the impact on sample photometric redshift posteriors is subdominant compared with other sources of error in the context of the DES Y3 analysis. The validity of this assumption for LSST is however questionable and will require further work.  }

Eq.~\eqref{eq:quasi_likelihood_ratio} is approximately independent of the modelling of the $\overline{w}_{\rm DM}$ term, assuming that we pick sufficiently thin redshift bins to `divide-out' the redshift-dependence of the dark matter clustering term $\overline{w}_{\rm DM}$. Based on the aforementioned independence assumption between redshift bins, the joint likelihood for all bins $i$ now reads: 
\begin{equation}
    p(\mathbf{\Gamma} | \mathbf{b^{b}}, \mathbf{b^{R}}, \mathbf{n^{R}}, \mathbf{n^{B}}) = \prod_{i = 1}^{N_{\rm bins}} p(\Gamma_{i} | b^{B}_i, b^{R}_i, n^{R}_i, n^{B}_i)
\end{equation}
The function that describes the set of ratios $b_i^{B}/b_i^{R}$ will be denoted as $C(z, \Delta \chi_{\bot})$ and depends both on redshift and the size of the annulus $\Delta \chi$. For a selected annulus size we will use the abbreviation $C(z)$. 

\section{The Composite Likelihood}
\label{subsec:composite_like}
To formulate a joint likelihood for the data vector of both galaxy positions and photometry, we use the composite likelihood ansatz \citep[e.g.][]{Varin11anoverview} that uses the product of marginal likelihoods for both the photometry $\mathbf{\hat{F}}$ and the vector of cross-correlation functions $\boldsymbol{\Gamma}$:
\begin{equation}
    p(\mathbf{\hat{F}}, \mathbf{\Gamma} | \overline{\mathbf{n}}^{\rm B, sys}, \overline{\mathbf{n}}^{\rm B}, \overline{\mathbf{n}}^{\rm R}, \mathbf{C(z)}) = p(\mathbf{\hat{F}} | \overline{\mathbf{n}}^{B})^{\upsilon_1} \, p(\mathbf{\Gamma} | \mathbf{\overline{n}^{\rm B, sys}}, \mathbf{C(z)})^{\upsilon_2} \, ,
    \label{eq:composite_like}
\end{equation}
\mmref{where $\boldsymbol{\upsilon}$ are weights 
that can be selected to improve the efficiency of the estimation \citep[see e.g.][]{Varin11anoverview} by increasing the influence of one part of the composite likelihood over the other. Furthermore the composite likelihood can be conditioned on auxiliary parameters such as the field. For simplicity we consider here only the simple case of $\boldsymbol{\upsilon} = 1$ and refer for an additional discussion to \S~\ref{sec:future_work}.}

We note that measurements of LSS and weak lensing often use galaxy samples that are selected by increasing redshift, to form tomographic bins. This analysis methodology can be incorporated into Eq.~\eqref{eq:composite_like} by replacing $\mathbf{\hat{F}}$ and $\mathbf{\Gamma}$ with the joint data vectors of the selected galaxy samples, which would include covariances between the $\mathbf{\Gamma}$ measurements for different tomographic bins. Furthermore the quality and number of available photometric bands can change for different spatial areas. Similarly we need to construct a joint data vector of $\mathbf{\hat{F}}$ and $\mathbf{\Gamma}$ that incorporates these covariances. This can be modelled either analytically \citep[e.g.][]{stoyan1994fractals, 2020arXiv200409542S} or by using spatial resampling techniques. 
In this work we will concentrate on the composite likelihood as given in Eq.~\eqref{eq:composite_like}.

\section{Model Evaluation and Parametrization of Systematics}
\label{sec:model_eval_param_sys}
Parameter inference is only a single step in a full statistical analysis and needs to be combined with additional analysis steps. We need to ensure that parameters can be uniquely inferred and the posterior does not exhibit flat regions or strong degeneracies, which can make the application of MCMC techniques difficult \citep[see e.g.][]{RePEc:ecm:emetrp:v:39:y:1971:i:3:p:577-91, Raue_joining_forces}. Furthermore, one needs to investigate the sensitivity of the results against changes in the prior and likelihood. Finally, one has to judge if the inferred posteriors are sensible in the context of the cosmological/astrophysical model and evaluate if the fitted model is a good representation of the observed data. The last step will be the topic of this section. \S~\ref{subsubsec:post_pred_tests} describes posterior predictive checks as a means to evaluate the goodness of fit of the model and in \S~\ref{subsubsec:parametrizing_systematics} we propose a method to parametrize systematics due to biased photometric likelihoods.


\subsection{Model Evaluation: Posterior Predictive Checks}
\label{subsubsec:post_pred_tests}
The idea of posterior predictive checks (PPC) is to simulate synthetic data from a fitted model, that is then compared with the original measurements to serve as an internal consistency check. For example there exist several approaches that allow us to estimate the quality of probability calibration based on the distribution of posterior predictive $p$-values \citep[e.g.][]{Gelman96posteriorpredictive}. This paper will only provide a short discussion of posterior predictive testing, which is still an area of active research. 
The basic idea of model checking is to investigate if data predicted by the fitted model \mmref{is representative of} the observed data. 

Starting from the composite likelihood defined in Eq.~\eqref{eq:composite_like}, the posterior predictive distribution reads 
\begin{equation}
    p(\mathcal{D}_{\rm rep} | \mathcal{D}) = \iint \mathrm{d}\mathbf{n^{B}} \, \mathrm{d}\mathbf{C(z)} \, p(\mathcal{D}^{\rm rep} | \mathbf{n^{B}}, \mathbf{C(z)}) p(\mathbf{n^{B}}, \mathbf{C(z)} | \mathcal{D}) \, ,
\end{equation}
where $\mathcal{D} = \{\Gamma(w^{\rm BR}, w^{\rm RR}), \hat{\mathbf{F}}\}$ and $\mathcal{D}^{\rm rep}$ denote the replicated, or predicted, measurements sampled from the fitted model. \rev{The posterior predictive distribution can then be compared with the original measurement to check for consistency. An agreement between the original measurement and the posterior predictive is an indicator for model consistency. This procedure can be formalized in a p-value. } 

We note that our model specifies only the ratio between $w^{\rm RB}$ and $w^{\rm RR}$. However, we can always sample replications for one of these quantities using the measurement of the other via the data transformation specified in Eq.~\eqref{eq:angcorr_ratio}. 


To avoid confusion, it is important to clarify that posterior predictive checking and model comparison, while often methodologically similar, have different goals. Posterior predictive checks aim to provide internal consistency tests for a given model and inference framework. Model comparison/combination has the goal to compare/combine multiple models based on some measure of fitting accuracy. However, model comparison and combination can also be based on posterior predictive accuracy \citep[see e.g.][]{2013arXiv1307.5928G}. In the development of new models and the evaluation of directions for improvement, both of these concepts work together.


\subsection{Parametrizing Systematics: Smoothing Kernel}
\label{subsubsec:parametrizing_systematics}
In \S\ref{subsec:testing_model}, we will discuss how miscalibrated likelihoods can lead to systematic biases and uncertainties in the deconvolution operation. Our goal in this section is to include a simple transformation into the model that parametrizes these systematics. A simple choice is the Gaussian convolution kernel, which modifies the sample redshift distribution as
\mmref{
\begin{equation}
    p(z | \mathbf{n^{\rm B, sys}}) = \int \mathrm{d}\boldsymbol{\tau} \, p(\boldsymbol{\tau}) \int p(z | \overline{z}, \boldsymbol{\tau}) p(\overline{z} | \mathbf{n^B}) \, \mathrm{d}\overline{z} \, .
\end{equation}}
Here $\mathbf{n^{\rm B, sys}}$ denotes the parameters that describe the sample redshift distribution after the convolution is applied to the original sample redshift distribution $p(\overline{z} | \mathbf{n^B})$, where the convolution kernel is a gaussian with standard deviation and shift in the mean $\boldsymbol{\tau} = (\Delta \mu, \Delta \sigma)$: 
\mmref{
\begin{equation}
    p(z | \overline{z}, \boldsymbol{\tau}) = \frac{1}{\sqrt{\left(2 \pi \Delta \sigma\right)^2}} \exp{\left[-\frac{1}{2} \left(\frac{z - \overline{z} + \Delta \mu}{\Delta \sigma}\right)^2\right]}
\end{equation}}
Assuming the same histogram parametrization for $p(\overline{z} | \mathbf{n^{B}})$ as for $p(z | \mathbf{n^{B, sys}})$, we see that this implies an affine transformation $\mathbf{n^{\rm B}} \xrightarrow[]{} \mathbf{A}(\tau) \cdot \mathbf{n^{\rm B}}$
 where the matrix $\mathbf{A}$ is given as
 \mmref{
\begin{equation}
\begin{split}
    A_{\beta \, j} &= \Phi\left(z_H^j \bigg| \frac{z_H^\beta - z_L^\beta}{2} + \Delta \mu, \Delta \sigma\right) - \Phi\left(z_L^j \bigg| \frac{z_H^\beta - z_L^\beta}{2} + \Delta\mu, \Delta \sigma\right) \\
     &= \frac{1}{2} \left(\erf{\left(\frac{z_H^j - \left(\frac{z_H^\beta - z_L^\beta}{2}\right) - \Delta \mu}{\Delta \sigma \sqrt{2}}\right)} - \erf{\left(\frac{z_L^j - \left(\frac{z_H^\beta - z_L^\beta}{2}\right) - \Delta \mu}{\Delta \sigma \sqrt{2}}\right)}\right) \, ,
\end{split}
\end{equation}
where $\Phi$ denotes the cumulative distribution function of a normal distribution and $\erf$ the error function. We  choose a Gaussian smoothing kernel since it has been shown that unbiased cosmological inference from measurements of weak lensing and LSS critically depends on accurate recovery on the mean and standard deviation of the photometric sample redshift distribution. Biases in both of these statistics can be parametrized using this kernel.} In contrast to the normal distribution, which exhibits a closed form solution under an affine transformation, the logit-normal does not have this property. However we empirically find that we can approximate the shape of the distribution after an affine transformation as a logit-normal distribution to good accuracy. For a given set of $\boldsymbol{\tau}$, we can find the updated parameter values by assumed density filtering.

While marginalization using sampling techniques is possible, we choose a computationally efficient approximation and marginalize over a discrete model set that consists of different smoothing sizes $\Delta \sigma_i$ and $\Delta \mu = 0$
\begin{equation}
    p(\mathbf{nz}^{B}, \mathbf{C(z)} | \boldsymbol{\Gamma}, \mathbf{\hat{F}}) = \sum_{\Delta \sigma_i} p(\Delta \sigma_i | \boldsymbol{\Gamma}, \mathbf{\hat{F}}) \, p(\mathbf{nz^B}, \mathbf{C(z)} | \boldsymbol{\Gamma}, \mathbf{\hat{F}}, \Delta \sigma_i)  \, .
\end{equation}
\mmref{This assumes that there is no systematic shift in the sample redshift distribution, and deviations from the true underlying distribution are due to miscalibrated but on average unbiased individual galaxy likelihoods. Furthermore, we will assume that $p(\Delta \sigma_i | \boldsymbol{\Gamma}, \mathbf{\hat{F}}) = p(\Delta \sigma_i)$, i.e., the smoothing size is a prior choice independent of the data that can be calibrated on simulations. For simplicity we will use a flat prior $p(\Delta \sigma)$ here. }

\subsection{Complete model summary}
Here we review and summarize our complete model. We review the structure of all components in \S~\ref{subsec:model_structure} and review the inference strategy in \S~\ref{subsubsec:model_inference}.
\subsubsection{Model Structure}
\label{subsec:model_structure}
\begin{figure}
   \centering  
   \includegraphics[scale=0.65]{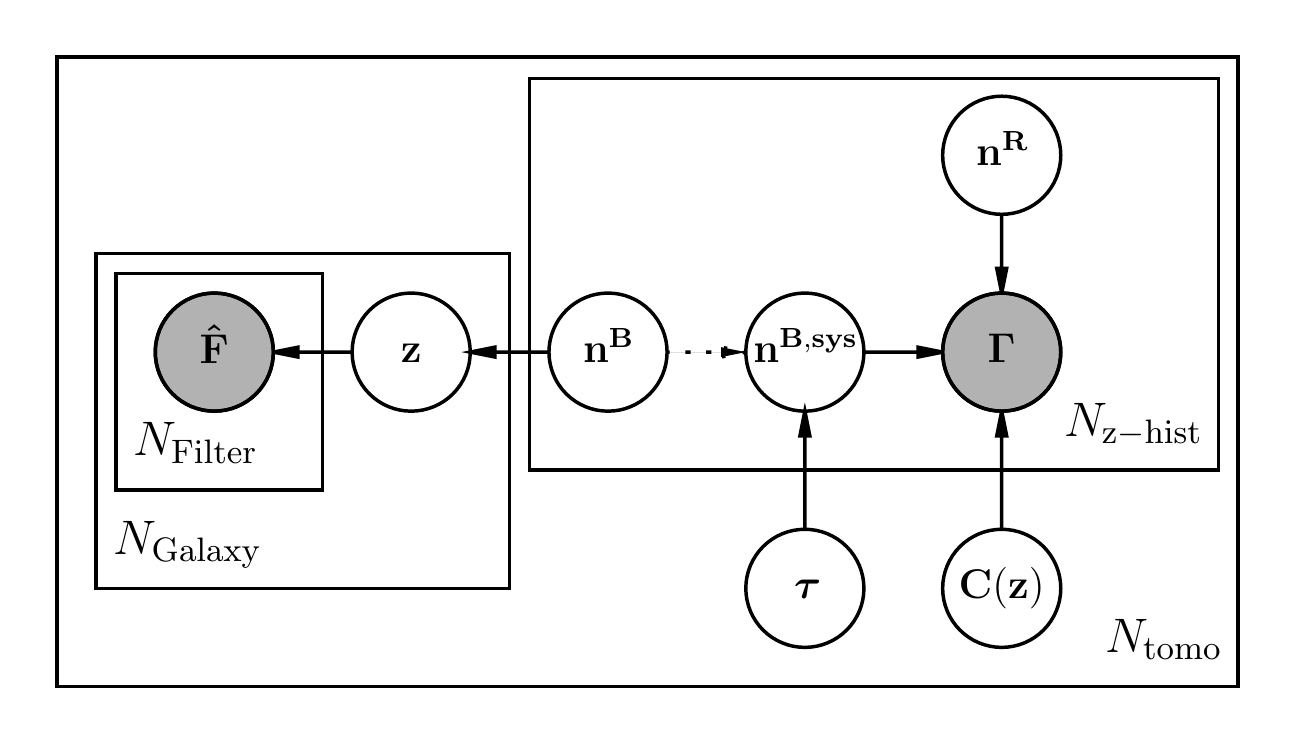}
  \caption{\label{fig:testgraph_simple} Directed graphical representation of the statistical model described in this paper. Empty/filled circles denote random variables that are latent/observed. $N_{(*)}$ denotes the dimensionality of the random variable. Boxes encapsulate random variables with the same dimensionality. Solid/dotted lines indicate random/deterministic relationships between random variables.   } 
\end{figure}
Fig.~\ref{fig:testgraph_simple} summarizes the joint inference strategy presented in the previous sections in a directed graphical model. Each random variable is denoted as a circle, probabilistic/deterministic relationships are denoted as solid/dotted lines. Boxes around random variables denote the dimensionality of the random variable. For example the color vector $\mathbf{f}_i$ is an $N_{\rm filter}$ dimensional random variable for $N_{\rm galaxies}$ in $N_{\rm tomo}$ tomographic bins. Filled circles denote observed random variables, in our case the photometry $\mathbf{\hat{F}}$ and the cross correlation ratios $\boldsymbol{\Gamma}$.

The graphical model is structured into three parts: the left part represents the photometric likelihood, the middle bullets describe our treatment of systematics, and the right part describes the clustering redshift likelihood, which depends on the spectroscopic redshift distribution and the redshift-dependent galaxy-dark matter bias ratio. 

The structure of the graph illustrates the construction of the model via the composite likelihood ansatz discussed in \S~\ref{subsec:composite_like}. It separates the two data sources $\mathbf{\hat{F}}$ and $\boldsymbol{\Gamma}$ in the left and right part of the graph. 
As we are mainly interested in performing inferences on the $\mathbf{n}^{\rm B}$ variables, we marginalize over the $z$ variables, which provides significant computational advantages. The mapping between $\mathbf{n}^{\rm B}$  and $\mathbf{n}^{\rm B, sys}$ takes the form of a deterministic transformation and is therefore indicated by a dotted line. 

While $\mathbf{n}^{R}$ is here treated as a random variable, its `shot noise' uncertainties are very small for the considered sample sizes of the spectroscopic sample. We therefore decided to fix its value to the maximum likelihood value, i.e., the histogram height.  

\subsubsection{Model Inference}
\label{subsubsec:model_inference}
The presented model consists of two likelihood terms and a deterministic transformation $\mathbf{n^{\rm B}} \rightarrow \mathbf{n^{\rm B, sys}}$. Starting with the photometric likelihood, we employ the inference scheme detailed in Appendices~\ref{par:ml_inference} and \ref{par:derive_laplace} that results in a posterior $p(\mathbf{n^{B}} | \mathbf{\hat{F}})$ defined in Eq.~\eqref{eq:logit_normal_post}. We then employ the transformation detailed in \S~\ref{subsubsec:parametrizing_systematics} to parametrize systematics in the inferred posterior from biased photometric likelihoods. This yields a systematics corrected posterior $p(\mathbf{n^{\rm B, sys}} | \mathbf{\hat{F}})$ using the methodology described in \S~\ref{subsubsec:parametrizing_systematics}. The final combination with the clustering likelihood term $p(\mathbf{n^{\rm B, sys}}, \mathbf{C(z)} | \mathbf{\hat{F}}, \boldsymbol{\Gamma}) \propto p(\boldsymbol{\Gamma} | \mathbf{n^{\rm B, sys}}, \mathbf{C(z)}) \, p(\mathbf{n^{\rm B, sys}} | \mathbf{\hat{F}})$ is then performed using a Monte Carlo Markov chain (MCMC) sampling approach. 

We update the parameters $(\mathbf{n^{\rm B, sys}}, \mathbf{C(z)})$ in two sampling blocks: the set of parameters that describe the redshift distribution of the base sample $\mathbf{n^{\rm B, sys}}$ and the parameters $\mathbf{c}$ that describe the evolution of the redshift-dependent galaxy-dark matter bias ratio $\mathbf{C(z)}$, \rev{where we do not impose a prior on the parameters $\mathbf{c}$}. 

Concretely, we iteratively sample from the conditional distributions $p(\mathbf{n^{B, sys}} | \mathbf{\hat{F}}, \boldsymbol{\Gamma}, \mathbf{c})$ and then from $p(\mathbf{c} | \mathbf{\hat{F}}, \boldsymbol{\hat{\Gamma}}, \mathbf{n^{\rm B, sys}})$. This means that we iteratively sample each parameter block in turn, while holding the other parameter block fixed. The sampling method that can be employed to update each parameter blocks is flexible\footnote{We tried several approaches like Hamiltonian MCMC or Elliptical Slice sampling. All approaches work well; we discuss here the structurally simplest scheme.}. We use a Metropolis-Hastings sampling scheme to \mmref{sample} the $\mathbf{c}$ parameters. To sample from the conditional $p(\mathbf{n^{B, sys}} | \mathbf{\hat{F}}, \boldsymbol{\Gamma}, \mathbf{c})$ we also employ a Metropolis scheme, however we perform the sampling not in terms of the $\mathbf{n^{\rm B, sys}}$ parameters, but in logit space, i.e. in terms of the $\mathbf{y}$ parameters that are connected with $\mathbf{n^{\rm B, sys}}$, or their normalized analog $\mathbf{\pi^{\rm sys}}$, via Eq.~\eqref{eq:additive_logistic}. In this way we can utilize proposal distributions that are defined in real space to sample a distribution defined on the simplex.  \mmreff{We reiterate that the posterior $p(\mathbf{n^B}|\mathbf{\hat{F}})$ has, in our framework, an analytical form and sampling is therefore very efficient. However if we include a treatment of systematics or a clustering redshift likelihood into the inference, we need to employ sampling approaches because the posterior has no longer a closed form solution.}     

\section{Forecast using Simulation Data}
\label{sec:forecast_ideal_data}
\begin{figure}
   \centering  
   \includegraphics[scale=0.65]{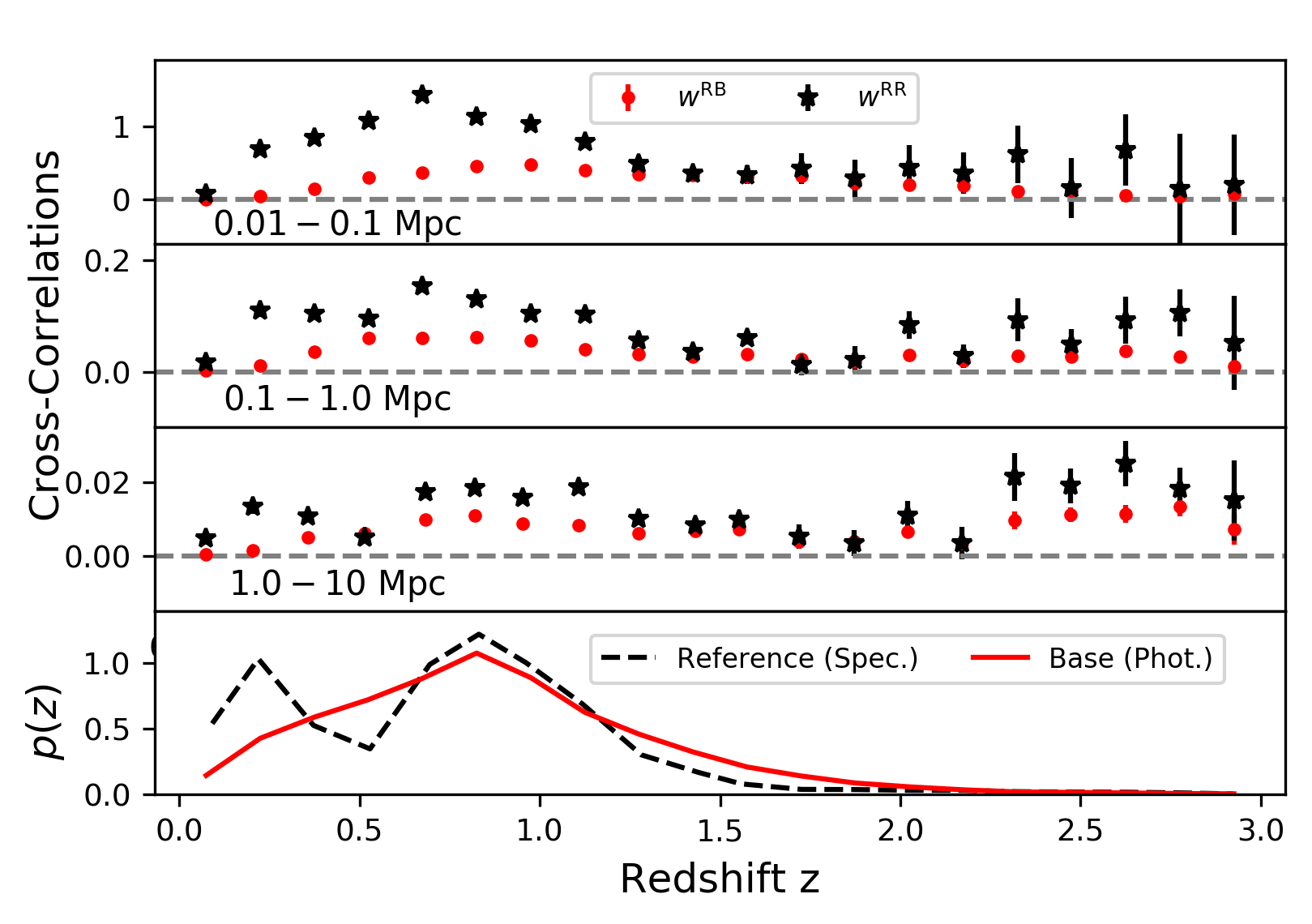}
  \caption{\label{fig:meas_wx_plot} The top three panels show cross-correlation measurements between the photometric base sample and the spectroscopic reference sample $w^{\rm RB}$ and \mmref{correlation function measurements} of the spectroscopic reference sample $w^{\rm RR}$ for 3 different annuli (see \S~\ref{sec:clustering_likelihood}) 
  as a function of redshift. The errorbars correspond to the $\pm 1\sigma$ measurement errors of the correlation function measurements.  The lowest panel plots the \mmref{true} sample redshift probability density function of the spectroscopic reference sample (`Spec.') and the photometric base sample (`Phot.'). \rev{The grey horizontal dashed lines in the three upper panels indicate the zero-line.}}  
\end{figure}
To demonstrate the effectiveness of our inference methodology, we consider an idealized setup which allows us to forecast the constraints on the sample redshift distribution that we can expect from a DESI-like spectroscopic survey overlapping with the LSST Y10 footprint. We assume in this section that the composite likelihood is well calibrated, both in the clustering and in the template fitting part.
This assumption will likely not hold in practice and we therefore study the impact of likelihood mis-specification on the inference in \S~\ref{subsec:testing_model}. In particular we will evaluate the performance of our methodology by comparing with science requirements of the first year (`LSST Y1') and the 10th year (`LSST Y10') of the LSST data release defined in \citet{2018arXiv180901669T}. \mmref{We note that we will utilize a mock simulation of galaxy photometry likelihoods in \S~\ref{subsec:applying_model} instead of SED likelihoods constructed on the simulated photometry, because the simulated photometry of the DC2 simulations showed a discontinuous color-redshift mapping, which induced an unrealistically large error in our template fitting results. In \S~\ref{subsec:testing_model}, which will discuss aspects of model checking and will not interpret results in the context of LSST science requirements, we will use both Machine Learning and Template Fitting methods described in \S~\ref{sec:pzsample}. \rev{We note that while the discussion so far focussed on a likelihood definition that utilized SED fitting, Machine Learning-based conditional density estimates can be used as a substitute as well. } }

\subsection{Measuring Cross-Correlations}
\label{subsec:meas_cross_correlations}
\begin{figure}
   \centering  
   \includegraphics[scale=0.5]{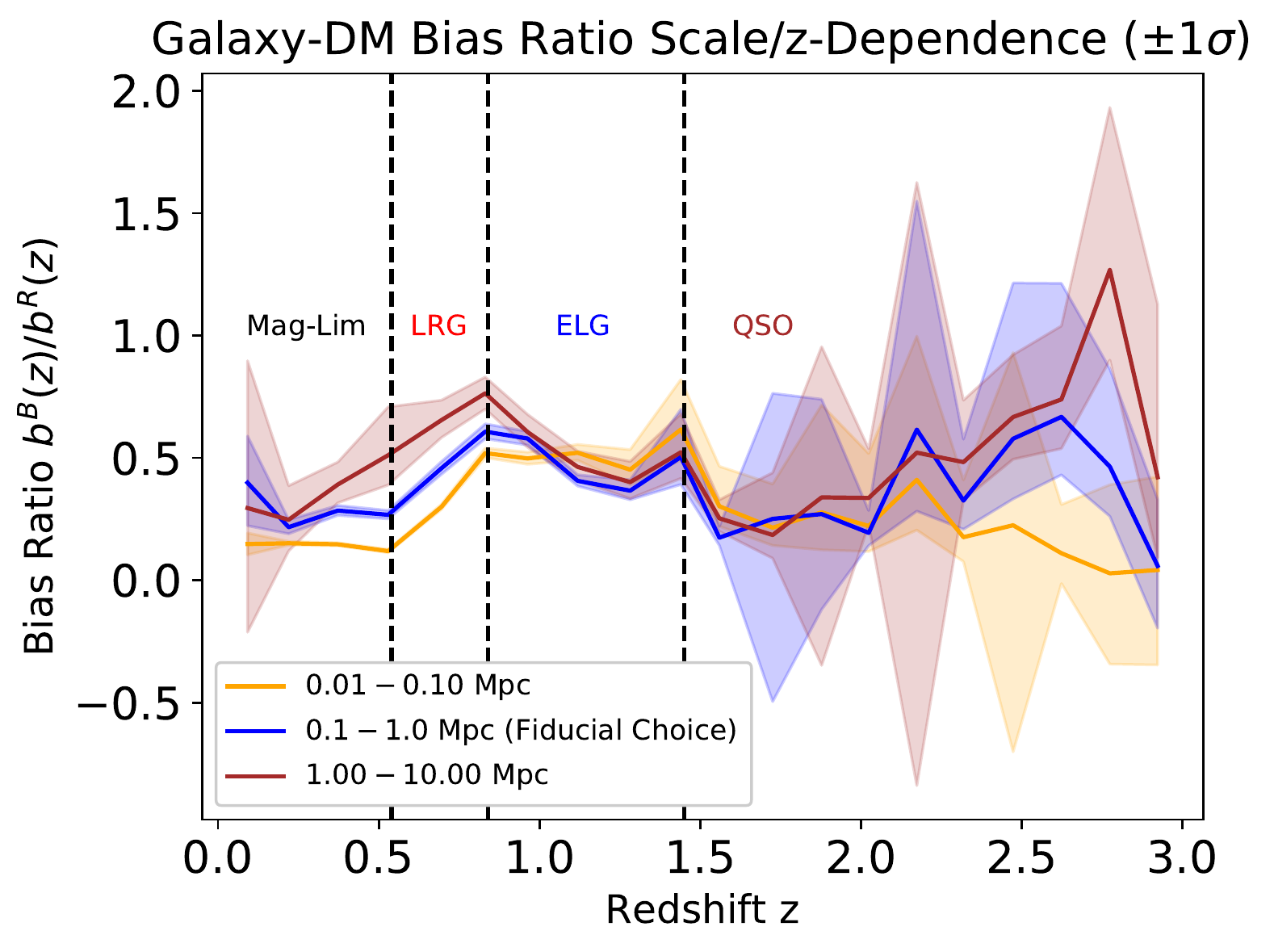}
  \caption{\label{fig:bias_model} Galaxy-dark matter bias ratio as a function of redshift for different scales. We show here the $[16, 84]$ percentiles, that correspond to $\pm 1\sigma$ for a Normal distribution.  The redshift ranges of the different spectroscopic subsamples are plotted by vertical lines.  Within these ranges, the bias ratio is a relatively smooth function of redshift, indicating a smooth redshift dependence of the galaxy-dark matter bias of the photometric sample. At the borders of these ranges, the bias ratio curves are discontinuous.  } 
\end{figure}

We use the software package `the-wizz'\footnote{\url{https://github.com/morriscb/the-wizz}} \citep{Morrison2016} to measure cross-correlations between the reference (spectroscopic) and base (photometric) samples $w^{\rm RB}$ and correlations of the reference sample $w^{\rm RR}$ in 20 equally spaced redshift bins \mmref{from $z \in (0, 3)$, corresponding to $3042$~Mpc comoving distance at the mean redshift of $\left\langle z \right\rangle = 0.88$}. Fig.~\ref{fig:meas_wx_plot} shows these measurements in the three top panels for three annuli (see \S~\ref{sec:clustering_likelihood})  
of $0.01 - 0.1 \, {\rm Mpc}$, $0.1 - 1.0 \, {\rm Mpc}$ and $1.0 - 10 \, {\rm Mpc}$.  
\markus{In the lowest panel we show the sample redshift probability density functions for the reference and base samples. The \mmref{correlations} $\mathbf{w^{\rm RR}}$ are larger than the cross-correlations $\mathbf{w^{\rm RB}}$ in all three panels, implying on average a lower than unity ratio $b^{\rm B}(z)/b^{\rm R}(z)$. The errorbars increase with redshift due to the decreasing number of galaxies, leading to a larger shot noise error. Consider the shape of $\mathbf{w^{\rm RR}}$ and $\mathbf{w^{\rm RB}}$ for the largest annuli $[1.0 - 10.] \, {\rm Mpc}$, in the low redshift range of $z<1.0$. We see that the measurements of $\mathbf{w^{\rm RB}}$ and $\mathbf{w^{\rm RR}}$ roughly resemble the shape of the reference and base sample redshift distributions (lowest panel), implying a roughly linear ratio $b^{B}(z)/b^{R}(z)$ in this redshift range (compare with Fig.~\ref{fig:bias_model}). For smaller annuli, \mmref{the change in slope} around $z=0.5$ is less pronounced, producing a step around $z=0.5$ in the galaxy-dark matter bias ratio. At the high-redshift tail, where the redshift distribution of the reference sample flattens out, we see that $\mathbf{w^{\rm RB}}$ and $\mathbf{w^{\rm RR}}$ are approximately equal. Here, the sample redshift distribution of the base sample is larger than the one of the reference sample. Since the QSO sample will have a larger galaxy-dark matter bias than the base sample, we can expect $b^{B}(z)/b^{R}(z) < 1.0$ at high redshift. }
In order to represent a $5000 \, {\rm deg}^2$ overlap between DESI and LSST Y10, using measurements obtained on the the $300 \, {\rm deg}^2$ CosmoDC2 simulations, we scale the error on these measurements by a factor of 4 \markus{in the following analyses}. 

\mmref{We would like to generate a cross-correlation likelihood that allows us full control over the imposed redshift-dependent galaxy-dark matter bias ratio model and that has roughly\footnote{\mmref{The uncertainty in the correlation function measurements will likely differ from this factor of four scaling 
in the real data. }} the correct width of the full DESI area. Furthermore, since the mean of the ratio distribution depends on both the mean and the variance of $\mathbf{w^{RB}}$ and $\mathbf{w^{RR}}$, scaling the measurement error will induce biases in the mean of this ratio and therefore in the reconstructed sample photometric redshift distribution.}

 To correct for possible biases that would occur when the measurements \mmref{errors} are naively scaled \mmref{and allow for better control over the redshift dependent galaxy-dark matter bias ratio}, we first fit the galaxy-dark matter bias ratio $C(z)$ to the original data within these 20 bins. 
We show the results of this fit in Fig.~\ref{fig:bias_model} for different ranges in physical distance and indicate the redshift range of the different spectroscopic samples by vertical lines, where these limits are meant to guide the eye and do not constitute sharp breaks (compare with Fig.~\ref{fig:merged_calibration}). We see that within these redshift ranges, $C(z)$ is a smooth function and can be fitted by a 3rd degree Chebychev polynomial $C(z) = \sum_{i = 1}^{3} c_i T_i(z)$. Here, $T_i(z)$ denote Chebychev functions and $c_i$ denotes the expansion coefficients. Within the three redshift ranges $\{[0.0, 0.5], [0.5, 0.8], [0.8, 3.0]\}$, we perform a \mmref{regression fit to the median of the $C(z)$ posterior due its heavy tails}. For the following analysis we select the median annuli of $0.1 - 1 \, {\rm Mpc}$, which provides good signal-to-noise, while being less sensitive to small scale effects, than the $0.01 - 0.1 \, {\rm Mpc}$ bin, for which accurate modelling of galaxy-dark matter bias will be more difficult. \mmref{However, we are still considering the non-linear regime, in which more work is needed to model the galaxy-dark matter bias.} 

We scale the \mmref{correlation functions} $w^{RR}$ and $w^{RB}$ defined in Eq.~\eqref{eq:angcorr_ratio} and forecast a new data vector for $w^{RB}$, while holding the measurement of $w^{RR}$ fixed. \mmref{This amounts to multiplying the ratio $\mathbf{w^{RB}}/\mathbf{w^{RR}}$ by a constant for each redshift bin that compensates for the difference in the mean of the reconstructed sample photometric redshift distribution before and after we impose the fitted redshift dependent galaxy-dark matter bias model and perform the scaling. In this way we ensure that our ratio distribution is self-consistent with the photometric sample redshift distribution. } We then use these adjusted measurements in the composite likelihood (Eq.~\ref{eq:composite_like}). \mmref{This correction is necessary because we would otherwise merely use noisy measurements with wrongly decreased errorbars, which will lead to biases in the probability calibration of any inference. Furthermore we want to have control over the underlying redshift dependent galaxy-dark matter bias model to eliminate any additional specification error. It should therefore merely be seen as an approximate forecast of the constraining power that cross-correlations will add to the composite likelihood and a demonstration of the inference methodology. It is not an accurate treatment of galaxy-dark matter bias or the correlation function measurements expected in the final LSST measurement. For this we would require a more realistic simulation of the DESI-like spectroscopic sample, the final area and a much better understanding of the galaxy-dark matter bias of each galaxy population, all of which are subjects of active investigation in the field.} \rev{Similarly, we note that the described construction effectively corrects for systematics in the correlation function measurements, which is an optimistic assumption that will not hold true in real data. }

\begin{figure*}
   \centering  
   \includegraphics[scale=0.8]{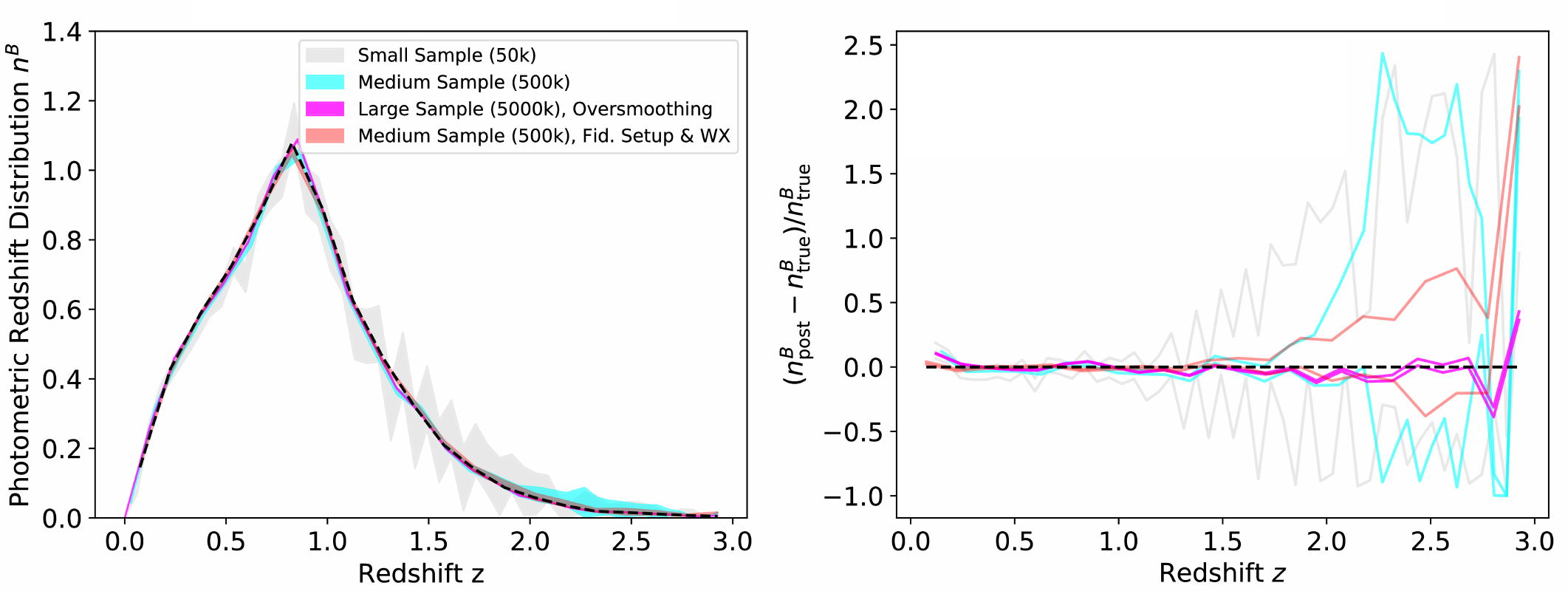}
  \caption{\label{fig:merged_graham_ideal} \textit{Left panel:} Posteriors of the sample redshift probability density function $p(z)$ of the photometric sample (short: photometric redshift distribution) parametrized by the parameters $\mathbf{n^B}$ for different setups listed in Tab.~\ref{tab:tableIdentifiers}. The x-axis shows the redshift value $z$, the y-axis the value of the $\mathbf{n^B}$ parameters. The errorbars are the $[16, 84]$ percentiles, which would correspond to $1 \sigma$ intervals for a normal distribution. The black dashed curve shows the spectroscopic redshift distribution in the binning used by the cross-correlation measurements. We consider four cases, and refer to Tab.~\ref{tab:tableIdentifiers} and \S~\ref{sec:forecast_ideal_data} for details on the experimental setup. We highlight a variance-dominated posterior `Small Sample (50k)', which shows a characteristic alternating, or `zig-zag' pattern, as well as a comparison between the cyan `Medium Sample (500k)' and `Medium Sample (500k), Fid. Setup + WX' posteriors. Here, the latter includes a cross-correlation `WX' data vector in its likelihood. This reduces the error especially in the high-redshift tail of the distributions. \textit{Right panel:} The y-axis shows the relative difference between the posterior of the photometric redshift distribution parametrized by the $\mathbf{n^{B}_{\rm post}}$ parameters and the spectroscopic redshift distribution $\mathbf{n^{B}_{\rm true}}$ (black dashed curve in the left panel).    } 
\end{figure*}
\subsection{Applying the Model}
\label{subsec:applying_model}
\begin{table*}
\centering
 \begin{tabular}{c c c c c c c} 
 \hline
 Abbreviation & Sample Size & Figure & Tikhonov Reg. $\alpha$ & Initial Binning & Effective Binning & WX \\ [0.5ex] 
 \hline\hline
Small Sample (50k) & 50k & Fig.~\ref{fig:merged_graham_ideal} & 0.1 & 50 & 50 & - \\ \hline
Medium Sample (500k) & 500k & Fig.~\ref{fig:merged_graham_ideal}  & 0.08 & 50 & 31 & - \\ \hline
Large Sample (5000k) Oversmoothing & 5000k & Fig.~\ref{fig:merged_graham_ideal} & 0.0001 & 25 & 25 & - \\ \hline
Medium Sample (500k), Fid. Setup \& WX & 500k & Fig.~\ref{fig:merged_graham_ideal} & 0.08 & 50 & 31 & \checkmark \\ \hline\hline
Tik. Regul. Low & 5000k & \mmref{Fig.~\ref{fig:boxplot_mean}} & 0.0001 & 50 & 25 & - \\ \hline
Tik. Regul. Medium & 5000k & \mmref{Fig.~\ref{fig:boxplot_mean}} & 1 & 50 & 25 & - \\ \hline
Tik. Regul. High & 5000k & \mmref{Fig.~\ref{fig:boxplot_mean}} & 10 & 50 & 25 & - \\ \hline
Oversmoothing & 5000k & \mmref{Fig.~\ref{fig:boxplot_mean}} & 0.0001 & 25 & 25 & - \\ \hline
Tik. Regul. Low + WX & 5000k & \mmref{Fig.~\ref{fig:boxplot_mean}} & 0.0001 & 50 & 25 & \checkmark \\ \hline
 \end{tabular}
 \caption{\label{tab:tableIdentifiers}  Summary of the different configurations that we test in this work. The first column lists the abbreviations, the second refers to the Figure where the setup is analysed. The next columns list the value of the Tikhonov regularization parameter $\alpha$ (see \S~\ref{subsub:tikhonov_regul}), the number of initial bins, the effective bin number after (potentially) applying merging bin regularization (see \S~\ref{subsubsection:merging_bin}) and an indicator if the composite likelihood includes the cross-correlation data (see \S~\ref{sec:clustering_likelihood}).   }
\end{table*}
For the photometric part of the composite likelihood we assume a redshift scaling of $\sigma(z) = 0.02 \, (1 + z)$, where $z$ denotes the true, or spectroscopic, redshift. This scaling is a photometric redshift performance benchmark for LSST frequently adopted in the literature \citep[e.g.][]{2020AJ....159..258G} and defined in the LSST science requirements document\footnote{\url{https://docushare.lsst.org/docushare/dsweb/Get/LPM-17} (page 4)}. We then generate a mock catalog by sampling values from the true sample redshift distribution and generate a catalog of mock likelihoods by scattering these values within this redshift error model. We reiterate that we assume here that the redshift likelihoods \mmref{constructed from the galaxies' photometry}, mimicked here by the aforementioned redshift error model, is perfectly known. We note that this is an idealized assumption that we impose to demonstrate the methodology described in the previous sections.

\mmref{Tab.~\ref{tab:tableIdentifiers}} summarizes the different configurations we use in this work. In particular we investigate posteriors obtained using several different sample sizes and regularization techniques. In particular, the first and second columns show the abbreviation used in the text and the corresponding figure. The generated sample size of the mock catalog is shown in the third column. The columns `Tikhonov Reg.~$\alpha$', `Initial Binning' and `Effective Binning' list the value of the Tikhonov regularization parameter $\alpha$ (see \S~\ref{subsub:tikhonov_regul}), as well as the used initial and effective bin number\footnote{As an approximate rule, one can expect a noisy deconvolution if no prior is applied, if the size of the bins is smaller than $\pm 1\sigma$ range of the individual galaxy redshift likelihoods for moderate sample sizes of the order $10^5$ galaxies. For our redshift range and photometric redshift scatter this would imply an effective number of bins of $30-40$.}, i.e., the histogram bin number after the merging bin regularization scheme.  The final column indicates whether the cross-correlation data vector is included in the composite likelihood. Fig.~\ref{fig:merged_graham_ideal} shows a selection of posteriors using setups from Tab.~\ref{tab:tableIdentifiers}. The left hand panel shows the obtained probability density function, the right panel the relative difference between these posteriors and the spectroscopic redshift distribution that is shown as the black dashed line in both panels. The error bars are the $[16, 84]$ percentiles, corresponding to a Gaussian $\pm 1 \sigma$ interval.  

The `Small Sample (50k)' setup highlights the noisy, i.e., variance-dominated deconvolution, where we clearly see the comparatively large and fluctuating errorbars in Fig.~\ref{fig:merged_graham_ideal}. We note that imposing a smoothing method will reduce these features. However this can come at the expense of additional biases as discussed later, and the characteristic covariance structure in the posterior is not {\em a priori} problematic, as long as draws from the posteriors are bounded and well-defined. Merging bin regularization exploits this anti-correlation structure to provide an `objective' regularization without the need to carefully motivate an external prior or smoothing model choice. 

An alternative that provides additional physical motivation is the inclusion of clustering redshift measurements into the composite likelihood. This can be seen by comparing the posteriors from the `Medium Sample (500k)' and the `Medium Sample (500k), Fid. Setup \& WX' cases. The corresponding results in Fig.~\ref{fig:merged_graham_ideal} show that for the same regularization, the inclusion of the cross-correlation data into the likelihood decreases the uncertainties, which is especially visible in the high-redshift tail.  
We note that these results are dependent on the chosen galaxy-dark matter bias model. As discussed in the beginning of this section, the parametrization used here is very flexible and the effective number of parameters can likely be reduced, if a more physical model is chosen. In this regard, we can view the presented reduction in the statistical error due to clustering redshifts as conservative. As mentioned previously, biases due to ill-motivated regularization choices play an important role, especially for large sample sizes, where the statistical error is small. We illustrate this here in the `Large Sample (5000k) Oversmoothing' case, by deliberately choosing a coarser binning without merging bin regularization. We clearly see that the statistical error is quite small with the bias dominating. 

In order to investigate the quality of probability calibration, we consider the posterior \mmref{distribution over the mean values of photometric sample redshift distributions drawn from the posterior of $\mathbf{n^{B}}$.\footnote{\mmref{Concretely, we draw a number of $n^{B}$ realizations that each parametrize a photometric sample redshift distribution and evaluate the mean on each of these distributions.}}} This is a reasonable choice, as it has been shown that accurate modelling of weak lensing and LSS critically depends on accurate recovery of the posterior mean.

\begin{figure}
   \centering  
   \includegraphics[scale=0.55]{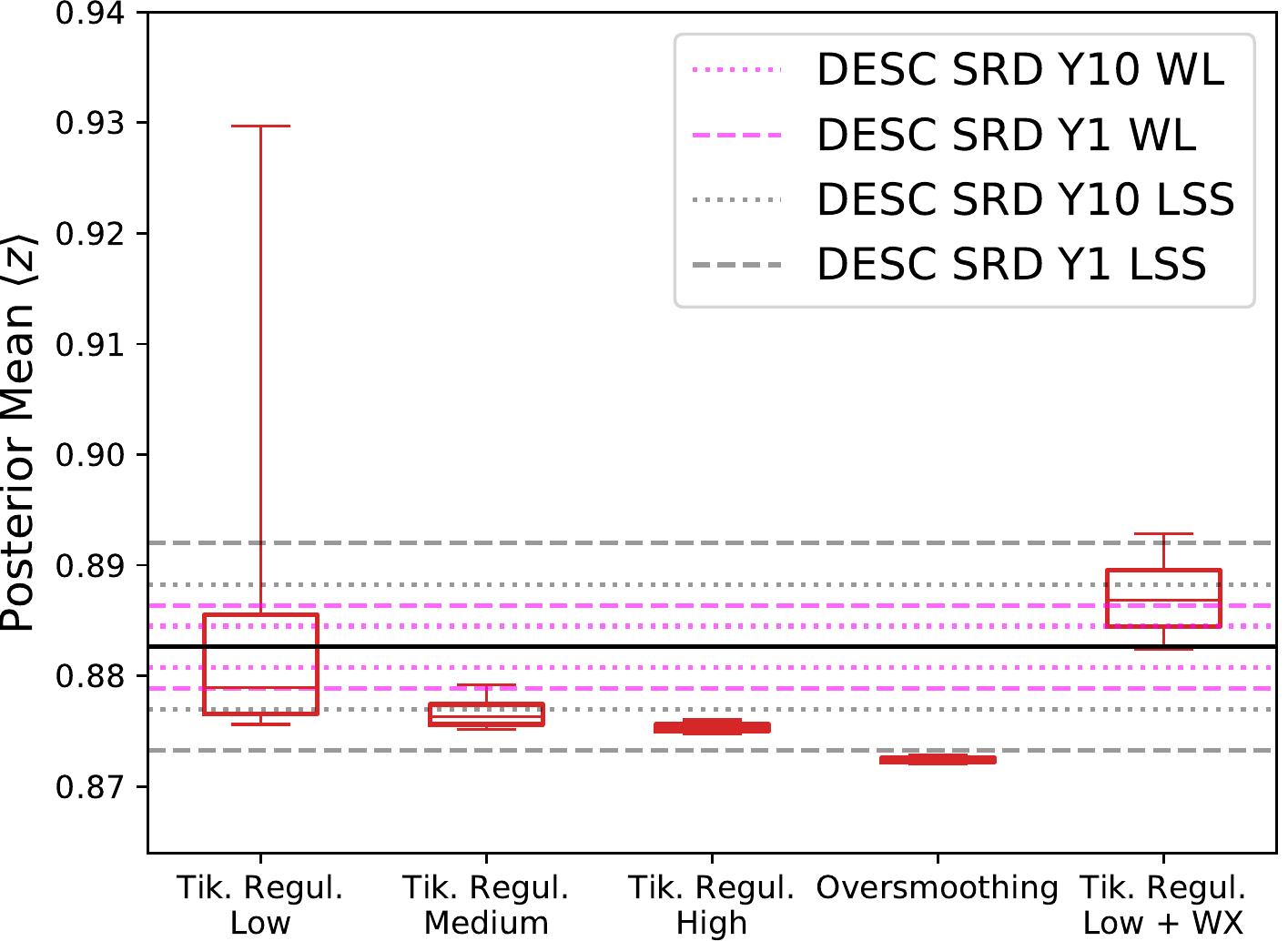}
  \caption{\label{fig:boxplot_mean} Boxplot illustration of the  mean of the posterior sample redshift probability density function of the photometric sample (short: posterior mean) for different experimental setups listed in Tab.~\ref{tab:tableIdentifiers} and detailed in \S~\ref{sec:forecast_ideal_data}. The x-axis lists the different scenarios, the y-axis the value of the posterior mean. The box shows the $[16, 84]$ percentiles, the vertical lines with horizontal edges (whiskers) show the $[2.5, 97.5]$ percentiles, corresponding to the $1\sigma$ and $2\sigma$ intervals for the normal distribution. The horizontal line in the box is the median. \mmref{The (dashed/dotted), (grey/magenta) lines correspond to the requirement on the uncertainty of the posterior mean as quoted in the LSST DESC Science Requirements Document \citep[DESC SRD,][]{2018arXiv180901669T} for (Y1/Y10) (LSS/WL) measurements.} The central, solid grey line is the mean of the true redshift distribution, shown as the dashed black line in the left panel of Fig.~\ref{fig:merged_graham_ideal}. All results have been obtained using a mock catalog of 5000k galaxies with photometric redshift scatter that is perfectly calibrated. We highlight the decrease in the statistical error and potential increase in systematic bias for larger regularization, going from the leftmost to the fourth case. The rightmost boxplot shows the impact of including clustering redshift information into the likelihood.     } 
\end{figure}

Fig.~\ref{fig:boxplot_mean} shows five boxplots that each visualize the distribution of the posterior mean that corresponds to a different setup under consideration. The box edges denote the $[16, 84]$ percentiles, and the definition of the whiskers, i.e., the thin vertical lines with short horizontal edges represent the $[2.5, 97.5]$ percentiles. The horizontal line within the box represents the median\footnote{We note that the original definition of the boxplot uses a different definition of the box size and the whiskers. We refer to \citet{boxplots_ref} for more details.}. The x-axis shows several different scenarios, as listed in Tab.~\ref{tab:tableIdentifiers}; the y-axis shows the value of the posterior mean. The middle solid \mmref{black line} corresponds to the mean of the true redshift distribution, shown as the dashed black line in the left panel of Fig.~\ref{fig:merged_graham_ideal}. We reiterate that all results have been obtained using a mock catalog containing 5000k galaxies. \mmref{The (dashed/dotted), (grey/magenta) horizontal lines represent the requirement values for (Y1/Y10), (LSS/WL) measurements as given in the LSST DESC Science Requirements Document \citep[DESC SRD][]{2018arXiv180901669T}.} \mmref{We note that the LSST DESC Science Requirements Documents considers a tomographic analysis and not a single bin, as we do here. We therefore restrict ourselves to a qualitative comparison. Furthermore it should be noted that higher order moments of the photometric sample redshift distribution will also correlate with cosmological parameters, especially for a clustering likelihood \citep[see e.g.][]{2020JCAP...03..044N, 2020arXiv200714989H}, and our metric is therefore bound to be incomplete. Redefining these metrics and requirements is the subject of ongoing work. }

We consider three scenarios: `Tik Regul. Low', `Tik. Regul. Medium' and Tik. Regul. High'. As can be seen from Tab.~\ref{tab:tableIdentifiers} these scenarios differ by their value of the Tikhonov regularization parameter $\alpha$. With increasing $\alpha$, the error bars decrease and the bias in the results increases. When comparing this with the `oversmoothing' results, we see the same pattern. This similarity in behaviors arises because both a large $\alpha$ and choosing large bins reduces the variance of each bin. 

Finally we show the impact of including the cross-correlation measurements into the data vector in the `Tik. Regul. Low + WX' scenario, which adds clustering \mmref{information} to the `Tik. Regul. Low' scenario. When comparing these two cases, we see that the distribution of the posterior mean is now symmetric and reasonably centered within the science requirements. In particular we note that the uncertainties are still much larger when compared with the previously considered, strongly regularized cases. This shows that while the effect of reducing the variance of the posteriors is similar when using regularization or including cross-correlation data into the composite likelihood, the posteriors can be much better calibrated in the latter case. Using a smoothing, or regularization, method essentially makes assumptions about the true shape of the distribution without strict data evidence. In contrast, adding cross-correlations to the composite likelihood adds this information in a physical, data-driven way. 

Another effect that needs consideration is the increase in the intrinsic estimator bias due to the `downsampling' of the probability density function to a lower resolution, e.g., by picking larger bin width or by imposing a different regularization or smoothing scheme. This loss in resolution implies that we inadvertently limit the accuracy with which small scale structure can be reconstructed in the density field along the line-of-sight. As demonstrated and studied in detail in \citet{2017MNRAS.466.2927R}, this effect can lead to biases in the cosmological parameter inference that are often small, but that would need scrutiny for upcoming data analyses. Since we gave a detailed description of this effect in \citet{2017MNRAS.466.2927R} including schemes to detect and mitigate these effects, we do not focus on it in  detail here. However, this effect can be illustrated for the current setup, since the merging bin regularization downsamples the resolution to a relatively coarse grid of 31 bins. Furthermore consider the redshift distribution of true redshifts discretized using the 20 bin grid used to obtain the cross-correlation measurements. Since we use this distribution, i.e. the black curve in Fig.~\ref{fig:merged_graham_ideal}, as a reference, we also have to consider its intrinsic discretization error. Concretely, when comparing the mean estimated from this curve with the sample mean, we obtain a difference in these values of 0.0079.  While this is of the same order as the Y1 science requirements in Fig.~\ref{fig:boxplot_mean}, Y10 requirements will necessitate an increase in sample size or the inclusion of cross-correlation constraints that will allow us to perform inference at a higher resolution. Due to the slow expected convergence of \markus{deconvolution} estimators with sample size \citep[see e.g.][]{doi:10.1080/01621459.1988.10478718}, it is likely that several orders of magnitude increase in sample size will be necessary. This is attainable for the large numbers of observed galaxies in LSST Y10, and our methodology can scale to large sample sizes. However, in order to reach the sample sizes that are expected for LSST observations, we need to develop an implementation that optimizes storage space and uses an efficient parallelization strategy, which is beyond the scope of this work.  

Alternatively, we could use a different scheme that employs a continuous model like logistic Gaussian processes \citep{2020MNRAS.491.4768R} or Dirichlet processes. The convergence of these density estimators will likely be better, however they will also require additional computational overhead in the inference. A detailed study of estimator convergence is needed to settle on a recommendation and prove significant improvement over the simple histogram scheme employed here.

Most importantly, however, it is likely that systematic errors due to the miscalibration and mis-specification of the composite likelihood, either by a suboptimal galaxy-dark matter bias model or due to miscalibrated SED likelihoods, will lead to an error budget that will dominate the aforementioned errors. If the mis-specification can be parametrized and marginalized over, the variance of the parameter posteriors will be increased, otherwise they will lead to biases in the resulting parameter posteriors. In the following section we will discuss these sources of error. We will showcase the usage of posterior predictive checks as a way to detect miscalibration and suggest procedures for consistent model checking and refinement. 

\begin{figure}
   \centering  
   \includegraphics[scale=0.55]{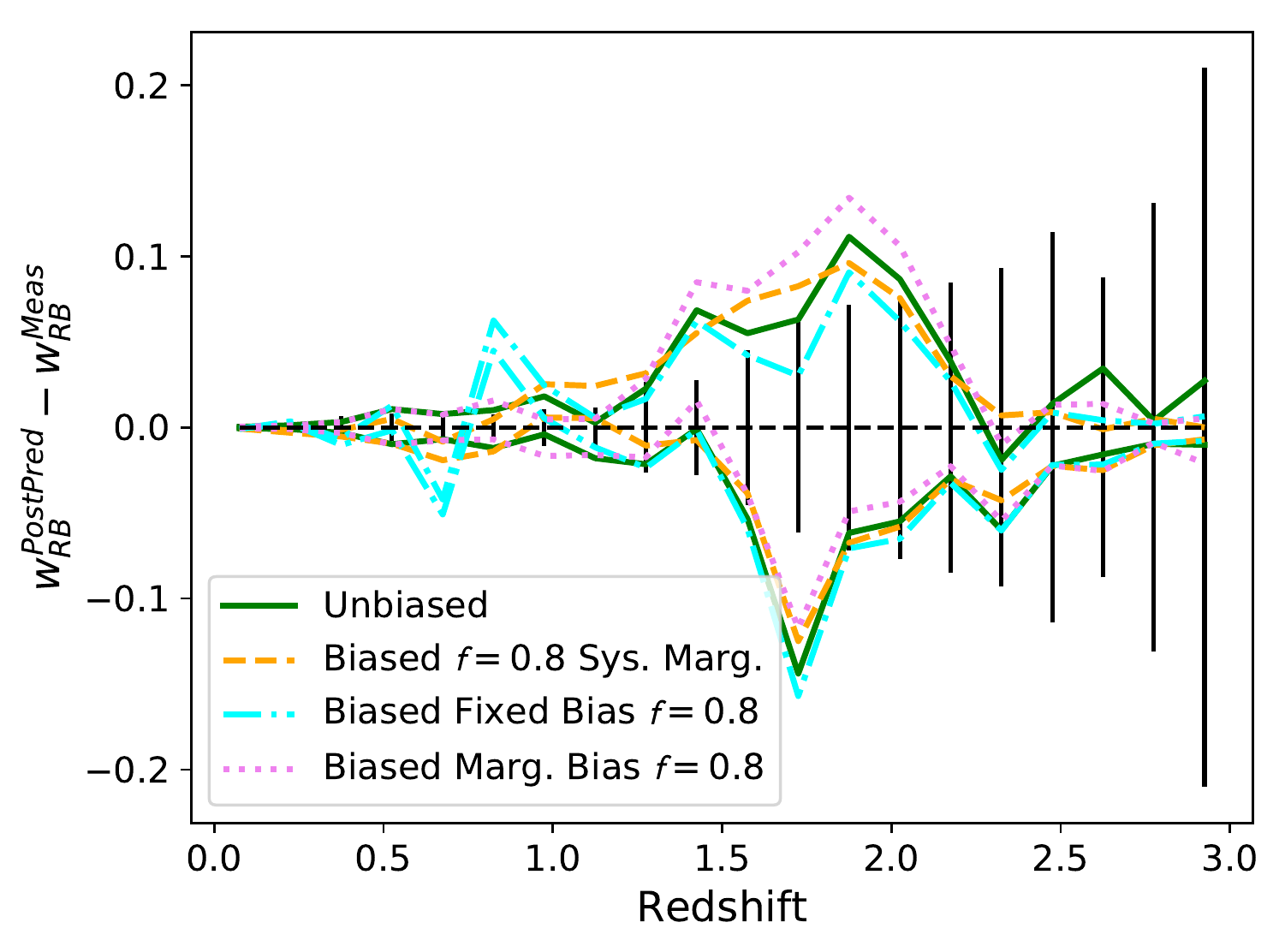}
  \caption{\label{fig:posterior_predictive} We plot the residual between replicated and true \markus{scale-averaged cross-correlation measurement $\mathbf{w^{\rm RB}}$ between a DESI-like spectroscopic reference (`R') and photometric base (`B') sample} 
  as a function of redshift. The horizontal line with errorbars shows the uncertainties in the original measurement. The green/magenta contours show the scenario where we marginalize over all \markus{parameters $\mathbf{c}$ that parametrize the redshift dependent galaxy-dark matter bias ratio function $C(z)$ (see \S~\ref{subsec:meas_cross_correlations})} 
  without the systematics kernel for the unbiased/biased ($f=0.8$) cases. The yellow/blue contours consider the biased scenario ($f=0.8$), but fix $C(z)$ and \mmref{do/do not} marginalize over the systematics kernel. The error bars and contours show the [5, 95] percentiles. We see that the replicated measurements do not show significant tension with the original measurements, if we either marginalize over the systematic (`Biased f=0.8 Sys. Marg.') or if we use a flexible redshift-dependent galaxy-dark matter bias model (`Biased Marg. Bias f=0.8'). Only if the form of the galaxy-dark matter bias is known to good precision -- in this case we hold its values fixed -- are PPCs using cross-correlations sensitive in detecting tensions. 
  }
\end{figure}

\subsection{Testing the Model}
\label{subsec:testing_model}
We discussed and showcased our inference methodology in the previous section using idealized data. To complement the discussion, this section highlights how miscalibrated likelihoods can lead to biases in the inferred sample redshift distribution and how posterior predictive checks can be used to detect these issues. 

To mimic well-calibrated photometric likelihoods we use conditional density estimates from the \textsc{FlexZboost} package \citep{2020A&C....3000362D}. We note that these conditional densities are not photometric likelihoods in the sense of Eq.~\eqref{eq:composite_like}. \markus{The free parameters in the photometric likelihood are the redshifts of the galaxies and parameters that describe properties of the Spectral Energy Distribution (SED). In conditional density estimation, non-physical parameters describe a flexible model that provides a mapping between photometry and redshift. This flexible model is then fitted to known calibration data.} 
Thus while in SED fitting the distribution of redshift constitutes a posterior distribution, conditional density estimation treats it as a predictive distribution (often without marginalizing over the modelling uncertainty). However since the goal of this subsection is to demonstrate potential systematic biases and uncertainties in the \markus{deconvolution} 
operation, this difference is not of great importance here.

\markus{In order to simulate the impact that a population of galaxies with inaccurately calibrated photometric redshift likelihoods has on the deconvolved redshift distribution, we consider a redshift range of $0.2 - 0.8$ and a total of 500k galaxies. We randomly substitute 80\% of the FlexZboost conditional density predictions with 
photometric likelihoods obtained using a template fitting run from the \textsc{BPZ} code by employing a k-nearest neighbor substitution in redshift. The result is a dataset in which a fraction of 80\% ($f=0.8$) of galaxies have a likelihood from the BPZ code, and only 20\% retain their conditional density predictions from the FlexZboost code. 
We picked this setup because the \textsc{BPZ} predictions within this redshift range, while being inferior to the \textsc{FlexZboost} predictions, still have an acceptable quality. We perform this experiment by selecting a range in redshift because we will perform posterior predictive checks (PPC) using cross-correlations and need to control where we would expect systematics. This will allow us to disentangle model misspecification issues from the systematics, e.g. from the `noisy' deconvolution, described previously. We note that the quality of photometric redshift likelihoods does not sharply change with redshift in this way, in real photometric samples. Instead photometric redshift quality is a complex function in color space that strongly depends e.g. on the quality of the photometry, the number of available bands, the amount of calibration data and the template set. Modelling this accurately is beyond the scope of this work and would require the measurement of cross-correlations in color cells and the extension of PPC to the full composite likelihood, that includes sampling the photometry of galaxies in these color cells. Using the mean of the FlexZboost conditional densities for the selection instead of the true redshifts would `smooth-out' the quality of the likelihoods as a function of redshift at the boundaries of the $0.2 - 0.8$ redshift interval. However this will also not be representative of the aforementioned difficulties. We therefore choose an unrealistically simple case that nonetheless illustrates the usefulness of PPC. Furthermore it allows us to highlight difficulties in their application in a controlled manner, by picking a fixed redshift range in which individual galaxy likelihoods are biased.     
}

In the spirit of PPC, we generate new cross-correlation measurements using the joint posterior of the sample redshift distribution parameters $\mathbf{n^{B}}$ and the parameters that govern the galaxy-dark matter bias ratio evolution $\mathbf{c}$, \markus{following the Chebychev basis expansion described in \S~\ref{subsec:meas_cross_correlations}. For simplicity we will deconvolve the redshift distribution on the same 20 bin redshift grid used in the cross-correlation data vector. For 500k galaxies, this leads to very small statistical errorbars in the deconvolution.   As a simplification we can then fix the $\mathbf{n^{B}}$ posterior to its maximum likelihood value. As mentioned in the previous section, this oversmoothing will lead to biases in the $\mathbf{n^{\rm B}}$ posteriors. However, since we will only perform a posterior predictive analysis with respect to the clustering likelihood, that is less constraining than the photometric likelihood, the systematics incurred by these simplifications and the underestimation of statistical error, are sub-dominant compared with the overall statistical error budget from the correlation function measurements. 
}

Fig.~\ref{fig:posterior_predictive} shows the sampled cross correlation measurements from the fitted joint model, in residual to the original measurements. We showcase four scenarios. In the unbiased case we use \textsc{FlexZBoost} PDFs and \mmref{marginalize} over all $\mathbf{c}$ parameters. Due to the good calibration of these Machine Learning-produced conditional distributions, we obtain very similar results compared with the previous section. The reason for this success is, of course, the representative training set that would not be available in a practical application. Furthermore we consider three scenarios with $f=0.8$. The scenario shown in yellow fixes the galaxy-dark matter bias parameters ($\mathbf{c}$), but marginalizes over the parameters of the systematics kernel. The blue/magenta lines fix/marginalize over the $\mathbf{c}$ parameters, but do not include the systematics kernel correction. 

As can be seen in Fig.~\ref{fig:posterior_predictive}, the treatment of galaxy-dark matter bias has a profound impact on the consistency between the replicated and original cross-correlation measurements. Within the errors, the results are consistent for all scenarios except the one without systematics kernel correction that uses a fixed $C(z)$ model. As shown in the yellow line, these biases can be corrected by the systematics kernel marginalization. This illustrates that if sufficient information about $C(z)$ is available, the clustering likelihood alone can allow for powerful posterior predictive checks. If this is not the case, consistency tests of redshift distributions with respect to clustering redshift measurements can be misleading. 
Provided sufficient information on the galaxy-dark matter bias, we can parametrize the biases in the deconvolved density estimate using, e.g., a convolution with a kernel function as described in \S~\ref{subsubsec:parametrizing_systematics}. We show these results as the yellow lines `Biased $f=0.8$ Sys. Marg'. Here we perform a discretized marginalization as described in \S~\ref{subsubsec:parametrizing_systematics}\markus{, by convolving the $\mathbf{n^{B}}$ vector with  a Gaussian kernel function of width 
$\Delta \sigma \in [0.001, 0.2]$ in 40 steps.} We see that this correction can compensate for the misspecified likelihoods even in the case of a fixed $C(z)$ model. The degeneracy between the redshift-dependent galaxy-dark matter bias ratio model and the $\mathbf{n^{B}}$ parameters highlights the importance of performing posterior predictive checks in color space to provide additional information on the redshift distribution. However, this requires careful modelling of SEDs and the development of a transparent, reproducible analysis framework that additionally includes tests for parameter degeneracies and a model comparison framework. This is beyond the scope of this work, but will be addressed in a future paper.

\section{Summary and Conclusions}
\label{sec:summary_and_conclus}


Accurate photometric redshift inference is one of the most important challenges in large area photometric surveys like LSST, DES, HSC, or KiDS. As discussed in detail in Appendix~\ref{sec:methods}, photometric redshift inference is, from a statistical point of view, a deconvolution problem, where an underlying true redshift distribution is convolved with an SED model-dependent error distribution given by the photometric likelihood. The deconvolution inference of sample redshift distribution is not new \citep[e.g.][]{2005MNRAS.359..237P, 2016MNRAS.460.4258L, 2020arXiv200712178M}, and spatial information has also been incorporated into the inference \citep[e.g.][]{2019arXiv191007127A, 2019MNRAS.483.2801S, 2019MNRAS.483.2487J, 2020MNRAS.491.4768R}.  We extended these prior works by developing a fast approximate inference scheme for deconvolution, that combines redshift information from both the photometry and the spatial distribution of galaxies in terms of a composite likelihood ansatz. We particularly provided a discussion on regularization techniques and the tradeoff between bias and variance in the Bayesian context for medium to large sample sizes. 

\mmref{In particular, our goal is to include the treatment of photometric redshift via the likelihood of the galaxies' photometry into the current cosmological inference framework, which is based on correlation functions. 
The main reason for our likelihood choice is to allow the easy integration into the likelihood inference framework based on two-point statistics of galaxy density and shear fields. 
This is more difficult for other approaches presented in \citet{2019arXiv191007127A} and \citet{2019MNRAS.483.2801S} since, in the currently demonstrated form, the redshift information from cross-correlating the overlapping spectroscopic sample is included via an estimator and not using a likelihood (that would depend on cosmological parameters).  While the works of  \citet{2005MNRAS.359..237P, 2016MNRAS.460.4258L, 2020arXiv200712178M} are structurally similar in terms of the treatment of the photometric likelihood, they do not discuss the effect of including clustering information. It is noteworthy that the early work by \citet{2005MNRAS.359..237P} provides an excellent, explicit discussion of regularization, which is the main difficulty in the photometric redshift problem, although not in the context of probability calibration and the aforementioned joint inference framework.   }   We summarized our approach to the challenges that arise in \markus{the estimation of redshift distributions for samples of galaxies in the context of photometric surveys.} 
Concretely, we considered the combination of photometric information with two-point statistics, the scalability and regularization of the deconvolution inference in the large sample scenario, and investigate the impact of systematics from misspecified individual galaxy photometric likelihoods, proposing parametrizations for these systematics. These achievements lay the foundations for future extensions that we will discuss in the next section. 

In \S~\ref{subsec:photoZestim} we described our inference methodology that is designed to facilitate inference on large galaxy catalogs to be expected in LSST. The scheme uses a Laplace Approximation in logit space and facilitates inference using an iterative scheme of expectation maximization update equations. This provides computational advantages over sampling approaches. Additionally this methodology facilitates fast joint inference with a cross-correlation data vector (see \S~\ref{sec:clustering_likelihood}) that we included in a composite likelihood ansatz. As highlighted in \S~\ref{subsec:composite_like}, this provides the possibility of additional extensions that include two-point statistics from cosmological weak lensing and galaxy-galaxy lensing measurements. As we discussed in \S~\ref{sec:towards_z_prob}, ensemble redshift distribution inference based on a photometric likelihood is a deconvolution problem, which requires regularization to yield bounded and well-defined results. In this context, we discussed a regularization scheme that consists of a combination of Tikhonov regularization with (more importantly) a scheme that merges neighboring bins to exploit the characteristic covariance structure in the deconvolved densities. In agreement with the findings of the original paper by \citet{Kuusela:220015} that proposed and applied this scheme to the Poisson inverse problem, we find that the `Merging Bin' scheme leads to better calibrated results as compared with Tikhonov regularization and with an oversmoothing scheme that selects a coarser redshift binning for the sample redshift histograms. 

In order to test and discuss the quality of our posterior inference, we used data from the CosmoDC2 simulations to generate a spectroscopic DESI-like sample and a photometric mock catalog, that uses an LSST-like photometric error model. This allowed us to test the impact of a spectroscopic calibration sample with an inhomogeneous galaxy population as a function of redshift. We found that the ratio between the redshift-dependent galaxy-dark matter bias of the photometric and the spectroscopic sample is a smooth function of redshift, if the spectroscopic calibration sample consists of a single galaxy population, and is discontinuous if the galaxy population strongly changes. We therefore employed a step-wise smooth function based on a Chebychev polynomial expansion to parametrize this ratio. 

In \S~\ref{sec:forecast_ideal_data} we performed a forecast of redshift inference performance on ideal data, assuming perfectly calibrated individual galaxy redshift likelihoods. We found that using the aforementioned merging bin regularization, we were able to produce accurate posteriors of ensemble redshift distributions. We reiterate that using other regularization schemes, like an overly large Tikhonov regularization parameter, or an oversmoothing approach that picks overly wide histogram bins, can lead to significant biases in the recovered posterior mean. \rev{We reiterate that the `Oversmoothing' method makes the likelihood insensitive to the within-bin shape of the true (unknown) sample redshift distribution, which induces the aforementioned biases. The basic need to regularize is present in both point estimation (Maximum Likelihood or Maximum A-Posteriori) and probabilistic inference, independent of the chosen method. It is a property of the conditioning of the mapping between parameter space and data space.}

When compared with the DESC science requirements for WL and large scale structure 
measurements in terms of the mean of the photometric sample redshift distribution, we found that we can meet the DESC SRD Y1 goals and remain consistent with the DESC SRD Y10 goals with 5000k galaxies, if cross-correlations are included in the joint composite likelihood. In practical applications, however, Spectral Energy Distribution (SED) templates for galaxies will be subject to modelling biases that cannot be well calibrated using spectroscopic data \citep[see e.g.][]{2020MNRAS.496.4769H}. We therefore proposed to use posterior predictive checks (PPC) as a means to evaluate the quality of our inference. Here, we compared replications of the data sampled from the fitted model with the original measurement to evaluate model goodness-of-fit. Specifically cross-correlation redshift inference is often used to calibrate photometric redshifts obtained using photometry \citep{2008ApJ...684...88N, 10.1093/mnras/stw3033, 2017arXiv171002517D}. In \S~\ref{subsec:testing_model} we demonstrated that PPC of cross-correlation measurements can detect systematic biases in the recovered sample redshift distribution if the galaxy-dark matter bias of the photometric and spectroscopic samples is known to sufficient accuracy. 

In order to parametrize potential biases in the sample redshift distribution posteriors caused by misspecified photometric likelihoods, particularly over-deconvolution effects that lead to overly narrow redshift distributions, we proposed a simple Gaussian filter that, as demonstrated in \S~\ref{subsec:testing_model}, was able to correct these biases. 

\section{Future Work}
\label{sec:future_work}
In future work, it will be important to extend the inference scheme developed in this paper. We plan to consider a range of extensions, e.g., iterated nested Laplace approximations \citep{bornkamp} in logit space, the usage of more flexible distributions that can be fitted using variational inference schemes, as well as the development of specialized subsampling MCMC schemes. The different techniques will be evaluated in combination with regularization approaches based on the quality of their probability coverage. Another extension, particularly to reduce the bias in the density estimation, is to consider other parametrizations for the deconvolved density either by employing density estimators with better mean squared error scaling like Kernel Density estimators, basis function expansions or using methods such as logistic Gaussian Processes \citep[e.g.][]{2020MNRAS.491.4768R}.  The combination of photometric and clustering information can be extended by connecting the modelling of SEDs and redshift-dependent galaxy-dark matter bias modelling via the luminosity function as shown in \citet{2018MNRAS.476.4649V}. This also has the potential to reduce the degeneracy between SED and redshift-dependent galaxy-dark matter bias systematics. Finally, we note that the data quality of the photometry will not be the same in all areas on the sky. In order to include these field-to-field variations into the composite likelihood framework, we can, for example, condition the likelihood on the field and include a corresponding data covariance into the likelihood by employing either resampling techniques \citep[see e.g.][]{10.5555/2556084} or using theoretical modelling \citep[e.g.][]{stoyan1994fractals, 2020arXiv200409542S}.

\rev{Another important area for future work would be the extension of our model checking framework towards the inclusion of the photometric likelihood to break degeneracies between redshift dependent galaxy-dark matter bias and the sample redshift distribution parameters. The usefulness of this approach will depend on degeneracies in the SED modelling e.g. between redshift and galaxy type.

Furthermore it will be vital to follow-up this work with more extensive studies on LSST-like simulations that improve upon the limitations of this work as mentioned throughout the paper. We highlight the importance of studying how a more realistic galaxy-dark matter bias modelling and treatment of calibration fields affects the results. } 

To conclude, we have presented an efficient photometric redshift inference framework that combines information from both the photometry and the spatial distribution of galaxies. The methodology is designed to scale well to large samples. We complement this framework with methods for regularization, model checking and redshift systematics parametrization. The forecasts we performed using CosmoDC2 data give us confidence that, \markus{with the additional improvements described here,} the methodology presented will 
enable accurate and well-calibrated redshift inference for LSST and other ongoing and future large area photometric surveys.
\mmref{
\section*{Software}
Besides software referenced directly in the text, we performed the analyses in this work using the following software packages: the python language \citep{CS-R9526}, scipy \citep{2020SciPy-NMeth}, numpy \citep{2020NumPy-Array}, jupyter notebook \citep{Kluyver:2016aa}, ipython \citep{4160251}, matplotlib \citep{4160265} and pandas \citep{mckinney-proc-scipy-2010}. }

{\color{black} \section*{Data Availability Statement}
The cosmoDC2 extragalactic catalog is publicly available at \url{https://portal.nersc.gov/project/lsst/cosmoDC2/_README.html}.
The ancillary catalogs (photo-z and DESI-like selection for cosmoDC2) and other derived data underlying this article will be shared on reasonable request to the corresponding author.
The source code that implements the algorithms presented in this article will be made available via Zenodo.

\section*{Acknowledgements}
This paper has undergone internal review in the LSST Dark Energy Science Collaboration. We would like to thank the internal reviewers David Alonso, Will Hartley and David Kirkby for their insightful comments. MMR thanks Mikael Kuusela and Ann B. Lee for useful discussions. 

MMR led and planned the project from the initial idea and motivation to the experimental design. He performed the analyses in interaction with the coauthors, and prepared the paper. CBM contributed by: development of clustering redshifts software, creation of clustering redshifts photometric and spectroscopic samples, creation of clustering redshift data products. SJS constructed the photometric redshift catalogs and contributed to writing the corresponding portions of the paper. SW provided feedback on the method development and manuscript. RM advised on motivation, scope, experimental design, and analysis, and contributed to the editing of the paper draft. YYM developed the access tools for cosmoDC2 and photo-z data products.}

MMR and RM are supported by DOE grant DE-SC0010118 and a grant from the Simons Foundation (Simons Investigator in Astrophysics, Award ID 620789). SJS acknowledges support from DOE grant DESC0009999 and NSF/AURA grant N56981C. This work is supported by the NSF AI Institute: Physics of the Future, NSF PHY- 2020295.
This material is based upon work supported in part by the National Science Foundation through Cooperative Agreement 1258333 managed by the Association of Universities for Research in Astronomy (AURA), and the Department of Energy under  Contract  No.  DE-AC02-76SF00515  with  the  SLAC  National  Accelerator Laboratory. Additional LSST funding comes from private donations, grants to universities, and in-kind support from LSSTC Institutional Members. SJS acknowledges support from DOE grant DE-SC0009999 and NSF/AURA grant N56981C. Support for YYM\ was provided by NASA through the NASA Hubble Fellowship grant no.\ HST-HF2-51441.001 awarded by the Space Telescope Science Institute, which is operated by the Association of Universities for Research in Astronomy, Incorporated, under NASA contract NAS5-26555.

The DESC acknowledges ongoing support from the Institut National de 
Physique Nucl\'eaire et de Physique des Particules in France; the 
Science \& Technology Facilities Council in the United Kingdom; and the
Department of Energy, the National Science Foundation, and the LSST 
Corporation in the United States.  DESC uses resources of the IN2P3 
Computing Center (CC-IN2P3--Lyon/Villeurbanne - France) funded by the 
Centre National de la Recherche Scientifique; the National Energy 
Research Scientific Computing Center, a DOE Office of Science User 
Facility supported by the Office of Science of the U.S.\ Department of
Energy under Contract No.\ DE-AC02-05CH11231; STFC DiRAC HPC Facilities, 
funded by UK BIS National E-infrastructure capital grants; and the UK 
particle physics grid, supported by the GridPP Collaboration.  This 
work was performed in part under DOE Contract DE-AC02-76SF00515.


\bibliographystyle{mnras}
\bibliography{biblio.bib} 

\begin{thebibliography}{}
\makeatletter
\relax
\def\mn@urlcharsother{\let\do\@makeother \do\$\do\&\do\#\do\^\do\_\do\%\do\~}
\def\mn@doi{\begingroup\mn@urlcharsother \@ifnextchar [ {\mn@doi@}
  {\mn@doi@[]}}
\def\mn@doi@[#1]#2{\def\@tempa{#1}\ifx\@tempa\@empty \href
  {http://dx.doi.org/#2} {doi:#2}\else \href {http://dx.doi.org/#2} {#1}\fi
  \endgroup}
\def\mn@eprint#1#2{\mn@eprint@#1:#2::\@nil}
\def\mn@eprint@arXiv#1{\href {http://arxiv.org/abs/#1} {{\tt arXiv:#1}}}
\def\mn@eprint@dblp#1{\href {http://dblp.uni-trier.de/rec/bibtex/#1.xml}
  {dblp:#1}}
\def\mn@eprint@#1:#2:#3:#4\@nil{\def\@tempa {#1}\def\@tempb {#2}\def\@tempc
  {#3}\ifx \@tempc \@empty \let \@tempc \@tempb \let \@tempb \@tempa \fi \ifx
  \@tempb \@empty \def\@tempb {arXiv}\fi \@ifundefined
  {mn@eprint@\@tempb}{\@tempb:\@tempc}{\expandafter \expandafter \csname
  mn@eprint@\@tempb\endcsname \expandafter{\@tempc}}}

\bibitem[\protect\citeauthoryear{{Abbott} et~al.,}{{Abbott}
  et~al.}{2018a}]{2018PhRvD..98d3526A}
{Abbott} T.~M.~C.,  et~al., 2018a, \mn@doi [\prd] {10.1103/PhysRevD.98.043526},
  \href {https://ui.adsabs.harvard.edu/abs/2018PhRvD..98d3526A} {98, 043526}

\bibitem[\protect\citeauthoryear{{Abbott} et~al.,}{{Abbott}
  et~al.}{2018b}]{2018ApJS..239...18A}
{Abbott} T.~M.~C.,  et~al., 2018b, \mn@doi [\apjs] {10.3847/1538-4365/aae9f0},
  \href {https://ui.adsabs.harvard.edu/\#abs/2018ApJS..239...18A} {239, 18}

\bibitem[\protect\citeauthoryear{{Aihara} et~al.,}{{Aihara}
  et~al.}{2018}]{2018PASJ...70S...4A}
{Aihara} H.,  et~al., 2018, \mn@doi [\pasj] {10.1093/pasj/psx066}, \href
  {https://ui.adsabs.harvard.edu/abs/2018PASJ...70S...4A} {70, S4}

\bibitem[\protect\citeauthoryear{{Alarcon} et~al.,}{{Alarcon}
  et~al.}{2020a}]{2020arXiv200711132A}
{Alarcon} A.,  et~al., 2020a, arXiv e-prints, \href
  {https://ui.adsabs.harvard.edu/abs/2020arXiv200711132A} {p. arXiv:2007.11132}

\bibitem[\protect\citeauthoryear{{Alarcon}, {S{\'a}nchez}, {Bernstein}  \&
  {Gazta{\~n}aga}}{{Alarcon} et~al.}{2020b}]{2019arXiv191007127A}
{Alarcon} A.,  {S{\'a}nchez} C.,  {Bernstein} G.~M.,   {Gazta{\~n}aga} E.,
  2020b, \mn@doi [\mnras] {10.1093/mnras/staa2478}, \href
  {https://ui.adsabs.harvard.edu/abs/2020MNRAS.498.2614A} {498, 2614}

\bibitem[\protect\citeauthoryear{{Albrecht} et~al.,}{{Albrecht}
  et~al.}{2006}]{dark_energy_taskforce}
{Albrecht} A.,  et~al., 2006, arXiv e-prints, \href
  {https://ui.adsabs.harvard.edu/abs/2006astro.ph..9591A} {pp
  astro--ph/0609591}

\bibitem[\protect\citeauthoryear{{Arnouts}, {Cristiani}, {Moscardini},
  {Matarrese}, {Lucchin}, {Fontana}  \& {Giallongo}}{{Arnouts}
  et~al.}{1999}]{1999MNRAS.310..540A}
{Arnouts} S.,  {Cristiani} S.,  {Moscardini} L.,  {Matarrese} S.,  {Lucchin}
  F.,  {Fontana} A.,   {Giallongo} E.,  1999, \mn@doi [\mnras]
  {10.1046/j.1365-8711.1999.02978.x}, \href
  {https://ui.adsabs.harvard.edu/\#abs/1999MNRAS.310..540A} {310, 540}

\bibitem[\protect\citeauthoryear{Atchison \& Shen}{Atchison \&
  Shen}{1980}]{10.1093/biomet/67.2.261}
Atchison J.,  Shen S.,  1980, \mn@doi [Biometrika] {10.1093/biomet/67.2.261},
  67, 261

\bibitem[\protect\citeauthoryear{{Ben{\'\i}tez}}{{Ben{\'\i}tez}}{2000}]{2000ApJ...536..571B}
{Ben{\'\i}tez} N.,  2000, \mn@doi [\apj] {10.1086/308947}, \href
  {https://ui.adsabs.harvard.edu/\#abs/2000ApJ...536..571B} {536, 571}

\bibitem[\protect\citeauthoryear{{Benjamin} et~al.,}{{Benjamin}
  et~al.}{2013}]{2013MNRAS.431.1547B}
{Benjamin} J.,  et~al., 2013, \mn@doi [\mnras] {10.1093/mnras/stt276}, \href
  {https://ui.adsabs.harvard.edu/\#abs/2013MNRAS.431.1547B} {431, 1547}

\bibitem[\protect\citeauthoryear{{Benson}}{{Benson}}{2012}]{2012NewA...17..175B}
{Benson} A.~J.,  2012, \mn@doi [\na] {10.1016/j.newast.2011.07.004}, \href
  {https://ui.adsabs.harvard.edu/abs/2012NewA...17..175B} {17, 175}

\bibitem[\protect\citeauthoryear{{Bernstein} \& {Huterer}}{{Bernstein} \&
  {Huterer}}{2010}]{2010MNRAS.401.1399B}
{Bernstein} G.,  {Huterer} D.,  2010, \mn@doi [\mnras]
  {10.1111/j.1365-2966.2009.15748.x}, \href
  {https://ui.adsabs.harvard.edu/abs/2010MNRAS.401.1399B} {401, 1399}

\bibitem[\protect\citeauthoryear{Bishop}{Bishop}{2006}]{10.5555/1162264}
Bishop C.~M.,  2006, Pattern Recognition and Machine Learning (Information
  Science and Statistics).
Springer-Verlag, Berlin, Heidelberg

\bibitem[\protect\citeauthoryear{{Bonnett}}{{Bonnett}}{2015}]{2015MNRAS.449.1043B}
{Bonnett} C.,  2015, \mn@doi [\mnras] {10.1093/mnras/stv230}, \href
  {https://ui.adsabs.harvard.edu/\#abs/2015MNRAS.449.1043B} {449, 1043}

\bibitem[\protect\citeauthoryear{Bornkamp}{Bornkamp}{2011}]{bornkamp}
Bornkamp B.,  2011, Journal of Computational and Graphical Statistics, 20, 656

\bibitem[\protect\citeauthoryear{Box \& Muller}{Box \& Muller}{1958}]{box1958}
Box G. E.~P.,  Muller M.~E.,  1958, \mn@doi [Ann. Math. Statist.]
  {10.1214/aoms/1177706645}, 29, 610

\bibitem[\protect\citeauthoryear{{Brown} et~al.,}{{Brown}
  et~al.}{2014}]{Brown:2014}
{Brown} M. J.~I.,  et~al., 2014, \mn@doi [\apjs] {10.1088/0067-0049/212/2/18},
  \href {https://ui.adsabs.harvard.edu/abs/2014ApJS..212...18B} {212, 18}

\bibitem[\protect\citeauthoryear{{Carrasco Kind} \& {Brunner}}{{Carrasco Kind}
  \& {Brunner}}{2013}]{2013MNRAS.432.1483C}
{Carrasco Kind} M.,  {Brunner} R.~J.,  2013, \mn@doi [\mnras]
  {10.1093/mnras/stt574}, \href
  {https://ui.adsabs.harvard.edu/\#abs/2013MNRAS.432.1483C} {432, 1483}

\bibitem[\protect\citeauthoryear{Carroll \& Hall}{Carroll \&
  Hall}{1988}]{doi:10.1080/01621459.1988.10478718}
Carroll R.~J.,  Hall P.,  1988, \mn@doi [Journal of the American Statistical
  Association] {10.1080/01621459.1988.10478718}, 83, 1184

\bibitem[\protect\citeauthoryear{{Cawthon} et~al.,}{{Cawthon}
  et~al.}{2020}]{2020arXiv201212826C}
{Cawthon} R.,  et~al., 2020, arXiv e-prints, \href
  {https://ui.adsabs.harvard.edu/abs/2020arXiv201212826C} {p. arXiv:2012.12826}

\bibitem[\protect\citeauthoryear{{Chang} et~al.,}{{Chang}
  et~al.}{2016}]{2016MNRAS.459.3203C}
{Chang} C.,  et~al., 2016, \mn@doi [\mnras] {10.1093/mnras/stw861}, \href
  {https://ui.adsabs.harvard.edu/abs/2016MNRAS.459.3203C} {459, 3203}

\bibitem[\protect\citeauthoryear{Chen \& Guestrin}{Chen \&
  Guestrin}{2016}]{Chen:16}
Chen T.,  Guestrin C.,  2016, in Proceedings of the 22Nd ACM SIGKDD
  International Conference on Knowled\ ge Discovery and Data Mining. KDD '16.
ACM, New York, NY, USA, pp 785--794, \mn@doi{10.1145/2939672.2939785}, \url
  {http://doi.acm.org/10.1145/2939672.2939785}

\bibitem[\protect\citeauthoryear{{Clerkin}, {Kirk}, {Lahav}, {Abdalla}  \&
  {Gazta{\~n}aga}}{{Clerkin} et~al.}{2015}]{2015MNRAS.448.1389C}
{Clerkin} L.,  {Kirk} D.,  {Lahav} O.,  {Abdalla} F.~B.,   {Gazta{\~n}aga} E.,
  2015, \mn@doi [\mnras] {10.1093/mnras/stu2754}, \href
  {https://ui.adsabs.harvard.edu/\#abs/2015MNRAS.448.1389C} {448, 1389}

\bibitem[\protect\citeauthoryear{{Collister} \& {Lahav}}{{Collister} \&
  {Lahav}}{2004}]{2004PASP..116..345C}
{Collister} A.~A.,  {Lahav} O.,  2004, \mn@doi [Publications of the
  Astronomical Society of the Pacific] {10.1086/383254}, \href
  {https://ui.adsabs.harvard.edu/\#abs/2004PASP..116..345C} {116, 345}

\bibitem[\protect\citeauthoryear{{Craig} \& {Brown}}{{Craig} \&
  {Brown}}{1986}]{1986ipag.book.....C}
{Craig} I.~J.~D.,  {Brown} J.~C.,  1986, {Inverse problems in astronomy. A
  guide to inversion strategies for remotely sensed data}

\bibitem[\protect\citeauthoryear{{DESI Collaboration} et~al.,}{{DESI
  Collaboration} et~al.}{2016}]{2016arXiv161100036D}
{DESI Collaboration} et~al., 2016, arXiv e-prints, \href
  {https://ui.adsabs.harvard.edu/abs/2016arXiv161100036D} {p. arXiv:1611.00036}

\bibitem[\protect\citeauthoryear{{Dalmasso}, {Pospisil}, {Lee}, {Izbicki},
  {Freeman}  \& {Malz}}{{Dalmasso} et~al.}{2020}]{2020A&C....3000362D}
{Dalmasso} N.,  {Pospisil} T.,  {Lee} A.~B.,  {Izbicki} R.,  {Freeman} P.~E.,
  {Malz} A.~I.,  2020, \mn@doi [Astronomy and Computing]
  {10.1016/j.ascom.2019.100362}, \href
  {https://ui.adsabs.harvard.edu/abs/2020A&C....3000362D} {30, 100362}

\bibitem[\protect\citeauthoryear{{Davis} et~al.,}{{Davis}
  et~al.}{2017}]{2017arXiv171002517D}
{Davis} C.,  et~al., 2017, arXiv e-prints, \href
  {https://ui.adsabs.harvard.edu/\#abs/2017arXiv171002517D} {p.
  arXiv:1710.02517}

\bibitem[\protect\citeauthoryear{Davison \& Hinkley}{Davison \&
  Hinkley}{2013}]{10.5555/2556084}
Davison A.~C.,  Hinkley D.~V.,  2013, Bootstrap Methods and Their Application.
Cambridge University Press, USA

\bibitem[\protect\citeauthoryear{{Feldmann} et~al.,}{{Feldmann}
  et~al.}{2006}]{2006MNRAS.372..565F}
{Feldmann} R.,  et~al., 2006, \mn@doi [\mnras]
  {10.1111/j.1365-2966.2006.10930.x}, \href
  {https://ui.adsabs.harvard.edu/\#abs/2006MNRAS.372..565F} {372, 565}

\bibitem[\protect\citeauthoryear{{Gatti} et~al.,}{{Gatti}
  et~al.}{2018}]{2018MNRAS.477.1664G}
{Gatti} M.,  et~al., 2018, \mn@doi [\mnras] {10.1093/mnras/sty466}, \href
  {https://ui.adsabs.harvard.edu/\#abs/2018MNRAS.477.1664G} {477, 1664}

\bibitem[\protect\citeauthoryear{{Gatti} et~al.,}{{Gatti}
  et~al.}{2020}]{2020arXiv201208569G}
{Gatti} M.,  et~al., 2020, arXiv e-prints, \href
  {https://ui.adsabs.harvard.edu/abs/2020arXiv201208569G} {p. arXiv:2012.08569}

\bibitem[\protect\citeauthoryear{Gelman, li Meng  \& Stern}{Gelman
  et~al.}{1996}]{Gelman96posteriorpredictive}
Gelman A.,  li Meng X.,   Stern H.,  1996, Statistica Sinica, pp 733--807

\bibitem[\protect\citeauthoryear{{Gelman}, {Hwang}  \& {Vehtari}}{{Gelman}
  et~al.}{2013}]{2013arXiv1307.5928G}
{Gelman} A.,  {Hwang} J.,   {Vehtari} A.,  2013, arXiv e-prints, \href
  {https://ui.adsabs.harvard.edu/\#abs/2013arXiv1307.5928G} {p.
  arXiv:1307.5928}

\bibitem[\protect\citeauthoryear{{Gerdes}, {Sypniewski}, {McKay}, {Hao},
  {Weis}, {Wechsler}  \& {Busha}}{{Gerdes} et~al.}{2010}]{2010ApJ...715..823G}
{Gerdes} D.~W.,  {Sypniewski} A.~J.,  {McKay} T.~A.,  {Hao} J.,  {Weis} M.~R.,
  {Wechsler} R.~H.,   {Busha} M.~T.,  2010, \mn@doi [\apj]
  {10.1088/0004-637X/715/2/823}, \href
  {https://ui.adsabs.harvard.edu/\#abs/2010ApJ...715..823G} {715, 823}

\bibitem[\protect\citeauthoryear{{Giblin} et~al.,}{{Giblin}
  et~al.}{2021}]{2021A&A...645A.105G}
{Giblin} B.,  et~al., 2021, \mn@doi [\aap] {10.1051/0004-6361/202038850}, \href
  {https://ui.adsabs.harvard.edu/abs/2021A&A...645A.105G} {645, A105}

\bibitem[\protect\citeauthoryear{{Graham} et~al.,}{{Graham}
  et~al.}{2020}]{2020AJ....159..258G}
{Graham} M.~L.,  et~al., 2020, \mn@doi [\aj] {10.3847/1538-3881/ab8a43}, \href
  {https://ui.adsabs.harvard.edu/abs/2020AJ....159..258G} {159, 258}

\bibitem[\protect\citeauthoryear{{Greisel}, {Seitz}, {Drory}, {Bender},
  {Saglia}  \& {Snigula}}{{Greisel} et~al.}{2015}]{2015MNRAS.451.1848G}
{Greisel} N.,  {Seitz} S.,  {Drory} N.,  {Bender} R.,  {Saglia} R.~P.,
  {Snigula} J.,  2015, \mn@doi [\mnras] {10.1093/mnras/stv1005}, \href
  {http://adsabs.harvard.edu/abs/2015MNRAS.451.1848G} {451, 1848}

\bibitem[\protect\citeauthoryear{{Hadzhiyska}, {Alonso}, {Nicola}  \&
  {Slosar}}{{Hadzhiyska} et~al.}{2020}]{2020arXiv200714989H}
{Hadzhiyska} B.,  {Alonso} D.,  {Nicola} A.,   {Slosar} A.,  2020, arXiv
  e-prints, \href {https://ui.adsabs.harvard.edu/abs/2020arXiv200714989H} {p.
  arXiv:2007.14989}

\bibitem[\protect\citeauthoryear{Hahn, Beutler, Sinha, Berlind, Ho  \&
  Hogg}{Hahn et~al.}{2019}]{10.1093/mnras/stz558}
Hahn C.,  Beutler F.,  Sinha M.,  Berlind A.,  Ho S.,   Hogg D.~W.,  2019,
  \mn@doi [\mnras] {10.1093/mnras/stz558}, 485, 2956

\bibitem[\protect\citeauthoryear{Harris et~al.,}{Harris
  et~al.}{2020}]{2020NumPy-Array}
Harris C.~R.,  et~al., 2020, \mn@doi [Nature] {10.1038/s41586-020-2649-2}, 585,
  357–362

\bibitem[\protect\citeauthoryear{{Hartley} et~al.,}{{Hartley}
  et~al.}{2020}]{2020MNRAS.496.4769H}
{Hartley} W.~G.,  et~al., 2020, \mn@doi [\mnras] {10.1093/mnras/staa1812},
  \href {https://ui.adsabs.harvard.edu/abs/2020MNRAS.496.4769H} {496, 4769}

\bibitem[\protect\citeauthoryear{{Hearin}, {Korytov}, {Kovacs}, {Benson},
  {Aung}, {Bradshaw}, {Campbell}  \& {LSST Dark Energy Science
  Collaboration}}{{Hearin} et~al.}{2020}]{2020MNRAS.495.5040H}
{Hearin} A.,  {Korytov} D.,  {Kovacs} E.,  {Benson} A.,  {Aung} H.,  {Bradshaw}
  C.,  {Campbell} D.,   {LSST Dark Energy Science Collaboration} 2020, \mn@doi
  [\mnras] {10.1093/mnras/staa1495}, \href
  {https://ui.adsabs.harvard.edu/abs/2020MNRAS.495.5040H} {495, 5040}

\bibitem[\protect\citeauthoryear{{Heitmann} et~al.,}{{Heitmann}
  et~al.}{2019}]{2019ApJS..245...16H}
{Heitmann} K.,  et~al., 2019, \mn@doi [\apjs] {10.3847/1538-4365/ab4da1}, \href
  {https://ui.adsabs.harvard.edu/abs/2019ApJS..245...16H} {245, 16}

\bibitem[\protect\citeauthoryear{{Heymans} et~al.,}{{Heymans}
  et~al.}{2020}]{2020arXiv200715632H}
{Heymans} C.,  et~al., 2020, arXiv e-prints, \href
  {https://ui.adsabs.harvard.edu/abs/2020arXiv200715632H} {p. arXiv:2007.15632}

\bibitem[\protect\citeauthoryear{{Hikage} et~al.,}{{Hikage}
  et~al.}{2019}]{2019PASJ...71...43H}
{Hikage} C.,  et~al., 2019, \mn@doi [\pasj] {10.1093/pasj/psz010}, \href
  {https://ui.adsabs.harvard.edu/abs/2019PASJ...71...43H} {71, 43}

\bibitem[\protect\citeauthoryear{{Hildebrandt} et~al.,}{{Hildebrandt}
  et~al.}{2017}]{2017MNRAS.465.1454H}
{Hildebrandt} H.,  et~al., 2017, \mn@doi [\mnras] {10.1093/mnras/stw2805},
  \href {https://ui.adsabs.harvard.edu/abs/2017MNRAS.465.1454H} {465, 1454}

\bibitem[\protect\citeauthoryear{{Hildebrandt} et~al.,}{{Hildebrandt}
  et~al.}{2021}]{2021A&A...647A.124H}
{Hildebrandt} H.,  et~al., 2021, \mn@doi [\aap] {10.1051/0004-6361/202039018},
  \href {https://ui.adsabs.harvard.edu/abs/2021A&A...647A.124H} {647, A124}

\bibitem[\protect\citeauthoryear{{Hoyle}}{{Hoyle}}{2016}]{2016A&C....16...34H}
{Hoyle} B.,  2016, \mn@doi [Astronomy and Computing]
  {10.1016/j.ascom.2016.03.006}, \href
  {https://ui.adsabs.harvard.edu/\#abs/2016A&C....16...34H} {16, 34}

\bibitem[\protect\citeauthoryear{{Hoyle} \& {Rau}}{{Hoyle} \&
  {Rau}}{2019}]{2018arXiv180202581H}
{Hoyle} B.,  {Rau} M.~M.,  2019, \mn@doi [\mnras] {10.1093/mnras/stz502}, \href
  {https://ui.adsabs.harvard.edu/abs/2019MNRAS.485.3642H} {485, 3642}

\bibitem[\protect\citeauthoryear{{Hoyle}, {Rau}, {Bonnett}, {Seitz}  \&
  {Weller}}{{Hoyle} et~al.}{2015}]{2015MNRAS.450..305H}
{Hoyle} B.,  {Rau} M.~M.,  {Bonnett} C.,  {Seitz} S.,   {Weller} J.,  2015,
  \mn@doi [\mnras] {10.1093/mnras/stv599}, \href
  {https://ui.adsabs.harvard.edu/\#abs/2015MNRAS.450..305H} {450, 305}

\bibitem[\protect\citeauthoryear{Hoyle et~al.,}{Hoyle
  et~al.}{2018}]{Hoyle_2018}
Hoyle B.,  et~al., 2018, \mn@doi [\mnras] {10.1093/mnras/sty957}, 478,
  592–610

\bibitem[\protect\citeauthoryear{{Hunter}}{{Hunter}}{2007}]{4160265}
{Hunter} J.~D.,  2007, \mn@doi [Computing in Science Engineering]
  {10.1109/MCSE.2007.55}, 9, 90

\bibitem[\protect\citeauthoryear{Huterer, Takada, Bernstein  \& Jain}{Huterer
  et~al.}{2006}]{10.1111/j.1365-2966.2005.09782.x}
Huterer D.,  Takada M.,  Bernstein G.,   Jain B.,  2006, \mn@doi [\mnras]
  {10.1111/j.1365-2966.2005.09782.x}, 366, 101

\bibitem[\protect\citeauthoryear{Huterer, Lin, Busha, Wechsler  \&
  Cunha}{Huterer et~al.}{2014}]{10.1093/mnras/stu1424}
Huterer D.,  Lin H.,  Busha M.~T.,  Wechsler R.~H.,   Cunha C.~E.,  2014,
  \mn@doi [\mnras] {10.1093/mnras/stu1424}, 444, 129

\bibitem[\protect\citeauthoryear{{Ilbert} et~al.,}{{Ilbert}
  et~al.}{2006}]{2006A&A...457..841I}
{Ilbert} O.,  et~al., 2006, \mn@doi [\aap] {10.1051/0004-6361:20065138}, \href
  {http://adsabs.harvard.edu/abs/2006A%26A...457..841I} {457, 841}

\bibitem[\protect\citeauthoryear{{Ivezi{\'c}} et~al.,}{{Ivezi{\'c}}
  et~al.}{2019}]{2019ApJ...873..111I}
{Ivezi{\'c}} {\v{Z}}.,  et~al., 2019, \mn@doi [\apj]
  {10.3847/1538-4357/ab042c}, \href
  {https://ui.adsabs.harvard.edu/abs/2019ApJ...873..111I} {873, 111}

\bibitem[\protect\citeauthoryear{Izbicki \& Lee}{Izbicki \&
  Lee}{2017}]{Izbicki:17}
Izbicki R.,  Lee A.~B.,  2017, \mn@doi [Electron. J. Statist.]
  {10.1214/17-EJS1302}, 11, 2800

\bibitem[\protect\citeauthoryear{Johnson et~al.,}{Johnson
  et~al.}{2016}]{10.1093/mnras/stw3033}
Johnson A.,  et~al., 2016, \mn@doi [\mnras] {10.1093/mnras/stw3033}, 465, 4118

\bibitem[\protect\citeauthoryear{{Jones} \& {Heavens}}{{Jones} \&
  {Heavens}}{2019}]{2019MNRAS.483.2487J}
{Jones} D.~M.,  {Heavens} A.~F.,  2019, \mn@doi [\mnras]
  {10.1093/mnras/sty3279}, \href
  {https://ui.adsabs.harvard.edu/\#abs/2019MNRAS.483.2487J} {483, 2487}

\bibitem[\protect\citeauthoryear{{Joudaki} et~al.,}{{Joudaki}
  et~al.}{2018}]{2018MNRAS.474.4894J}
{Joudaki} S.,  et~al., 2018, \mn@doi [\mnras] {10.1093/mnras/stx2820}, \href
  {https://ui.adsabs.harvard.edu/abs/2018MNRAS.474.4894J} {474, 4894}

\bibitem[\protect\citeauthoryear{{Joudaki} et~al.,}{{Joudaki}
  et~al.}{2020}]{2020A&A...638L...1J}
{Joudaki} S.,  et~al., 2020, \mn@doi [\aap] {10.1051/0004-6361/201936154},
  \href {https://ui.adsabs.harvard.edu/abs/2020A&A...638L...1J} {638, L1}

\bibitem[\protect\citeauthoryear{{Kalmbach} \& {Connolly}}{{Kalmbach} \&
  {Connolly}}{2017}]{Kalmbach:2017}
{Kalmbach} J.~B.,  {Connolly} A.~J.,  2017, \mn@doi [\aj]
  {10.3847/1538-3881/aa9933}, \href
  {https://ui.adsabs.harvard.edu/abs/2017AJ....154..277K} {154, 277}

\bibitem[\protect\citeauthoryear{Kluyver et~al.,}{Kluyver
  et~al.}{2016}]{Kluyver:2016aa}
Kluyver T.,  et~al., 2016, in Loizides F.,  Schmidt B.,  eds, Positioning and
  Power in Academic Publishing: Players, Agents and Agendas. pp 87 -- 90

\bibitem[\protect\citeauthoryear{{Korytov} et~al.,}{{Korytov}
  et~al.}{2019}]{cosmodc2:2019}
{Korytov} D.,  et~al., 2019, \mn@doi [\apjs] {10.3847/1538-4365/ab510c}, \href
  {https://ui.adsabs.harvard.edu/abs/2019ApJS..245...26K} {245, 26}

\bibitem[\protect\citeauthoryear{Kress}{Kress}{1998}]{kress1998numerical}
Kress R.,  1998, Numerical Analysis.
Graduate Texts in Mathematics, Springer New York, \url
  {https://books.google.com.na/books?id=R6182rh0tKEC}

\bibitem[\protect\citeauthoryear{Kuusela}{Kuusela}{2016}]{Kuusela:220015}
Kuusela M.~J.,  2016, PhD thesis, Lausanne, EPFL,
  \mn@doi{10.5075/epfl-thesis-7118}, \url
  {http://infoscience.epfl.ch/record/220015}

\bibitem[\protect\citeauthoryear{{Laureijs} et~al.,}{{Laureijs}
  et~al.}{2011}]{2011arXiv1110.3193L}
{Laureijs} R.,  et~al., 2011, arXiv e-prints, \href
  {https://ui.adsabs.harvard.edu/\#abs/2011arXiv1110.3193L} {p.
  arXiv:1110.3193}

\bibitem[\protect\citeauthoryear{{Leistedt}, {Mortlock}  \&
  {Peiris}}{{Leistedt} et~al.}{2016}]{2016MNRAS.460.4258L}
{Leistedt} B.,  {Mortlock} D.~J.,   {Peiris} H.~V.,  2016, \mn@doi [\mnras]
  {10.1093/mnras/stw1304}, \href
  {http://adsabs.harvard.edu/abs/2016MNRAS.460.4258L} {460, 4258}

\bibitem[\protect\citeauthoryear{{Ma}, {Hu}  \& {Huterer}}{{Ma}
  et~al.}{2006}]{2006ApJ...636...21M}
{Ma} Z.,  {Hu} W.,   {Huterer} D.,  2006, \mn@doi [\apj] {10.1086/497068},
  \href {https://ui.adsabs.harvard.edu/abs/2006ApJ...636...21M} {636, 21}

\bibitem[\protect\citeauthoryear{{Malz} \& {Hogg}}{{Malz} \&
  {Hogg}}{2020}]{2020arXiv200712178M}
{Malz} A.~I.,  {Hogg} D.~W.,  2020, arXiv e-prints, \href
  {https://ui.adsabs.harvard.edu/abs/2020arXiv200712178M} {p. arXiv:2007.12178}

\bibitem[\protect\citeauthoryear{{Mandelbaum}}{{Mandelbaum}}{2018}]{2018ARA&A..56..393M}
{Mandelbaum} R.,  2018, \mn@doi [\araa] {10.1146/annurev-astro-081817-051928},
  \href {https://ui.adsabs.harvard.edu/abs/2018ARA&A..56..393M} {56, 393}

\bibitem[\protect\citeauthoryear{{Matarrese}, {Coles}, {Lucchin}  \&
  {Moscardini}}{{Matarrese} et~al.}{1997}]{1997MNRAS.286..115M}
{Matarrese} S.,  {Coles} P.,  {Lucchin} F.,   {Moscardini} L.,  1997, \mn@doi
  [\mnras] {10.1093/mnras/286.1.115}, \href
  {https://ui.adsabs.harvard.edu/abs/1997MNRAS.286..115M} {286, 115}

\bibitem[\protect\citeauthoryear{{McLeod}, {Balan}  \& {Abdalla}}{{McLeod}
  et~al.}{2017}]{2017MNRAS.466.3558M}
{McLeod} M.,  {Balan} S.~T.,   {Abdalla} F.~B.,  2017, \mn@doi [\mnras]
  {10.1093/mnras/stw2989}, \href
  {https://ui.adsabs.harvard.edu/\#abs/2017MNRAS.466.3558M} {466, 3558}

\bibitem[\protect\citeauthoryear{{McQuinn} \& {White}}{{McQuinn} \&
  {White}}{2013}]{2013MNRAS.433.2857M}
{McQuinn} M.,  {White} M.,  2013, \mn@doi [\mnras] {10.1093/mnras/stt914},
  \href {https://ui.adsabs.harvard.edu/\#abs/2013MNRAS.433.2857M} {433, 2857}

\bibitem[\protect\citeauthoryear{Meister}{Meister}{2009}]{meister2009deconvolution}
Meister A.,  2009, Deconvolution Problems in Nonparametric Statistics.
Lecture Notes in Statistics, Springer Berlin Heidelberg, \url
  {https://books.google.de/books?id=ItGkJPZQj-MC}

\bibitem[\protect\citeauthoryear{{M{\'e}nard}, {Scranton}, {Schmidt},
  {Morrison}, {Jeong}, {Budavari}  \& {Rahman}}{{M{\'e}nard}
  et~al.}{2013}]{2013arXiv1303.4722M}
{M{\'e}nard} B.,  {Scranton} R.,  {Schmidt} S.,  {Morrison} C.,  {Jeong} D.,
  {Budavari} T.,   {Rahman} M.,  2013, arXiv e-prints, \href
  {https://ui.adsabs.harvard.edu/\#abs/2013arXiv1303.4722M} {p.
  arXiv:1303.4722}

\bibitem[\protect\citeauthoryear{{Morrison}, {Hildebrandt}, {Schmidt},
  {Baldry}, {Bilicki}, {Choi}, {Erben}  \& {Schneider}}{{Morrison}
  et~al.}{2017}]{Morrison2016}
{Morrison} C.~B.,  {Hildebrandt} H.,  {Schmidt} S.~J.,  {Baldry} I.~K.,
  {Bilicki} M.,  {Choi} A.,  {Erben} T.,   {Schneider} P.,  2017, \mn@doi
  [\mnras] {10.1093/mnras/stx342}, \href
  {https://ui.adsabs.harvard.edu/abs/2017MNRAS.467.3576M} {467, 3576}

\bibitem[\protect\citeauthoryear{{Myles} et~al.,}{{Myles}
  et~al.}{2020}]{2020arXiv201208566M}
{Myles} J.,  et~al., 2020, arXiv e-prints, \href
  {https://ui.adsabs.harvard.edu/abs/2020arXiv201208566M} {p. arXiv:2012.08566}

\bibitem[\protect\citeauthoryear{Neal \& Hinton}{Neal \&
  Hinton}{1993}]{Neal93anew}
Neal R.~M.,  Hinton G.~E.,  1993, in Learning in Graphical Models. Kluwer
  Academic Publishers, pp 355--368

\bibitem[\protect\citeauthoryear{{Newman}}{{Newman}}{2008}]{2008ApJ...684...88N}
{Newman} J.~A.,  2008, \mn@doi [\apj] {10.1086/589982}, \href
  {https://ui.adsabs.harvard.edu/\#abs/2008ApJ...684...88N} {684, 88}

\bibitem[\protect\citeauthoryear{{Newman} et~al.,}{{Newman}
  et~al.}{2015}]{2015APh....63...81N}
{Newman} J.~A.,  et~al., 2015, \mn@doi [Astroparticle Physics]
  {10.1016/j.astropartphys.2014.06.007}, \href
  {https://ui.adsabs.harvard.edu/\#abs/2015APh....63...81N} {63, 81}

\bibitem[\protect\citeauthoryear{{Nicola} et~al.,}{{Nicola}
  et~al.}{2020}]{2020JCAP...03..044N}
{Nicola} A.,  et~al., 2020, \mn@doi [\jcap] {10.1088/1475-7516/2020/03/044},
  \href {https://ui.adsabs.harvard.edu/abs/2020JCAP...03..044N} {2020, 044}

\bibitem[\protect\citeauthoryear{{Padmanabhan} et~al.,}{{Padmanabhan}
  et~al.}{2005}]{2005MNRAS.359..237P}
{Padmanabhan} N.,  et~al., 2005, \mn@doi [\mnras]
  {10.1111/j.1365-2966.2005.08915.x}, \href
  {https://ui.adsabs.harvard.edu/abs/2005MNRAS.359..237P} {359, 237}

\bibitem[\protect\citeauthoryear{Pawitan}{Pawitan}{2001}]{pawitan2001all}
Pawitan Y.,  2001, In All Likelihood: Statistical Modelling and Inference Using
  Likelihood.
Oxford science publications, OUP Oxford, \url
  {https://books.google.com/books?id=M-3pSCVxV5oC}

\bibitem[\protect\citeauthoryear{{Perez} \& {Granger}}{{Perez} \&
  {Granger}}{2007}]{4160251}
{Perez} F.,  {Granger} B.~E.,  2007, \mn@doi [Computing in Science Engineering]
  {10.1109/MCSE.2007.53}, 9, 21

\bibitem[\protect\citeauthoryear{{Prat} et~al.,}{{Prat}
  et~al.}{2018}]{2018MNRAS.473.1667P}
{Prat} J.,  et~al., 2018, \mn@doi [\mnras] {10.1093/mnras/stx2430}, \href
  {https://ui.adsabs.harvard.edu/abs/2018MNRAS.473.1667P} {473, 1667}

\bibitem[\protect\citeauthoryear{{Prat} et~al.,}{{Prat}
  et~al.}{2019}]{2019MNRAS.487.1363P}
{Prat} J.,  et~al., 2019, \mn@doi [\mnras] {10.1093/mnras/stz1309}, \href
  {https://ui.adsabs.harvard.edu/abs/2019MNRAS.487.1363P} {487, 1363}

\bibitem[\protect\citeauthoryear{{Quiroz}, {Villani}, {Kohn}, {Tran}  \&
  {Dang}}{{Quiroz} et~al.}{2018}]{2018arXiv180708409Q}
{Quiroz} M.,  {Villani} M.,  {Kohn} R.,  {Tran} M.-N.,   {Dang} K.-D.,  2018,
  arXiv e-prints, \href {https://ui.adsabs.harvard.edu/abs/2018arXiv180708409Q}
  {p. arXiv:1807.08409}

\bibitem[\protect\citeauthoryear{Raccanelli, Rahman  \& Kovetz}{Raccanelli
  et~al.}{2017}]{10.1093/mnras/stx691}
Raccanelli A.,  Rahman M.,   Kovetz E.~D.,  2017, \mn@doi [\mnras]
  {10.1093/mnras/stx691}, 468, 3650

\bibitem[\protect\citeauthoryear{Ranganathan}{Ranganathan}{2004}]{Ranganathan04assumeddensity}
Ranganathan A.,  2004, Assumed Density Filtering,
  \url{http://www.ananth.in/Notes_files/adf.pdf}

\bibitem[\protect\citeauthoryear{{Rau}, {Seitz}, {Brimioulle}, {Frank},
  {Friedrich}, {Gruen}  \& {Hoyle}}{{Rau} et~al.}{2015}]{2015MNRAS.452.3710R}
{Rau} M.~M.,  {Seitz} S.,  {Brimioulle} F.,  {Frank} E.,  {Friedrich} O.,
  {Gruen} D.,   {Hoyle} B.,  2015, \mn@doi [\mnras] {10.1093/mnras/stv1567},
  \href {https://ui.adsabs.harvard.edu/\#abs/2015MNRAS.452.3710R} {452, 3710}

\bibitem[\protect\citeauthoryear{{Rau}, {Hoyle}, {Paech}  \& {Seitz}}{{Rau}
  et~al.}{2017}]{2017MNRAS.466.2927R}
{Rau} M.~M.,  {Hoyle} B.,  {Paech} K.,   {Seitz} S.,  2017, \mn@doi [\mnras]
  {10.1093/mnras/stw3338}, \href
  {https://ui.adsabs.harvard.edu/\#abs/2017MNRAS.466.2927R} {466, 2927}

\bibitem[\protect\citeauthoryear{{Rau}, {Wilson}  \& {Mandelbaum}}{{Rau}
  et~al.}{2020}]{2020MNRAS.491.4768R}
{Rau} M.~M.,  {Wilson} S.,   {Mandelbaum} R.,  2020, \mn@doi [\mnras]
  {10.1093/mnras/stz3295}, \href
  {https://ui.adsabs.harvard.edu/abs/2020MNRAS.491.4768R} {491, 4768}

\bibitem[\protect\citeauthoryear{Raue, Kreutz, Theis  \& Timmer}{Raue
  et~al.}{2013}]{Raue_joining_forces}
Raue A.,  Kreutz C.,  Theis F.,   Timmer J.,  2013, \mn@doi [Philosophical
  transactions. Series A, Mathematical, physical, and engineering sciences]
  {10.1098/rsta.2011.0544}, 371, 20110544

\bibitem[\protect\citeauthoryear{Rothenberg}{Rothenberg}{1971}]{RePEc:ecm:emetrp:v:39:y:1971:i:3:p:577-91}
Rothenberg T.~J.,  1971, Econometrica, 39, 577

\bibitem[\protect\citeauthoryear{{Salvato}, {Ilbert}  \& {Hoyle}}{{Salvato}
  et~al.}{2019}]{2019NatAs...3..212S}
{Salvato} M.,  {Ilbert} O.,   {Hoyle} B.,  2019, \mn@doi [Nature Astronomy]
  {10.1038/s41550-018-0478-0}, \href
  {https://ui.adsabs.harvard.edu/abs/2019NatAs...3..212S} {3, 212}

\bibitem[\protect\citeauthoryear{{S{\'a}nchez} \& {Bernstein}}{{S{\'a}nchez} \&
  {Bernstein}}{2019}]{2019MNRAS.483.2801S}
{S{\'a}nchez} C.,  {Bernstein} G.~M.,  2019, \mn@doi [\mnras]
  {10.1093/mnras/sty3222}, \href
  {https://ui.adsabs.harvard.edu/\#abs/2019MNRAS.483.2801S} {483, 2801}

\bibitem[\protect\citeauthoryear{{S{\'a}nchez}, {Raveri}, {Alarcon}  \&
  {Bernstein}}{{S{\'a}nchez} et~al.}{2020}]{2020arXiv200409542S}
{S{\'a}nchez} C.,  {Raveri} M.,  {Alarcon} A.,   {Bernstein} G.~M.,  2020,
  \mn@doi [\mnras] {10.1093/mnras/staa2542}, \href
  {https://ui.adsabs.harvard.edu/abs/2020MNRAS.tmp.2467S} {}

\bibitem[\protect\citeauthoryear{{S{\'a}nchez} et~al.,}{{S{\'a}nchez}
  et~al.}{2021}]{2021arXiv210513542S}
{S{\'a}nchez} C.,  et~al., 2021, arXiv e-prints, \href
  {https://ui.adsabs.harvard.edu/abs/2021arXiv210513542S} {p. arXiv:2105.13542}

\bibitem[\protect\citeauthoryear{{Schmidt}, {M{\'e}nard}, {Scranton},
  {Morrison}  \& {McBride}}{{Schmidt} et~al.}{2013}]{2013MNRAS.431.3307S}
{Schmidt} S.~J.,  {M{\'e}nard} B.,  {Scranton} R.,  {Morrison} C.,   {McBride}
  C.~K.,  2013, \mn@doi [\mnras] {10.1093/mnras/stt410}, \href
  {https://ui.adsabs.harvard.edu/abs/2013MNRAS.431.3307S} {431, 3307}

\bibitem[\protect\citeauthoryear{{Scottez} et~al.,}{{Scottez}
  et~al.}{2016}]{2016MNRAS.462.1683S}
{Scottez} V.,  et~al., 2016, \mn@doi [\mnras] {10.1093/mnras/stw1500}, \href
  {https://ui.adsabs.harvard.edu/\#abs/2016MNRAS.462.1683S} {462, 1683}

\bibitem[\protect\citeauthoryear{{Scranton} et~al.,}{{Scranton}
  et~al.}{2005}]{2005ApJ...633..589S}
{Scranton} R.,  et~al., 2005, \mn@doi [\apj] {10.1086/431358}, \href
  {https://ui.adsabs.harvard.edu/\#abs/2005ApJ...633..589S} {633, 589}

\bibitem[\protect\citeauthoryear{{Simon} \& {Hilbert}}{{Simon} \&
  {Hilbert}}{2018}]{2018A&A...613A..15S}
{Simon} P.,  {Hilbert} S.,  2018, \mn@doi [\aap] {10.1051/0004-6361/201732248},
  \href {https://ui.adsabs.harvard.edu/abs/2018A&A...613A..15S} {613, A15}

\bibitem[\protect\citeauthoryear{{Speagle} \& {Eisenstein}}{{Speagle} \&
  {Eisenstein}}{2015}]{2015arXiv151008073S}
{Speagle} J.~S.,  {Eisenstein} D.~J.,  2015, arXiv e-prints, \href
  {https://ui.adsabs.harvard.edu/\#abs/2015arXiv151008073S} {p.
  arXiv:1510.08073}

\bibitem[\protect\citeauthoryear{{Spergel} et~al.,}{{Spergel}
  et~al.}{2015}]{2015arXiv150303757S}
{Spergel} D.,  et~al., 2015, arXiv e-prints, \href
  {https://ui.adsabs.harvard.edu/\#abs/2015arXiv150303757S} {p.
  arXiv:1503.03757}

\bibitem[\protect\citeauthoryear{{St{\"o}lzner}, {Joachimi}, {Korn},
  {Hildebrandt}  \& {Wright}}{{St{\"o}lzner}
  et~al.}{2020}]{2020arXiv201207707S}
{St{\"o}lzner} B.,  {Joachimi} B.,  {Korn} A.,  {Hildebrandt} H.,   {Wright}
  A.~H.,  2020, arXiv e-prints, \href
  {https://ui.adsabs.harvard.edu/abs/2020arXiv201207707S} {p. arXiv:2012.07707}

\bibitem[\protect\citeauthoryear{Stoyan \& Stoyan}{Stoyan \&
  Stoyan}{1994}]{stoyan1994fractals}
Stoyan D.,  Stoyan H.,  1994, Fractals, Random Shapes and Point Fields: Methods
  of Geometrical Statistics.
Wiley Series in Probability and Statistics, Wiley, \url
  {https://books.google.com/books?id=Dw3vAAAAMAAJ}

\bibitem[\protect\citeauthoryear{{Tagliaferri}, {Longo}, {Andreon},
  {Capozziello}, {Donalek}  \& {Giordano}}{{Tagliaferri}
  et~al.}{2003}]{2003LNCS.2859..226T}
{Tagliaferri} R.,  {Longo} G.,  {Andreon} S.,  {Capozziello} S.,  {Donalek} C.,
    {Giordano} G.,  2003, {Neural Networks for Photometric Redshifts
  Evaluation}.
pp 226--234, \mn@doi{10.1007/978-3-540-45216-4_26}

\bibitem[\protect\citeauthoryear{{Tanaka} et~al.,}{{Tanaka}
  et~al.}{2018}]{2018PASJ...70S...9T}
{Tanaka} M.,  et~al., 2018, \mn@doi [\pasj] {10.1093/pasj/psx077}, \href
  {https://ui.adsabs.harvard.edu/abs/2018PASJ...70S...9T} {70, S9}

\bibitem[\protect\citeauthoryear{{The LSST Dark Energy Science Collaboration}
  et~al.,}{{The LSST Dark Energy Science Collaboration}
  et~al.}{2018}]{2018arXiv180901669T}
{The LSST Dark Energy Science Collaboration} et~al., 2018, arXiv e-prints,
  \href {https://ui.adsabs.harvard.edu/\#abs/2018arXiv180901669T} {p.
  arXiv:1809.01669}

\bibitem[\protect\citeauthoryear{Uitert et~al.,}{Uitert
  et~al.}{2017}]{UitertKids}
Uitert E.,  et~al., 2017, \mn@doi [\mnras] {10.1093/mnras/sty551}, 476

\bibitem[\protect\citeauthoryear{Varin, Reid  \& Firth}{Varin
  et~al.}{2011}]{Varin11anoverview}
Varin C.,  Reid N.,   Firth D.,  2011, Statist. Sinica, pp 5--42

\bibitem[\protect\citeauthoryear{Virtanen et~al.,}{Virtanen
  et~al.}{2020}]{2020SciPy-NMeth}
Virtanen P.,  et~al., 2020, \mn@doi [Nature Methods]
  {10.1038/s41592-019-0686-2}, \href {https://rdcu.be/b08Wh} {17, 261}

\bibitem[\protect\citeauthoryear{{W}es {M}c{K}inney}{{W}es
  {M}c{K}inney}{2010}]{mckinney-proc-scipy-2010}
{W}es {M}c{K}inney 2010, in {S}t\'efan van~der {W}alt {J}arrod {M}illman eds,
  {P}roceedings of the 9th {P}ython in {S}cience {C}onference. pp 56 -- 61,
  \mn@doi{10.25080/Majora-92bf1922-00a}

\bibitem[\protect\citeauthoryear{Wickham \& Stryjewski}{Wickham \&
  Stryjewski}{2012}]{boxplots_ref}
Wickham H.,  Stryjewski L.,  2012, Technical report, 40 years of boxplots.
had.co.nz

\bibitem[\protect\citeauthoryear{{Zhou} et~al.,}{{Zhou}
  et~al.}{2020a}]{2020arXiv200106018Z}
{Zhou} R.,  et~al., 2020a, arXiv e-prints, \href
  {https://ui.adsabs.harvard.edu/abs/2020arXiv200106018Z} {p. arXiv:2001.06018}

\bibitem[\protect\citeauthoryear{{Zhou} et~al.,}{{Zhou}
  et~al.}{2020b}]{2020RNAAS...4..181Z}
{Zhou} R.,  et~al., 2020b, \mn@doi [Research Notes of the American Astronomical
  Society] {10.3847/2515-5172/abc0f4}, \href
  {https://ui.adsabs.harvard.edu/abs/2020RNAAS...4..181Z} {4, 181}

\bibitem[\protect\citeauthoryear{{van Daalen} \& {White}}{{van Daalen} \&
  {White}}{2018}]{2018MNRAS.476.4649V}
{van Daalen} M.~P.,  {White} M.,  2018, \mn@doi [\mnras]
  {10.1093/mnras/sty545}, \href
  {https://ui.adsabs.harvard.edu/abs/2018MNRAS.476.4649V} {476, 4649}

\bibitem[\protect\citeauthoryear{van Rossum}{van Rossum}{1995}]{CS-R9526}
van Rossum G.,  1995, Technical Report CS-R9526, Python tutorial.
Centrum voor Wiskunde en Informatica (CWI), Amsterdam

\bibitem[\protect\citeauthoryear{{van den Busch} et~al.,}{{van den Busch}
  et~al.}{2020}]{2020A&A...642A.200V}
{van den Busch} J.~L.,  et~al., 2020, \mn@doi [\aap]
  {10.1051/0004-6361/202038835}, \href
  {https://ui.adsabs.harvard.edu/abs/2020A&A...642A.200V} {642, A200}

\makeatother
\end{thebibliography}

\appendix
\rev{
\section{Introduction to Deconvolution Problems} 
\label{sec:methods}
As we will see in the following subsections, the photometric redshift problem is a deconvolution problem, where the redshift distribution of a sample of photometrically observed galaxies is inferred from their noisy photometric measurements. To give the reader an intuitive understanding of deconvolution problems, we present a short introduction into the classical deconvolution problem. A similar description in the context of photometric redshift estimation can be found in \citet{2005MNRAS.359..237P}. 

Consider three vectors of random variables $\mathbf{Z}$, $\mathbf{Z}^{p}$ and $\boldsymbol{\epsilon}$ with dimension $N_{\rm gal}$, which denotes the photometric sample size. $\mathbf{Z}$ and  $\mathbf{Z}^{p}$ denote the true and photometric redshifts of the galaxies in the sample and $\boldsymbol{\epsilon}$ the residual error between both quantities. The additive noise model that connects these random variables is given as: 
\begin{equation}
    \mathbf{Z}^{p} = \mathbf{Z} + \boldsymbol{\epsilon} \, . 
\end{equation}
The probability densities\footnote{\mmref{Note that the probability densities $p_z$ and $p_z^{p}$ are both redshift distributions of samples of galaxies. They differ since ($p_z^{p}$, $p_z$) denotes the sample distribution of (photometric, true or spectroscopic) galaxy redshifts. Thus $p_z^{p}$ would be broader, since the error in the redshift, drawn from  $p_{\epsilon}$, is convolved with the true redshift. }} associated with these random variables are: 
\begin{align}
    Z_j &\sim p_z \\
    Z_j^p &\sim p_z^{p} \\
    \epsilon_j &\sim p_{\epsilon} \, ,
\end{align}
where $j \in \{1, \dots, N^{\rm gal}\}$ and `$\sim$' connects the realization of a random variable on the left hand side with the probability density function (PDF) on the right hand side from which this realization is drawn. 

The random variable $\epsilon_j$ is assumed to be identically and independently distributed, as well as independent of $Z_j$. These assumptions do not hold in the photometric redshift scenario, as the noise very clearly depends on the color, and therefore redshift, of the galaxy. However, in the following toy model, we adopt these assumptions for simplicity. The theory can be easily extended towards input-dependent noise \citep[see e.g.][]{meister2009deconvolution} without changing the intuition presented in this section. 

In order to derive an estimator for $p_z$, we use the convolution theorem\footnote{A Fourier-based approach is not necessary. Concretely, the likelihood framework presented in the following section works in real space. A Fourier description for the classical deconvolution problem is, however, analytically tractable and provides a clear picture of the nature of the problem and the importance of regularization.} that connects the PDF of the sum of independent random variables with the convolution of their densities. 
We can therefore write: 
\begin{equation}
p_z^{p} = p_z * p_{\epsilon} = \int p_z(z^{p} - z) p_{\epsilon}(z) \mathrm{d}z = \int p_{\epsilon}(z^p - z) p_z(z) \, \mathrm{d}z \, ,
\end{equation}

The Fourier transform of a probability distribution is the characteristic function. We will denote the characteristic functions of ($p_z$, $p_z^p$, $p_{\epsilon}$) as ($p_z^{\rm ft}$, $p_z^{\rm p, ft}$, $p_{\epsilon}^{\rm ft}$).  Given a sample drawn from a PDF, e.g., the sample of photometric redshifts of $N_{\rm gal}$ galaxies, we can estimate $p_z^{\rm p, ft}$ as\footnote{\mmref{Here, $\hat{}$ denotes an estimator for the respective function. }}
\begin{equation}
    \hat{p}_z^{\rm p, ft}(t) = \frac{1}{N_{\rm gal}} \sum_{j = 1}^{N_{\rm gal}} \exp{\left(i t Z^{p}_j \right)} \, .
\end{equation}
The argument $t$ of the characteristic function could be interpreted as a kind of redshift-frequency if we treat the redshift of a galaxy as a `time parameter'.
Under the assumption of independence between $\mathbf{Z}$ and $\mathbf{\epsilon}$ we can write: 
\begin{equation}
    p_z^{\rm p, ft} (t) = \hat{p}_z^{\rm ft}(t) \hat{p}_{\epsilon}^{\rm ft}(t) = p_z^{\rm ft}(t) p_{\epsilon}^{\rm ft}(t) \, .
\end{equation}
Therefore an estimator for $p_z^{\rm ft}(t)$ is given as:
\begin{equation}
    \hat{p}_z^{ft}(t) = \frac{1}{p^{\rm ft}_{\epsilon}(t) N_{\rm gal}} \sum_{j = 1}^{N_{\rm gal}} \exp{\left(i t Z_j^{p} \right)} \, ,
\end{equation}
where we assume $p^{\rm ft}_{\epsilon}(t)$ is known and nonzero everywhere. 
We note that this estimator is consistent and unbiased \citep{meister2009deconvolution}. The error term $p_{\epsilon}^{\rm ft}(t)$ here acts as a `filter' to weight down small scale modes in the distribution. However, we note that this term $1/p_{\epsilon}^{\rm ft}$ can become large when $p_{\epsilon}^{\rm ft}$ is small. 

As a consequence the inverse Fourier transform 
\begin{equation}
    p_{z}(z) = \frac{1}{2 \pi} \int \exp{\left(-i t z\right)} \, \hat{p}_z^{ft}(t) \, \mathrm{d}t \, ,
\end{equation}
is neither integrable nor square integrable. 
Loosely speaking this implies that the parameter space that describes the shape of $p_z(z)$ does not have to be bounded. We will see this effect also for the more complex model considered in the later sections of this work. We reiterate that while the estimator of $\hat{p}_{z}^{\rm ft}$ has very desirable properties, the inverse transformation is not well defined, hence deconvolution problems are part of a larger class of `inverse problems'.

In order to obtain well-defined results, we therefore have to perform regularization either by regulating the shape/parametrization of $p_z^{p}$ (e.g. using Kernel methods), projecting $p_z$ onto a suitable basis like wavelet functions or by directly restricting the $1/p_{\epsilon}^{\rm ft}$ term, as implemented in a Ridge method \citep[e.g.][\S~2.2.3]{meister2009deconvolution}. 
We will not discuss the details of these methods and refer to the literature for a more detailed explanation \citep[e.g.][]{meister2009deconvolution}. It is, however, instructive to study the functional form of one of these regularized estimators. Making the ansatz of a kernel density estimate for the photometric redshift PDF, one can show that the deconvolved density $\hat{p}_z$ can be estimated as \citep[e.g.][]{meister2009deconvolution}:
\begin{equation}
    \hat{p}_z(z) = \frac{1}{2 \pi} \int \exp{\left(-i t z \right)} \left(\frac{K^{\rm ft}(t b)}{p_{\epsilon}^{\rm ft}(t)}\right) \frac{1}{N_{\rm gal}} \sum_{j = 1}^{N_{\rm gal}} \exp{\left(i t Z_j^p\right)}  \mathrm{d}t \, ,
    \label{eq:kde_classical_convolution}
\end{equation}
where $b$ denotes the bandwidth and $K^{\rm ft}$ the fourier transform of the kernel function that enters the kernel density estimation ansatz for $p_z^{p}$. We see that by restricting the shape of  the density $p_z^{p}$ to a kernel density estimate whose smoothness is governed by the parameter $b$, we regularize the $1/p_\epsilon^{\rm ft}(t)$ term by a multiplicative factor, that renders the inverse Fourier transformation both integrable and square integrable assuming bounded, compactly supported and non-vanishing $K^{\rm ft}$. The bandwidth parameter governs the tradeoff between the bias, or `smoothness', of the density estimate, and its variance. Choosing a larger bandwidth washes out small scale noise in the reconstructed density.  In the limit of vanishing bandwidth, we would again obtain an ill-posed inverse Fourier transformation.

While the considered toy model of the photometric redshift problem is analytically tractable, it does not describe the realistic situation. Besides the relatively simple extension towards a galaxy-dependent photometric noise, the noise distribution $p_{\epsilon}$ is, in photometric redshift estimation, given as a joint likelihood between the photometry of all galaxies in the sample, that depends on the additional parameters that enter the SED modelling. The redshift of each galaxy is a parameter that enters its likelihood and the sample redshift distribution is its prior. Furthermore, the model does not properly account for the spatial distribution of galaxies. 

We finally note that inverse problems like the classical deconvolution problem also appear in several other scenarios like the measurements of shapes, where the point-spread function (PSF) of galaxies convolves the galaxies' light profiles and leads to a loss of information.

}
\section{Deriving the E-M update equations}
\label{par:ml_inference}
\mmref{We start the discussion with an intuitive motivation for the theoretical foundation of the E-M algorithm. Assume a `system' that consists of hidden variables $Y$ and observed variables $Z$. We wish to find a set of parameters $\theta$ that maximize the joint distribution of both variables given $\theta$. We know from statistical physics that the free energy of this system $F(p, \theta)$ should be minimized and depends on the distribution of hidden variables, or states, $p(y)$ and the parameters of the conditional $p(y | z, \theta)$. The E-M algorithm performs this minimization iteratively, where we assume an initial choice for $\theta$. In the $E$-step, we choose a distribution $p(y)$, while holding $\theta$ fixed, so that $F(p, \theta^{\rm old})$ is minimized. In the subsequent $M$-step we hold $p$ fixed, but choose $\theta$ in a way that  $F(p^{\rm old}, \theta)$ is minimized. This procedure is iterated until the free energy does not change much with additional iterations, i.e., the scheme converges. In practical calculations, the connection with the variational free energy is not often used, but it is a useful concept to build up an intuitive understanding of the method. We refer the interested reader to \citet{Neal93anew} for a more detailed explanation. In the following we will describe the derivation of the E-M algorithm in the concrete context of finding the maximum likelihood solution of our photometric likelihood. For that we will use a different notation, however the intuition remains unchanged, if we associate the free energy (up to a sign) with the term $\mathcal{L}(q, \boldsymbol{\pi})$ in Eq.~\eqref{eq:em_def}.}

To derive the Expectation-Maximization algorithm\footnote{The interested reader will find the following derivation in analogy to the derivation to the E-M update equations for the Gaussian Mixture model \citep[see][]{10.5555/1162264}.}, we first introduce the parameter vector $\bm{\zeta}_i$ for each galaxy $i$ that is a $N_{\rm tot}$ dimensional vector to indicate bin membership\footnote{We are referring to bins as defined in Eq.~\eqref{eq:prior_def}.} in the `1-hot encoding' scheme.  
\markus{This means that for each galaxy, we have a $N_{\rm tot}$ dimensional binary vector, where (`0'/`1') indicates that the galaxy (resides/does not reside) in the respective redshift bin.} For the remainder of this section we will work with the normed histogram bin heights $\bm{\pi} = \mathbf{n}^B  \Delta z$ 
\mmref{that parametrize the prior distribution over $z-\alpha$ as defined in Eq.~\ref{eq:prior_def}}. The prior distribution over $\bm{\overline{\zeta}}$ \footnote{Here, $\bm{\overline{\zeta}}$ denotes the collection of $\bm{\zeta}$ vectors of all galaxies.}  is then given as
\begin{equation}
    p(\bm{\overline{\zeta}} | \bm{\pi} ) \propto \prod_{k = 1}^{N_{\rm tot}} \prod_{n = 1}^{N_{\rm gal}} \pi_k^{\zeta_{n, k}} \, ,
    \label{eq:cond_zeta_pi}
\end{equation}
where $\sum_{k = 1}^{N_{\rm tot}} \pi_k = 1$.  Using $\bm{\overline{\zeta}}$ we can write the conditional distribution of the measured photometry given $\bm{\overline{\zeta}}$ as
\begin{equation}
p(\hat{\mathbf{F}} | \bm{\overline{\zeta}}, \bm{\pi}) \propto \prod_{\beta = 1}^{N_{\rm gal}} \prod_{k = 1}^{N_{\rm tot}} \left(\int_{z_L^k}^{z_R^k} \, \mathrm{d}z_\beta \, \int_{\alpha_L^k}^{\alpha_R^k} \, \mathrm{d}\alpha_\beta \,  p(\mathbf{f}_{\beta} | \mathcal{T}(z_{\beta}, \alpha_{\beta}), \Sigma_\beta)\right)^{\zeta_{\rm \beta, k}} \, .
\label{eq:data_term}
\end{equation}
It is important at this point to note that a marginalization over the parameter vectors $\mathbf{\zeta}_i$ for all galaxies will yield the second, i.e. the likelihood, term in Eq.~\eqref{eq:post_eq_nb}. 

To derive the iterative optimization scheme we first consider the decomposition of the posterior as
\begin{equation}
    \log{p(\boldsymbol{\pi} | \mathbf{\hat{F}})} = \mathcal{L}(q, \boldsymbol{\pi}) + KL(q || p) + \log{p(\boldsymbol{\pi})} - \log{p(\mathbf{\hat{F}})} \, ,
    \label{eq:em_def}
\end{equation}
where 
\begin{align}
    \mathcal{L}(q, \mathbf{n^{B}}) &= \sum_{\bm{\overline{\zeta}}} q(\bm{\overline{\zeta}}) \log{\left(\frac{p(\mathbf{\hat{F}}, \bm{\overline{\zeta}})}{q(\bm{\overline{\zeta}})} \right)} \\
    KL(q || p) &= - \sum_{\bm{\overline{\zeta}}} q(\bm{\overline{\zeta}}) \log{\left(\frac{p(\bm{\overline{\zeta}} | \mathbf{\hat{F}}, \boldsymbol{\pi})}{q(\overline{\bm{\zeta}})}\right)}
\end{align}
This decomposition implies an iterative scheme to maximize $\log{p(\boldsymbol{\pi} | \mathbf{\hat{F}})}$. Given an initial parameter vector $\boldsymbol{\pi}^{\rm old}$, we first minimize $KL(q || p)$ in the `E-step' which directly implies $q(\overline{\bm{\zeta}}) = p(\bm{\overline{\zeta}} | \mathbf{\hat{F}}, \boldsymbol{\pi})$. In the `M'-step we fix the distribution $q(\overline{\bm{\zeta}})$ and maximize $\mathcal{L}(q, \boldsymbol{\pi})$. This maximization directive is then given as
\begin{equation}
    \mathcal{L}(q, \boldsymbol{\pi}) = S\left(\boldsymbol{\pi}, \boldsymbol{\pi^{\rm old}}\right) = \sum_{\bm{\overline{\zeta}}} p(\bm{\overline{\zeta}} | \mathbf{\hat{F}}, \boldsymbol{\pi^{\rm old}}) \log{p(\mathbf{\hat{F}}, \bm{\overline{\zeta}} | \boldsymbol{\pi})} + {\rm const.} \, ,
\end{equation}
which is the expectation of the data log-likelihood with respect to $\bm{\overline{\zeta}}$. After a new parameter vector $\boldsymbol{\pi^{\rm new}}$ is obtained, we continue with the `E'-step holding $\boldsymbol{\pi^{\rm new}}$ fixed. This process is continued until convergence.
In the following we will derive the corresponding update equations.

\subparagraph*{E-step:}
Given an old parameter vector $\mathbf{\pi}_{\rm old}$ we evaluate 
\begin{align}
    p(\bm{\overline{\zeta}} | \mathbf{\hat{F}}, &\bm{\pi}_{\rm old}) \propto \\
    &\prod_{\beta = 1}^{N_{\rm gal}} \prod_{j = 1}^{N_{\rm tot}} \left( \mathbf{\pi}_{\rm j, old}  \int_{z_L^k}^{z_R^k} \, \mathrm{d}z_\beta \, \int_{\alpha_L^k}^{\alpha_R^k} \, \mathrm{d}\alpha_\beta \,  p(\mathbf{f}_{\beta} | \mathcal{T}(z_{\beta}, \alpha_{\beta}), \Sigma_\beta)\right)^{\zeta_{\rm \beta j}} \, .
\end{align}

\subparagraph*{M-step:}
In the maximization step of the algorithm we want to maximize the expected data log-likelihood with respect to the parameter $\bm{\overline{\zeta}}$. Given the updated posterior $p(\bm{\overline{\zeta}} | \mathbf{\hat{F}}, \bm{\pi}_{\rm old})$ this expectation is given as: 
\begin{equation}
\begin{split}
&S\left(\bm{\pi}_{\rm new}, \bm{\pi}_{\rm old}\right) = \sum_{\beta = 1}^{N_{\rm gal}} \sum_{j = 1}^{N_{\rm tot}} E\left[\zeta_{\rm \beta \, j}\right] \times \\ &\left(\log{\left(\pi_{\rm \beta j}\right)} + \log{\left( \int_{z_L^j}^{z_R^j} \, \mathrm{d}z_\beta \, \int_{\alpha_L^j}^{\alpha_R^j} \, \mathrm{d}\alpha_\beta \,  p(\mathbf{f}_{\beta} | \mathcal{T}(z_{\beta}, \alpha_{\beta}), \Sigma_\beta)\right) }\right)
\end{split}\, , 
\end{equation}
where 
\begin{equation}
\begin{split}
    E\left[\zeta_{\rm \beta \, j}\right]  &=   \frac{\sum_{\zeta_{n, k}} \zeta^{n k} p( \zeta_{n k} | \mathbf{\hat{F}}, \bm{\pi}_{\rm old})}{ \sum_{\zeta_{n k}} p(\zeta_{n, k} | \mathbf{\hat{F}}, \bm{\pi}_{\rm old})}  \\
    &= \frac{\pi_{\rm i, old} \int_{z_L^i}^{z_R^i} \, \mathrm{d}z^{\beta} \, \int_{\alpha_L^i}^{\alpha_R^i} \, \mathrm{d}\alpha^{\beta} \, p(\mathbf{f}_{\beta} | \mathcal{T}(z_{\beta}, \alpha_{\beta}), \Sigma_\beta) }{  \sum_j \pi_{\rm j, old}   \int_{z_L^j}^{z_R^j} \, \mathrm{d}z^{\beta} \, \int_{\alpha_L^j}^{\alpha_R^j} \, \mathrm{d}\alpha^{\beta} \, p(\mathbf{f}_{\beta} | \mathcal{T}(z_{\beta}, \alpha_{\beta}), \Sigma_\beta)}
\end{split}
\end{equation}
We optimize $S\left(\mathbf{nz_{\rm z, t, new}}, \mathbf{nz_{\rm z, t, old}}\right)$ under the constraint $\sum_k \pi_{k} = 1$ using the Lagrange multiplier formalism:
\begin{equation}
    \overline{S}\left(\bm{\pi_{\rm new}}, \bm{\pi_{\rm old}}\right) = S\left(\bm{\pi_{\rm new}}, \bm{\pi_{\rm old}}\right) + \lambda \left( \sum_k \pi_{k}  - 1\right)
\end{equation}
Equating $\nabla_{\bm{\pi}} \overline{S}\left(\bm{\pi_{\rm new}}, \bm{\pi_{\rm old}}\right) == \mathbf{0}$, performing a summation over $k$, and using the summation constraint of $\bm{\pi}$, we obtain
\begin{equation}
    -\lambda = N_{\rm gal} \, .
\end{equation}
This leads to the update equations for the E-M scheme that are iterated until we reach convergence in $\bm{\pi}$\footnote{\mmref{In practice} we would iterate until the log-likelihood changes only by an extremely small amount.}:
\begin{align}
N_k^{t} &= \pi_k^{t - 1} \sum_{i = 1}^{N_{\rm gal}} \left(\frac{\int_{z_L^k}^{z_R^k} \mathrm{d}z_i \, \int_{\alpha_L^k}^{\alpha_R^k} \mathrm{d}\alpha_i \, p(\mathbf{\hat{f}_i} | \mathcal{T}(z_i, \alpha_i), \Sigma_i)}{\sum_{j = 1}^{N_{\rm tot}} \pi_j^{t - 1} \int_{z_L^j}^{z_R^j} \mathrm{d}z_i \int_{\alpha_L^j}^{\alpha_R^j} \mathrm{d}\alpha_i \, p(\mathbf{\hat{f}_i} | \mathcal{T}(z_i, \alpha_i), \Sigma_i)}\right) \\
\pi_k^{t} &= \frac{N_k^{t-1}}{\sum_k N_k^{t-1}} \, .
\end{align}

\markus{In appendix~\ref{par:derive_laplace} we will derive a Laplace approximation to the posterior based on this optimization scheme. We apply and discuss this scheme in \S~\ref{subsec:photometric_likelihood} and \S~\ref{sec:forecast_ideal_data}. }

\section{Deriving the Laplace Approximation}
\label{par:derive_laplace}
 \markus{In the previous \mmref{appendix} we derived an iterative scheme to obtain maximum likelihood estimates of the vector of normed histogram heights $\boldsymbol{\pi}_{\rm ML}$ based on the E-M algorithm.} 
 We note that the E-M algorithm is guaranteed to produce a maximum likelihood estimate $\boldsymbol{\pi}_{\rm ML}$ that lies on the simplex. The direct application of the Laplace approximation will effectively estimate Gaussian errors on the values. Applying this approximation around $\pi_{\rm ML}$ will lead to posteriors that reach to negative values, i.e. the posterior draws are not guaranteed to lie on the simplex. 
To extend the Laplace approximation to random variables that lie on the simplex, we first consider a mapping from simplex space to $\mathbb{R}^{N_{\rm bins} - 1}$.
This mapping is realized by the additive logistic transformation. Assume $\mathbf{y} \in \mathbb{R}^{N_{\rm bins} - 1}$, we define the function
\begin{equation}
    \boldsymbol{\pi}(\mathbf{y}) = \left(\frac{e^{y_1}}{1 + \sum_{i = 1}^{N_{\rm bins} - 1} e^{y_i}}, \dots, \frac{e^{y_{\rm N_{\rm bins} - 1}}}{1 + \sum_{i = 1}^{N_{\rm bins} - 1} e^{y_i}}, \frac{1}{1 + \sum_{i = 1}^{N_{\rm bins} - 1} e^{y_i}}\right)^{T} \, ,
    \label{eq:additive_logistic}
\end{equation}
with its inverse
\begin{equation}
    \mathbf{y}(\boldsymbol{\pi}) = \left[ \log{(\pi_1/\pi_{N_{\rm bins}})}, \dots, \log{(\pi_{N_{\rm bins} - 1}/\pi_{N_{\rm bins}})} \right] \, .
\end{equation}

We see that the transformed variables $\mathbf{y}$ are now defined in real space and we perform the Laplace approximation as usual. Assuming a flat prior in logistic space, we can directly utilize the invariance of the Maximum Likelihood estimate under variable transformations \citep[see e.g.][]{pawitan2001all} and approximate the posterior 
\begin{equation}
    p(\mathbf{y} | \mathbf{\hat{F}}) = \mathcal{N}(\mathbf{y} | \boldsymbol{\mu}_{\rm ML}, \boldsymbol{\Sigma}) \, ,
\end{equation}
where 
\begin{equation}
    \boldsymbol{\mu}_{\rm y \, ML} = \mathbf{y}(\boldsymbol{\pi}_{\rm ML}) \, ,
\end{equation}
and 
\begin{equation}
    \boldsymbol{\Sigma_y} = -\left.\mathbf{H}^{-1}\right\vert_{\mathbf{y} = \mathbf{y}_{\rm ML}} \, .
\end{equation}
Here $\mathbf{H}$ is the hessian of the log-likelihood (the second term in Eq.~\eqref{eq:post_eq_nb}) as a function of $\mathbf{y}$ evaluated at $\mathbf{y}(\boldsymbol{\pi}_{\rm ML})$.

The components of the hessian are given as
\begin{align}
    H_{a z} &= - \sum_{i = 1}^{N_{\rm gal}} \frac{\left(\sum_{j}^{N_{\rm tot}} \left(\frac{\partial \pi_j}{\partial y_z}\right) I_{i j}\right) \left(\sum_{j = 1}^{N_{\rm tot}} \left(\frac{\partial \pi_j}{\partial y_{a}}\right) I_{i j}\right)}{\left(\sum_{j = 1}^{N_{\rm tot}} \pi_j I_{i j}\right)^2} \\ &+ \sum_{i = 1}^{N_{\rm gal}} \frac{1}{\left(\sum_{j = 1}^{N_{\rm tot}} \pi_j I_{i j}\right)} \sum_{j = 1}^{N_{\rm tot}} \left(\frac{\partial^{2} \pi_j}{\partial y_{a} \partial y_z}\right) I_{i j} \, ,
\end{align}
where
\begin{equation}
    I_{i j} = \int_{z_L^j}^{z_R^j} \mathrm{d}z_i \, \int_{\alpha_L^j}^{\alpha_R^j} \mathrm{d}\alpha_i \,  p(\mathbf{\hat{f}_i} | \mathcal{T}(z_i, \alpha_i), \Sigma_i) \, .
\end{equation}
The first and second order derivatives are then evaluated to
\begin{align}
\frac{\partial \pi^{i}}{\partial y_j} = \left\{\begin{array}{ll} \pi_i (1 - \pi_i), & i = j \wedge i < N_D \\
         -\pi_i \pi_j, & i \neq j \wedge i < N_D \\
    -\frac{\pi_j}{1 + \sum_{z = 1}^{D-1} \exp{y_z}}, & i = N_D\end{array}\right.
\end{align}
and 
\begin{align}
    \frac{\partial^2 \pi^{i}}{\partial y_\alpha \partial y_j} = \left\{\begin{array}{ll} \frac{\partial \pi_i}{\partial y_\alpha} - 2 \pi_i \frac{\partial \pi_i}{\partial y_\alpha}, & i = j \wedge i < N_D \\
         -\frac{\partial \pi_i}{\partial y_\alpha} \pi_j - \pi_i \frac{\partial \pi_j}{\partial y_\alpha}, & i \neq j \wedge i < N_D \\
    \frac{\pi_j \pi_\alpha - \frac{\partial \pi_j}{\partial y_\alpha}}{1 + \sum_{z = 1}^{D-1} \exp{y_z}}, & i = N_D\end{array}\right.
\end{align}
Transformed into probability, or simplex, space, this posterior is then identified as a logit-normal distribution 
\begin{align}
    &p(\boldsymbol{\pi} | \mathbf{\hat{F}}) \approx \frac{1}{\sqrt{|2 \pi \boldsymbol{\Sigma}_{\mathbf{y}}|}}  \frac{1}{\prod_{i = 1}^{N_{\rm bins}} \pi_i}\\ 
    &\exp{\left(-0.5 \left(\log{\left(\frac{\boldsymbol{\pi}_{\rm -N_{\rm bins}}}{\pi_{N_{\rm bins}}}\right)} - \boldsymbol{\mu}_{\rm y, ML}\right) \boldsymbol{\Sigma}_{\mathbf{y}}^{-1} \left(\log{\left(\frac{\boldsymbol{\pi}_{\rm -N_{\rm bins}}}{\pi_{N_{\rm bins}}}\right)} - \boldsymbol{\mu}_{\rm y, ML}\right)\right)} \, .
\end{align}
We note that the logit-normal is a probability distribution on the simplex, just as the Dirichlet. In fact, the Dirichlet can be approximated well by a logit-normal \citep{10.1093/biomet/67.2.261}. However the logit-normal allows for a more complex covariance structure.
\markus{The scheme developed in this appendix is applied and analysed in \S~\ref{subsec:photometric_likelihood} and \S~\ref{sec:forecast_ideal_data}. }




\bsp	
\label{lastpage}
\end{document}